\documentclass[12pt,a4paper]{article} 
\pdfoutput=1
\usepackage{graphicx}
\usepackage{etex}
\usepackage{jheppub}
\usepackage{epstopdf}
\usepackage{epsfig}
\usepackage{dcolumn}  
\usepackage{bm}    
\usepackage{amssymb} 
\usepackage{amsmath,bm}
\usepackage{amsfonts}    
\usepackage{slashed}  
\usepackage{youngtab}
\usepackage[mathscr]{euscript}
\usepackage{epsfig}
\usepackage[utf8]{inputenc}
\usepackage{tikz}
\usetikzlibrary{tikzmark,calc,,arrows,shapes,decorations.pathreplacing}
\usepackage{xcolor}
\usepackage{mathrsfs}
\usepackage{verbatim}
\usepackage{cases}
\usepackage{bbding}
\usepackage{pifont}
\usepackage{ulem}
\usepackage{dcolumn}
\usepackage{graphicx}
\usepackage{pgf}
\usepackage{pstricks}
\usepackage{pst-node}
\usepackage[all]{xy}
\usepackage[figuresright]{rotating}

\usepackage{titlesec}

\titleclass{\subsubsubsection}{straight}[\subsection]

\newcounter{subsubsubsection}[subsubsection]
\renewcommand\thesubsubsubsection{\thesubsubsection.\arabic{subsubsubsection}}

\titleformat{\subsubsubsection}
  {\normalfont\normalsize\bfseries}{\thesubsubsubsection}{1em}{}
\titlespacing*{\subsubsubsection}
{0pt}{3.25ex plus 1ex minus .2ex}{1.5ex plus .2ex}

\makeatletter
\renewcommand\paragraph{\@startsection{paragraph}{5}{\z@}%
  {3.25ex \@plus1ex \@minus.2ex}%
  {-1em}%
  {\normalfont\normalsize\bfseries}}
\renewcommand\subparagraph{\@startsection{subparagraph}{6}{\parindent}%
  {3.25ex \@plus1ex \@minus .2ex}%
  {-1em}%
  {\normalfont\normalsize\bfseries}}
\def\toclevel@subsubsubsection{4}
\def\toclevel@paragraph{5}
\def\toclevel@paragraph{6}
\def\l@subsubsubsection{\@dottedtocline{4}{7em}{4em}}
\def\l@paragraph{\@dottedtocline{5}{10em}{5em}}
\def\l@subparagraph{\@dottedtocline{6}{14em}{6em}}
\makeatother

\usepackage{young}
\usepackage{graphicx,rotating,booktabs}

\usepackage{varwidth}

\newdimen\tableauside\tableauside=1.0ex
\newdimen\tableaurule\tableaurule=0.4pt
\newdimen\tableaustep
\def\phantomhrule#1{\hbox{\vbox to0pt{\hrule height\tableaurule width#1\vss}}}
\def\phantomvrule#1{\vbox{\hbox to0pt{\vrule width\tableaurule height#1\hss}}}
\def\sqr{\vbox{%
		\phantomhrule\tableaustep
		\hbox{\phantomvrule\tableaustep\kern\tableaustep\phantomvrule\tableaustep}%
		\hbox{\vbox{\phantomhrule\tableauside}\kern-\tableaurule}}}
\def\squares#1{\hbox{\count0=#1\noindent\loop\sqr
		\advance\count0 by-1 \ifnum\count0>0\repeat}}
\def\tableau#1{\vcenter{\offinterlineskip
		\tableaustep=\tableauside\advance\tableaustep by-\tableaurule
		\kern\normallineskip\hbox
		{\kern\normallineskip\vbox
			{\gettableau#1 0 }%
			\kern\normallineskip\kern\tableaurule}%
		\kern\normallineskip\kern\tableaurule}}
\def\gettableau#1 {\ifnum#1=0\let\next=\null\else
	{{\tiny\yng(1)}}s{#1}\let\next=\gettableau\fi\next}

\renewcommand{\(}{\left(}
\renewcommand{\)}{\right)}

\newcommand{\includeCroppedPdf}[2][]{%
    \IfFileExists{./#2-crop.pdf}{}{%
        \immediate\write18{pdfcrop #2 #2-crop.pdf}}%
    \includegraphics[#1]{#2-crop.pdf}}

\newcommand{\be}{ \begin{equation}}
\newcommand{\ee}{\end{equation}}
\newcommand{\bea}[1]{\begin{eqnarray}\label{#1} }
\newcommand{\eea}{\end{eqnarray}}

\def\ZZZ{{\hskip-3pt\hbox{ Z\kern-1.6mm Z}}}
\def\zzz{{\hskip-3pt\hbox{ z\kern-1mm z}}}

\newcommand{\gl}{\mu}

\def\bal#1\eal{\begin{align}#1\end{align}}

\renewcommand{\(}{\left(}
\renewcommand{\)}{\right)}

\def\one{{\hbox{ 1\kern-.8mm l}}}
\def\zero{{\hbox{ 0\kern-1.5mm 0}}}

\def\x{r}
\def\y{s}
\def\P{{\bf Q}}
\def\gl2{{$\mathfrak{gl}_2$}}

\def\e{\,{\rm e}}

\def\gl2{{$\mathfrak{gl}_2$}}
\def\sl2{{$\mathfrak{sl}_2$}}

\def\>{\rangle}
\def\<{\langle}

\title{Gluing affine Yangians with bi-fundamentals
}

\author{Wei Li} 
\affiliation{$^a$ Institute of Theoretical Physics, Chinese Academy of Sciences\\
\hspace*{0.3cm}100190 Beijing, P.R.\ China}

\emailAdd{weili@mail.itp.ac.cn}

\abstract{The affine Yangian of $\mathfrak{gl}_1$ is isomorphic to the universal enveloping algebra of $\mathcal{W}_{1+\infty}$ and can serve as a building block in the construction of new vertex operator algebras.
In \cite{Li:2019nna}, a two-parameter family generalization of $\mathcal{N}=2$ supersymmetric $\mathcal{W}_{\infty}$ algebra was constructed by ``gluing" two 
affine Yangians of $\mathfrak{gl}_1$ using operators that transform as $(\square, \overline{\square})$ and $(\overline{\square}, \square)$ w.r.t.\ the two affine Yangians. 
In this paper we realize a similar (but non-isomorphic) two-parameter gluing construction where the gluing operators transform as $(\square, \square)$ and $(\overline{\square}, \overline{\square})$ w.r.t.\ the two affine Yangians. 
The corresponding representation space consists of pairs of plane partitions connected by a common leg whose cross-section takes the shape of Young diagrams, offering a more transparent geometric picture than the previous construction. }

\begin{document}

\normalem

\setcounter{tocdepth}{2}
\maketitle

\makeatletter
\g@addto@macro\bfseries{\boldmath}
\makeatother

\section{Introduction}

The affine Yangian of $\mathfrak{gl}_1$ is isomorphic to the universal enveloping algebra of the $\mathcal{W}_{1+\infty}[\lambda]$ algebra
\begin{equation}\label{iso}
\textrm{affine Yangian of } \mathfrak{gl}_1 = \textrm{UEA}(\mathcal{W}_{1+\infty}[\lambda])\,,
\end{equation}
where the $\mathcal{W}_{1+\infty}[\lambda]$ algebra is the vertex operator algebra that is generated by one field per spin for $s=1,2,\cdots,\infty$.
This rather non-trivial isomorphism --- between a Yangian algebra and a higher spin algebra --- has been useful along both directions in various contexts. 
All these applications are based on the fact that the affine Yangian of $\mathfrak{gl}_1$ has a very natural representation in terms of plane partitions. 

The affine Yangian of $\mathfrak{gl}_1$ is constructed by \cite{SV} and \cite{Maulik:2012wi} in an attempt to prove the AGT correspondence \cite{Alday:2009aq} and its higher spin generalization \cite{Wyllard:2009hg}, i.e.\ the correspondence between the Nekrasov instanton partition function of 4D $\mathcal{N}=2$ $\mathfrak{su}(N)$ gauge theory and the conformal blocks of $\mathcal{W}_N$ algebra. 
The crucial link is the action of the affine Yangian of $\mathfrak{gl}_1$ on the intersection cohomology of the instanton moduli space, which is spanned by $N$-tuple Young diagrams \cite{SV}. These are in turn isomorphic to plane partition representations, see e.g.\ \cite{Gaberdiel:2017dbk}.
The correspondence \cite{Alday:2009aq} then follows due to the isomorphism.

On the other hand, the $\mathcal{W}_{\infty}$ algebra and its finite truncation $\mathcal{W}_{N}$ algebra appear 
in many interesting 2D CFTs, e.g.\ as the symmetry algebra of 
the CFT dual \cite{Gaberdiel:2010pz} of Vasiliev higher spin gravity in AdS$_3$ \cite{Vasiliev:1999ba}, and as the subalgebra of the CFT dual of string theory in AdS$_3$ in the tensionless limit \cite{Gaberdiel:2014cha}. 
By the isomorphism (\ref{iso}), one can also use plane partitions to characterize representations of these $\mathcal{W}$ algebras.
This has both conceptual and computational advantages over the traditional coset representations, see e.g.\ \cite{Datta:2016cmw}.
\bigskip

Given the usefulness of the isomorphism (\ref{iso}) explained above, we would like to generalize it. 
There are two directions. 
The first is to find similar isomorphisms for known affine Yangians (l.h.s.) or for known VOAs with higher spin fields (r.h.s.).
The second is to have a more systematic method to construct isomorphisms between new affine Yangians and new VOAs.
As we will see, useful for both is the construction that ``glues" multiple copies of (\ref{iso}). 

Along the first direction, one obvious question is how to supersymmetrize the isomorphism (\ref{iso}), i.e.\ to construct affine Yangian type algebras isomorphic to supersymmetric $\mathcal{W}_{\infty}$ algebras.
The affine Yangian algebra that is isomorphic to the UEA of the $\mathcal{N}=2$ supersymmetric $\mathcal{W}_{\infty}$ algebra (tensored with a $\mathfrak{u}(1)$ factor) can be constructed by gluing two copies of affine Yangian of $\mathfrak{gl}_1$, $\mathcal{Y}$ and $\hat{\mathcal{Y}}$, using additional operators that transform as 
$(\square, \overline{\square})$ and $(\overline{\square}, {\square})$ w.r.t.\ $\mathcal{Y}$ and $\hat{\mathcal{Y}}$ \cite{Gaberdiel:2017hcn, Gaberdiel:2018nbs}.

The corresponding representation is a pair of plane partitions (one for each copy of affine Yangian of $\mathfrak{gl}_1$) connected by a shared ``internal" direction.
The additional gluing operators create/annihilate building blocks of non-trivial states along the internal leg, whose 
cross-sections are either $\square$ at the end of $\mathcal{Y}$ and $\overline{\square}$ at the end of $\hat{\mathcal{Y}}$, or vice versa.
The resulting representation space of the glued algebra is a pair of plane partitions whose asymptotics along the shared directions are correlated as $[\textrm{R},\bar{\textrm{R}}]$ or $[\textrm{R},\bar{\textrm{R}}^t]$ --- called twin plane partitions in \cite{Li:2019nna,Gaberdiel:2018nbs}.
(Here $\textrm{R}$ is a Young diagram, $\bar{\textrm{R}}$ is its conjugate, and $\textrm{R}^t$ is its transpose. The shapes of these Young diagrams give the asymptotics of the plane partitions along the ``shared" direction.)
All the relations of the glued algebra can then be determined by studying the action of the algebra on these twin plane partitions \cite{Gaberdiel:2018nbs}.  
One can then check that it reproduces the correct $\mathcal{W}^{\mathcal{N}=2}$ charges.
\medskip

In \cite{Gaberdiel:2017hcn, Gaberdiel:2018nbs}, the input from the $\mathcal{W}^{\mathcal{N}=2}_{\infty}$ algebra (in particular, its decomposition in terms of two $\mathcal{W}_{1+\infty}$ subalgebras) inspired the gluing construction and helped fix the relevant parameters. 
However, the gluing construction itself, in particular, the procedure of fixing all the algebraic relations using constraints from their actions on the pair of plane partitions, is independent of the $\mathcal{W}^{\mathcal{N}=2}_{\infty}$ algebra and can be used to construct isomorphisms between new affine Yangians and new VOAs.

In particular, the construction for the affine Yangian corresponding to $\mathcal{W}^{\mathcal{N}=2}_{\infty}$ algebra has four main ingredients \cite{Gaberdiel:2017hcn, Gaberdiel:2018nbs}: 
\begin{enumerate}
\item It has two copies of  affine Yangians of $\mathfrak{gl}_1$ --- $\mathcal{Y}$ and $\hat{\mathcal{Y}}$.
\item The gluing operators transform as bimodules $[\square, \overline{\square}]$ and $[\overline{\square}, {\square}]$ w.r.t.\ $\mathcal{Y}$ and $\hat{\mathcal{Y}}$.
\item The parameters of $\mathcal{Y}$ and $\hat{\mathcal{Y}}$ are related by $
h_i=\hat{h}_i$.
\item The conformal dimension of the gluing operators are $\frac{3}{2}$, corresponding to that of the supercharges $G^{\pm}$.
\end{enumerate}
Note the hierarchy $(1)\rightarrow (2)\rightarrow (3)$ of the three items above, e.g.\ changing (1) would necessitate the change of (2) and (3).
On the other hand, (4) is independent from (2) and (3).

To use the gluing procedure of \cite{Gaberdiel:2017hcn, Gaberdiel:2018nbs} to construct isomorphisms between new affine Yangians and new VOAs, one simply modified these ingredients while retaining the essential feature of the procedure, i.e. fixing the glued algebra using constraints from its action on the corresponding set of plane partition type representations.

In \cite{Li:2019nna}, the ingredients (3) and (4) were relaxed.
The goal was to construct the most general glued algebras under the condition (1) and (2) using the gluing procedure of \cite{Gaberdiel:2017hcn, Gaberdiel:2018nbs}.
The result is a two-parameter generalization of the algebra obtained in  \cite{Gaberdiel:2017hcn, Gaberdiel:2018nbs}. 
(The corresponding VOAs are a two-parameter generalization of the $\mathcal{W}^{\mathcal{N}=2}_{\infty}$ algebra.)
In particular, for (3), the relation between the parameters of $\mathcal{Y} $ and $\hat{\mathcal{Y}}$ now falls into three cases;
and for (4), the conformal dimension of the gluing operators can now take all positive integer or half-integer values. 
Note that these constraints follow purely from the geometry of the twin plane partitions. 
We will call the constructions in \cite{Gaberdiel:2017hcn, Gaberdiel:2018nbs, Li:2019nna} box-antibox constructions.
\bigskip

In this paper, we would like to further relax item (2) in the list, namely, to consider gluing operators that transform as other types of bimodules of the two affine Yangians of $\mathfrak{gl}_1$.
The most obvious generalizations are gluing operators that transform as $[\square, \square]$ and $[\overline{\square}, \overline{\square}]$ w.r.t.\ $\mathcal{Y}$ and $\hat{\mathcal{Y}}$. 

We will first construct the algebra glued by operators that transform as $[\square, \square]$ w.r.t.\ $\mathcal{Y}$ and $\hat{\mathcal{Y}}$. 
The main motivation is that its corresponding plane partition representations are much easier to visualize than the one in \cite{Gaberdiel:2018nbs, Li:2019nna}.
As shown in \cite{Gaberdiel:2018nbs}, in term of the plane partitions, the $[\square,\overline{\square}]$ gluing operator  creates (on vacuum) an internal leg that looks like a long row of single boxes from the view point of $\mathcal{Y}$ and a ``high wall" from the view point of $\hat{\mathcal{Y}}$. 
This makes further development of the glued algebra, in particular the matching of the microscopic action of the glued algebra to some physical systems, a bit unwieldy.
For example,  when trying to connect to topological strings, it is a bit unnatural to interpret the pairs of plane partitions with ``walls" as the complements of melting crystals.

In the current new construction, the $[\square,\square]$ gluing operator  simply creates an internal leg that looks like a long row of boxes from both left and right plane partitions, namely, there is no need for the rows of boxes to undergo the transition to the ``higher walls" somewhere along the internal leg, as one moves from the left plane partition to the right one.
The representations look precisely like (the complement of) a crystal with two corners and the connecting edge partially melted.
 
The second motivation is that the new choice $[\square,\square]$ allows parameters that are outside the construction of \cite{Li:2019nna}.
In fact, the change in item (2) necessitates the changes in item (3).
We will see that the relations between the parameters in the two affine Yangians of $\mathfrak{gl}_1$ again fall into three cases, but different from those in \cite{Li:2019nna}.
This also allows more possibilities when trying to match the glued algebra with known algebras.
For example, if one wants to reproduce the affine Yangian of $\mathfrak{gl}_2$ by gluing two affine Yangians of $\mathfrak{gl}_1$, the parameters between the two $\mathcal{Y}$'s satisfy  one of the three relations dictated by the $[\square,\square]$ construction, not the one in the box-antibox construction.
We emphasize that the algebras constructed in this paper are not isomorphic to any of the algebras constructed in  \cite{Gaberdiel:2017hcn, Gaberdiel:2018nbs, Li:2019nna}.

We then apply the same procedure of \cite{Li:2019nna} and construct the most general family of glued algebras with the $[\square,\square]$ gluing operator.
The main content of this paper is the derivation of individual relations based on their actions on pairs of plane partitions glued by $[\square, \square]$ operators. 
Although the procedure follows closely \cite{Li:2019nna}, we see lots of difference in the detailed mechanisms. And certain aspects are simpler than the counterparts in the box-antibox construction in \cite{Li:2019nna}. 

Then we show that one can extend the algebra glued by $[\square,\square]$ operators with additional gluing operators that transform as $[\overline{\square}, \overline{\square}]$ w.r.t.\ the two affine Yangians of $\mathfrak{gl}_1$.
This can be achieved without introducing ``high walls" connecting the two plane partitions, but by allowing the 
$[\overline{\square}, \overline{\square}]$ gluing operators to act  appropriately on the pairs of plane partitions glued by $[\square, \square]$ operators.
The extended algebra can then be determined by analyzing their actions on these pairs of plane partitions.   
Finally, we explain the relation between the algebras constructed in this paper with those in the box-antibox construction in \cite{Li:2019nna}.
\bigskip

The plan of this paper is as follows. 
Section 2 is a review on the affine Yangian of $\mathfrak{gl}_1$, its plane partition representations, and the box-antibox gluing construction. 
In section 3, we explain the ingredients in the gluing construction using $[\square,\square]$ operators, in particular, the relevant plane partition type representations and their charge functions. 
In section 4 we use the ingredients in section 3 to fix the algebra glued by $[\square,\square]$ operators.
In section 5 we extend the algebra constructed in section 4 by additional operators that transform as $[\overline{\square},\overline{\square}]$ w.r.t.\ the two affine Yangians of $\mathfrak{gl}_1$.
The final section explains the relation between the algebras constructed in this paper with those in the box-antibox construction in \cite{Li:2019nna} and discusses future problems.
Appendix A contains long computations used in section 5.

\section{Review on gluing of affine Yangians of $\mathfrak{gl}_1$}
\label{sec:review}

\subsection{Affine Yangian of $\mathfrak{gl}_1$, $\mathcal{W}_{1+\infty}$ algebra, and plane partitions}

\subsubsection{Affine Yangian of $\mathfrak{gl}_1$}

The affine Yangian of $\mathfrak{gl}_1$ is an associative algebra defined by the following OPE-like equations
\begin{equation}\label{bosonicdef}
\begin{aligned}
\begin{aligned}
\psi(z)\, e(w) &  \sim  \varphi_3(\Delta)\, e(w)\, \psi(z)\,, \\ 
\psi(z)\, f(w) & \sim \varphi_3^{-1}(\Delta)\, f(w)\, \psi(z) \,,
\end{aligned}
&\qquad
\begin{aligned}
e(z)\, e(w) & \sim    \varphi_3(\Delta)\, e(w)\, e(z)\,, \\
 f(z)\, f(w) &  \sim    \varphi_3^{-1}(\Delta)\, f(w)\, f(z) \,,
\end{aligned}\\
[e(z)\,, f(w)]  & \sim  - \frac{1}{h_1h_2h_3}\, \frac{\psi(z) - \psi(w)}{z-w} \ , 
\end{aligned}
\end{equation}
where the three fields $\{e,\psi,f\}$ have the mode expansions
\begin{equation}\label{generating}
e(z) = \sum_{j=0}^{\infty} \, \frac{e_j}{z^{j+1}} \ , \qquad 
f(z) = \sum_{j=0}^{\infty} \, \frac{f_j}{z^{j+1}} \ , \qquad 
\psi(z)  = 1 + h_1h_2h_3 \, \sum_{j=0}^{\infty} \frac{\psi_j}{z^{j+1}} \ ;
\end{equation}
and throughout this paper
\begin{equation}
\Delta\equiv z-w\,,
\end{equation}
and $\varphi_3(u)$ is a cubic rational function 
\begin{equation}\label{varphidef}
\varphi_3(u) \equiv \frac{(u+h_1) (u+h_2) (u+h_3)}{(u-h_1) (u-h_2) (u-h_3)}   \ ,
\end{equation}
where the triplet $(h_1,h_2,h_3)$ obeys
\begin{equation}\label{sumh}
h_1+h_2+h_3=0 \ .
\end{equation}
Since the algebra (\ref{bosonicdef}) is invariant under permutation of $(h_1, h_2,h_3)$, sometimes it is convenient to use the $\mathcal{S}_3$ invariant parameters 
\begin{equation}
\sigma_2 \equiv h_1 h_2+h_2 h_3 +h_3 h_1 \qquad \textrm{and}\qquad \sigma_3 \equiv h_1 h_2 h_3 \,.
\end{equation}
The $\sim$ sign in (\ref{bosonicdef}) and throughout the paper means identity up to terms of $z^m w^n$ with $m,n \geq0$.
Finally, the relations (\ref{bosonicdef}) need to be supplemented by the initial conditions
\begin{equation}\label{initial}
\begin{aligned}
&[\psi_0,e_r] = 0 \ , \qquad [\psi_1,e_r] = 0 \ , \qquad [\psi_2,e_r] = 2\, e_r \ , \\
&[\psi_0,f_r] = 0 \ , \qquad [\psi_1,f_r] = 0 \ , \qquad [\psi_2,f_r] = - 2\, f_r \ ;
\end{aligned}
\end{equation}
and the Serre relations
\begin{equation}\label{Serre}
\begin{aligned}
&\sum_{\pi \in \mathcal{S}_3}\, \bigl(z_{\pi(1)} - 2 z_{\pi(2)} + z_{\pi(3)} \bigr)\, e(z_{\pi(1)})\, e(z_{\pi(2)})\, e(z_{\pi(3)}) \sim 0 \,,\\
&\sum_{\pi \in 
\mathcal{S}_3}\, \bigl(z_{\pi(1)} - 2 z_{\pi(2)} + z_{\pi(3)} \bigr)\, f(z_{\pi(1)})\, f(z_{\pi(2)})\, f(z_{\pi(3)}) \sim 0 \ .
\end{aligned}
\end{equation}

The relations (\ref{bosonicdef}) can be translated in terms of modes in the following way.
Take the first equation for example. 
Moving the denominator of $\varphi_3(\Delta)$ to the l.h.s., we have
\begin{equation}\label{psie}
( \Delta^3 + \sigma_2 \, \Delta - \sigma_3 )  \, \psi(z)\, e(w) \sim ( \Delta ^3 + \sigma_2 \, \Delta + \sigma_3 ) \, e(w) \, \psi(z) \ , 
\end{equation}
which gives
\begin{eqnarray}
\sigma_3 \{e_j,e_k \}  & = & [e_{j+3},e_k] - 3  [e_{j+2},e_{k+1}] + 3  [e_{j+1},e_{k+2}]  - [e_{j},e_{k+3}]   \nonumber \\
& & \ + \sigma_2 ( [e_{j+1},e_{k}] -   [e_{j},e_{k+1}] ) \ . \label{Y1} 
\end{eqnarray}
The other equations in (\ref{bosonicdef}) and the Serre relations (\ref{Serre}) can be translated in a similar way. For the complete list, see e.g.\ \cite{Prochazka:2015deb, Gaberdiel:2017dbk}.
\medskip

Note that although $\{e(z),\psi(z),f(z)\}$ are not holomorphic fields in a 2D CFT, the relations (\ref{bosonicdef}) are similar to OPE relations because (1) they are only defined up to regular terms, and (2) they are merely rewritings of the corresponding mode relations. Therefore, throughout this paper, we will call relations like (\ref{bosonicdef}) OPE relations, to distinguish them from their corresponding equations in terms of modes. 
They have the form
\begin{equation}\label{exampleab}
a(z) \, b(w)\sim \frac{N(\Delta)}{D(\Delta)} \,b(w) \, a(z) \,, 
\end{equation}
in which the numerator $N(\Delta)$ and the denominator $D(\Delta)$ might share some common factor $f(\Delta)$.\footnote{
In the affine Yangian of $\mathfrak{gl}_1$ (\ref{bosonicdef}), this doesn't happen; but we see examples of this in \cite{Gaberdiel:2017hcn,Gaberdiel:2018nbs,Li:2019nna} and later in this paper.}
We emphasize that one shouldn't cancel these $f(\Delta)$ factors, since  (\ref{exampleab}) should be understood as 
\begin{equation}\label{exampleab2}
D(\Delta)\, a(z)\, b(w) \sim N(\Delta) \, b(w)\, a(z)\,,
\end{equation}
which can then be translated into an equation in terms of modes $a_j$ and $b_j$. 
Removing a common factor $f(\Delta)$ (if present) from $D(\Delta)$ on l.h.s.\ and $N(\Delta)$ on r.h.s.\ of (\ref{exampleab2}) would change the corresponding relation in terms of modes. 
\medskip

It is easy to check that the affine Yangian algebra (\ref{bosonicdef}) has an automorphism
\begin{equation}\label{auto}
\begin{aligned}
&h_i \rightarrow \alpha\, h_i\,, \quad
u\rightarrow  \alpha \, u \,,\quad
\psi(u)\rightarrow \psi(u)\,,\quad e(u)\rightarrow \alpha^{-2}\, e(u)\,, \quad f(u)\rightarrow \alpha^{-2}\, f(u)\,.
\end{aligned}
\end{equation}
In terms of the modes, this is
\begin{equation}\label{automodes}
\begin{aligned}
&h_i \rightarrow \alpha\, h_i\,, \quad
&\psi_j \rightarrow \alpha^{j-2} \psi_j \,,\quad e_j \rightarrow \alpha^{j-1} e_j \,,\quad f_j \rightarrow \alpha^{j-1} f_j \,.
\end{aligned}
\end{equation}
In addition, there is an automorphism induced by the spectral shift $u\rightarrow u+a$ in $\{e(u),\psi(u),f(u)\}$.
In particular, under this shift,
\begin{equation}\label{spectral}
\psi_0\rightarrow \psi_0 \qquad\textrm{and} \qquad \psi_1\rightarrow \psi_1-a \psi_0\,.
\end{equation}

From the initial condition (\ref{initial}) and the (mode version of the) defining relations (\ref{bosonicdef}), we see that the affine Yangian of $\mathfrak{gl}_1$ has two central elements $\psi_0$ and $\psi_1$, where $\psi_1$ is subject to the spectral shift automorphism (\ref{spectral}).
Therefore one can label an affine Yangian of $\mathfrak{gl}_1$ by $(h_1,h_2,h_3,\psi_0)$ subject to (\ref{sumh}) and the automorphism
\begin{equation}\label{automode01}
h_i\rightarrow \alpha h_i \qquad\textrm{and}\qquad \psi_0\rightarrow \alpha^{-2} \psi_0 \,,
\end{equation}
which follows from (\ref{automodes}).
Namely, there are two independent parameters.

\subsubsection{$\mathcal{W}_{1+\infty}$ algebra}

The affine Yangian of $\mathfrak{gl}_1$ is isomorphic to the universal enveloping algebra of $\mathcal{W}_{1+\infty}[\lambda]$ algebra:
\begin{equation}\label{iso1}
\textrm{affine Yangian of } \mathfrak{gl}_1 = \textrm{UEA}(\mathcal{W}_{1+\infty}[\lambda])\,,
\end{equation}
where the $\mathcal{W}_{1+\infty}[\lambda]$ algebra is the vertex operator algebra that is generated by one field per spin for $s=1,2,\cdots,\infty$ and is labeled by two parameters: the central charge $c$ and the 't Hooft coupling $\lambda$.
These should be mapped to the two parameters of the affine Yangian of $\mathfrak{gl}_1$, i.e.\ $(h_1, h_2, h_3, \psi_0)$ subject to (\ref{sumh}) and (\ref{automode01}).

The isomorphism (\ref{iso1}) was suggested in \cite{Prochazka:2015deb} and proven in \cite{Gaberdiel:2017dbk}.
Theoretically, one should be able to derive the isomorphism (\ref{iso1}) as a classical (i.e.\ $q\rightarrow 1$) limit of the isomorphism between the quantum toroidal $\mathfrak{gl}_1$ and the $q$-deformed $\mathcal{W}_{1+\infty}[\lambda]$ algebra \cite{Miki, Feigin:2010qea}.
However, since the detailed map from the quantum toroidal $\mathfrak{gl}_1$ to the affine Yangian of $\mathfrak{gl}_1$ is rather involved (see \cite{Tsymbaliuk:2014fvq}), one can simply prove (\ref{iso1}) directly \cite{Gaberdiel:2017dbk}.

Recall that the $\mathcal{W}_{1+\infty}[\lambda]$ algebra above has a very rigid structure: once the OPE relations for the lower spin fields (up to OPEs between two spin-three fields) are known, all the OPEs with higher spin fields are fixed by Jacobi identities \cite{Gaberdiel:2012ku, Linshaw:2017tvv}. 
Therefore one only needs to know the map (\ref{iso1}) up to spin-$4$ field. 
Moreover, for each spin $s$, we only need to fix $W^{(s)}_{m}$ with $|m|\leq s$ to deduce $W^{(s)}_m$ for all $m\in \mathbb{Z}$ using the $\mathcal{W}_{1+\infty}$ algebraic relations.
To summarize, we only need to know the $W^{(s)}_{m}$ for $1\leq s \leq 4$ and $|m|\leq s$ in terms of $(e_j,\psi_j, f_i)$ modes.
They are
\begin{equation}\label{Wepsif}
W^{(s)}_{-1} \sim e_{s-1} \qquad \qquad W^{(s)}_{0} \sim \psi_{s} \qquad  \qquad W^{(s)}_{1} \sim f_{s-1} \ , 
\end{equation}
for $W^{(s)}_m$ with $m=-1,0,1$, where we have omitted the lower terms. 
The modes $W^{(s)}_m$ with $|m|\geq 2$ are composed of products of $(e_j,\psi_j, f_j)$ modes. For details see \cite{Gaberdiel:2017dbk}.

The map for the modes $W^{(s)}_{m}$ for $1\leq s \leq 4$ and $|m|\leq s$ in terms of $(e_j,\psi_j, f_i)$ modes then allows us to fix the map between the parameters of the two algebras  \cite{Gaberdiel:2017dbk}:
\begin{equation}\label{h123}
h_1 =  -\sqrt{\frac{N+k+1}{N+k}} \ , \quad h_2 =  \sqrt{\frac{N+k}{N+k+1}} \ , \quad h_3 = \frac{1}{\sqrt{(N+k)(N+k+1)}} \ , 
\end{equation}
and
\begin{equation}\label{psi0}
\psi_0 = N  \ ,
\end{equation}
in terms of the coset parameter $(N,k)$.\footnote{See \cite{Prochazka:2017qum} for a more elegant, $\mathcal{S}_3$ invariant, map.}
Note that one can still applies the automorphism (\ref{automode01}) on (\ref{h123}) and (\ref{psi0}).

\subsubsection{Plane partition representations}

The affine Yangian of $\mathfrak{gl}_1$ has a natural representation in terms of the plane partitions $\{\Lambda\}$:
\begin{equation}\label{Yaction}
\begin{aligned}
\psi(z)|\Lambda \rangle & = \Psi_{\Lambda}(z)|\Lambda \rangle  \ ,\\
e(z) | \Lambda \rangle & =  \sum_{ \square \in {\rm Add}(\Lambda)}\frac{\Big[ -  \frac{1}{\sigma_3} {\rm Res}_{w = h(\square)} \Psi_{\Lambda}(w) \Big]^{\frac{1}{2}}}{ z - h(\square) } 
| \Lambda + \square \rangle \ , \\
f(z) | \Lambda \rangle & =  \sum_{ \square \in {\rm Rem}(\Lambda)}\frac{\Big[ +  \frac{1}{\sigma_3} {\rm Res}_{w = h(\square)} \Psi_{\Lambda}(w) \Big]^{\frac{1}{2}}}{ z - h(\square) } | \Lambda - \square \rangle \ . 
\end{aligned}
\end{equation}
Each plane partition is an eigenfunction of the operator $\psi(u)$, whose eigenvalue $\Psi_{\Lambda}(u)$ (called charge function here) receives a contribution from each box\footnote{
Note that in this paper, $\square$ is used to denote both a one-box Young diagram (when describing the representations of the affine Yangian of $\mathfrak{gl}_1$) and a 3D box (as the building block of the plane partitions).
The meaning should be clear from the context.
} $\square$ in the plane partition $\Lambda$:
\begin{equation}\label{psieig}
\Psi_\Lambda(z) \equiv \psi_0(z) \, \prod_{ \square \in \Lambda} \varphi_3(z - h(\square) ) \ , 
\end{equation}
where 
\begin{equation}\label{psi0def}
\psi_0(z)\equiv 1 + \frac{\psi_0 \sigma_3}{z} 
\end{equation}
is the vacuum contribution, $\varphi_3$ is the cubic rational function defined in (\ref{varphidef}), and finally 
\begin{equation}\label{hbox}
h(\square) \equiv h_1 x_1(\square) + h_2 x_2(\square) +  h_3 x_3(\square) 
\end{equation}
is the coordinate function.
Note that knowing (\ref{Yaction}) and (\ref{psieig}) would allow us to reproduce the algebra (\ref{bosonicdef}).
\medskip

A plane partition representation is labeled by three Young diagrams $(\lambda_1,\lambda_2,\lambda_3)$, which are the asymptotic shapes of the plane partition along the three directions $(x_1,x_2,x_3)$.
The ground state of the representation $(\lambda_1,\lambda_2,\lambda_3)$ is the minimal plane partition configuration with the given asymptotics $(\lambda_1,\lambda_2,\lambda_3)$; and its descendants are those plane partitions with additional boxes, created by the operator $e(u)$.
The simplest representation is the vacuum module, with $(\lambda_1,\lambda_2,\lambda_3)=(0,0,0)$.

For all plane partition representations, the character of affine Yangian of $\mathfrak{gl}_1$ is given by the generating function of the plane partition configurations with the given asymptotics $(\lambda_1,\lambda_2,\lambda_3)$. 
For example, the character of the vacuum representation is the MacMahon function. 
\medskip

By the isomorphism (\ref{iso1}), the set of plane partitions also serve as representations for the $\mathcal{W}_{1+\infty}$ algebra. 
First of all, the traditional coset representation can be translated into plane partitions.
In terms of the coset realization $\frac{\mathfrak{su}(N)_{k}\oplus \mathfrak{su}(N)_1}{\mathfrak{su}(N)_{k+1}}$ (tensored with an additional $\mathfrak{u}(1)$), the representations of $\mathcal{W}_{1+\infty}$ algebra are labeled by two Young diagrams $(\nu_{+},\nu_{-})$, where $\nu_+$ transforms under the $\mathfrak{su}(N)_{k}$ in the numerator and $\nu_-$ transforms under the $\mathfrak{su}(N)_{k+1}$ in the denominator. 
In terms of the plane partitions, they are the representations with asymptotics
\begin{equation}\label{WtoPP}
(\lambda_1,\lambda_2,\lambda_3)=(\nu_+,\nu_-,0)\,.
\end{equation}
Namely, the coset representations correspond to plane partition representations with a trivial asymptotics along (at least) one direction. 
Since the $\mathcal{W}_{1+\infty}[\lambda]$ algebra has an $\mathcal{S}_3$ automorphism \cite{Gaberdiel:2012ku}, one can use it to rotate $(\nu_+,\nu_-,0)$ into $(0,\nu_+,\nu_-)$ etc. 
Taking the fusion products of these representations can then produce plane partition representations with non-trivial asymptotics along all three directions.
Finally, by the isomorphism (\ref{iso1}), the $\mathcal{W}_{1+\infty}$ character of a representation is also given by the generating function of its plane partition configurations. 
\medskip

The advantage of using  the affine Yangian of $\mathfrak{gl}_1$ (i.e.\ $\mathcal{Y}$)  to describe the $\mathcal{W}_{1+\infty}$ algebra is due to the plane partition representations.
For example, the $\mathcal{W}_{1+\infty}$ characters can be easily computed via generating functions of plane partitions.
Moreover, the action of $\mathcal{Y}$ on plane partitions (\ref{Yaction}) is very easy to describe and visualize, which makes it an important tool in constructing bigger algebras with individual $\mathcal{Y}$ as building blocks.

\subsection{Gluing construction using $[\square, \overline{\square}]$ and $[\overline{\square}, {\square}]$ gluing operators}

Given the usefulness of the isomorphism (\ref{iso1}) explained above, we would like to generalize it. 
There are two directions. 
The first is to find similar isomorphisms for known affine Yangians (l.h.s.) or for known VOAs with higher spin fields (r.h.s.).
The second is to have a more systematic method to construct isomorphisms between new affine Yangians and new VOAs.
As we will see, useful in both is the construction that ``glues" multiple copies of (\ref{iso1}). 

\subsubsection{$\mathcal{N}=2$ supersymmetric $\mathcal{W}_{\infty}$ algebra}

One example in the first direction is the analogue of isomorphism (\ref{iso1}) for the $\mathcal{N}=2$ supersymmetric $\mathcal{W}_{\infty}$ algebra with the spectrum of one $\mathcal{N}=2$ multiplet $\{W^{(s)0},W^{(s)+},W^{(s)-},W^{(s)1}\}$ for each spin $s=1,2,\dots,\infty$. 
As conjectured and partially checked in \cite{Gaberdiel:2017hcn}, all the bosonic fields $\{W^{(s)0},W^{(s)1}\}$ (after adding an additional spin-one field) together form two mutually commuting $\mathcal{W}_{1+\infty}$ algebras and the fermionic fields have non-trivial OPEs with both of them:
\begin{equation}\label{N2decompW}
\begin{aligned}
\mathfrak{u}(1)\oplus \mathcal{W}^{\mathcal{N}=2}_{\infty}[c,\lambda] &= \mathcal{W}_{1+\infty}[c_1,\lambda_1]\oplus  \mathcal{W}_{1+\infty}[c_2,\lambda_2]\oplus \bigoplus^{\infty} _{s=1} W^{(s)+} \oplus \bigoplus^{\infty} _{s=1} W^{(s)-}\,.
\end{aligned}
\end{equation}

As proposed in \cite{Gaberdiel:2017hcn}, this can be translated into a decomposition in terms of the affine Yangian of $\mathfrak{gl}_1$ in the following way.
First, the two $\mathcal{W}_{1+\infty}$ subalgebras in (\ref{N2decompW}) correspond to two affine Yangians of $\mathfrak{gl}_1$, $\mathcal{Y}$ given by (\ref{bosonicdef}) and $\hat{\mathcal{Y}}$ given by the hatted version of (\ref{bosonicdef}), with their parameters denoted collectively by $q$ and $\hat{q}$, respectively:
\begin{equation}\label{N2bosonic}
\mathcal{W}_{1+\infty}[c_1,\lambda_1]\oplus  \mathcal{W}_{1+\infty}[c_2,\lambda_2] \quad \longrightarrow\quad \mathcal{Y}(q)\oplus \hat{\mathcal{Y}}(\hat{q}) \,.
\end{equation}
Using the map (\ref{h123}), one can show that 
the parameters of $\mathcal{Y}$ and $\hat{\mathcal{Y}}$ are related by 
\begin{equation}\label{YYhatparameterN2}
h_i=\hat{h}_i \qquad\textrm{and} \qquad h_1\, h_3\, \psi_0+\hat{h}_1\, \hat{h}_3\, \hat{\psi}_0=-1\,,
\end{equation}
for details see \cite{Gaberdiel:2017hcn}.

Second, the fermionic fields in (\ref{N2decompW}) transform as bimodules w.r.t.\ the two $\mathcal{W}_{1+\infty}$ algebras in (\ref{N2decompW}).
In particular, the pair of supercharges $G^{+}$ and  $G^{-}$ transform as the  bimodules $[(0,\square), (0,\overline{\square})]$ and $[(0,\overline{\square}), (0,{\square})]$, respectively \cite{Gaberdiel:2017hcn}. 
They can be translated to the plane partition representations using (\ref{WtoPP})
\begin{equation}\label{bimodule}  
[(0, {\square},0)\,\, ,\,\,  (0,{\overline{\square}},0)  ]=:[\square, \overline{\square}]
\qquad \textrm{and}\qquad 
 [(0, \overline{\square},0) \,\, ,\,\,  (0,{\square},0)]=:[\overline{\square}, {\square}] \,,
\end{equation}
respectively.
We will use the shorthand $[\square,\overline{\square}]$ and $[\overline{\square},{\square}]$ to denote these two bimodules of $\mathcal{Y}$ and $\hat{\mathcal{Y}}$.
They interact with both $\mathcal{Y}$ and $\hat{\mathcal{Y}}$ and in  this sense can ``glue" them together to form a bigger algebra.\footnote{
See appendix G of \cite{Prochazka:2017qum} for a discussion on the gluing of two $\mathcal{W}_{1+\infty}$ algebras at particular parameters to form the $\mathcal{N}=2$ supersymmetric $\mathcal{W}_3$ algebra (i.e.\ a truncation of the $\mathcal{N}=2$ supersymmetric $\mathcal{W}_{\infty}$ algebra in this paper down to spin-$3$) in terms of the $\mathcal{W}$ algebra basis.}

Finally, since all fermionic fields in (\ref{N2decompW}) are $G^{\pm}$ descendants of the bottom components $W^{(s)0}$ (which are bosonic and hence in (\ref{N2bosonic})), they can all be generated by OPEs of $G^{\pm}$ and $W$ fields in (\ref{N2bosonic}).
In the corresponding Yangian algebra, they can all be obtained by interaction of the gluing operators in  (\ref{bimodule}) and the two bosonic subalgebras (\ref{N2bosonic}). 
Namely, using (\ref{bimodule}) to glue (\ref{N2bosonic}) can produce the analogue of isomorphism (\ref{iso1}) for the $\mathcal{N}=2$ supersymmetric $\mathcal{W}_{\infty}$:
\begin{equation}\label{N2decompY}
\textrm{UEA}(\mathfrak{u}(1)\oplus \mathcal{W}^{\mathcal{N}=2}_{\infty}[c,\lambda])=\mathcal{Y}(q)\oplus \hat{\mathcal{Y}}(\hat{q}) \oplus [\square, \overline{\square}]   \oplus [\overline{\square}, {\square}]  \,.
\end{equation}

\subsubsection{Gluing operators and twin plane partitions}

We now explain how to determine the algebraic relations for the r.h.s.\ of (\ref{N2decompY}), by gluing $\mathcal{Y}$ and $\hat{\mathcal{Y}}$ using the gluing operators $[\square, \overline{\square}]$ and $[\overline{\square}, {\square}]$.
The main strategy is to use the constraints from the algebra's action on the pairs of plane partitions. 

The starting point is a pair of affine Yangians of $\mathfrak{gl}_1$, $\mathcal{Y}$ given by (\ref{bosonicdef}) and $\hat{\mathcal{Y}}$ given by the hatted version of (\ref{bosonicdef}).
The affine Yangian of $\mathfrak{gl}_1$ (\ref{bosonicdef}) has a natural representation on the set of plane partitions via (\ref{Yaction}). 
The hatted version $\hat{\mathcal{Y}}$ acts on an independent set of plane partitions.
To distinguish the two, we use $\square$ to denote the boxes in the plane partition acted upon by $\mathcal{Y}$ (also called ``left" plane partition here), and use $\hat{\square}$ for the boxes in the ``right" plane partitions, acted upon by $\hat{\mathcal{Y}}$:
\begin{equation}\label{N2BBleftreview}
\begin{aligned}
&{{\square}}: \qquad \qquad e:\, \textrm{creation} \qquad  \psi:\, \textrm{charge}  \qquad  f:\, \textrm{annihilation} \,,\\
&\hat{{{\square}}}: \qquad \qquad \hat{e}:\, \textrm{creation} \qquad  \hat{\psi}:\, \textrm{charge}  \qquad  \hat{f}:\, \textrm{annihilation} \,.
\end{aligned}
\end{equation}

\begin{figure}[h!]
\begin{center}
\includegraphics[width=0.8\textwidth]{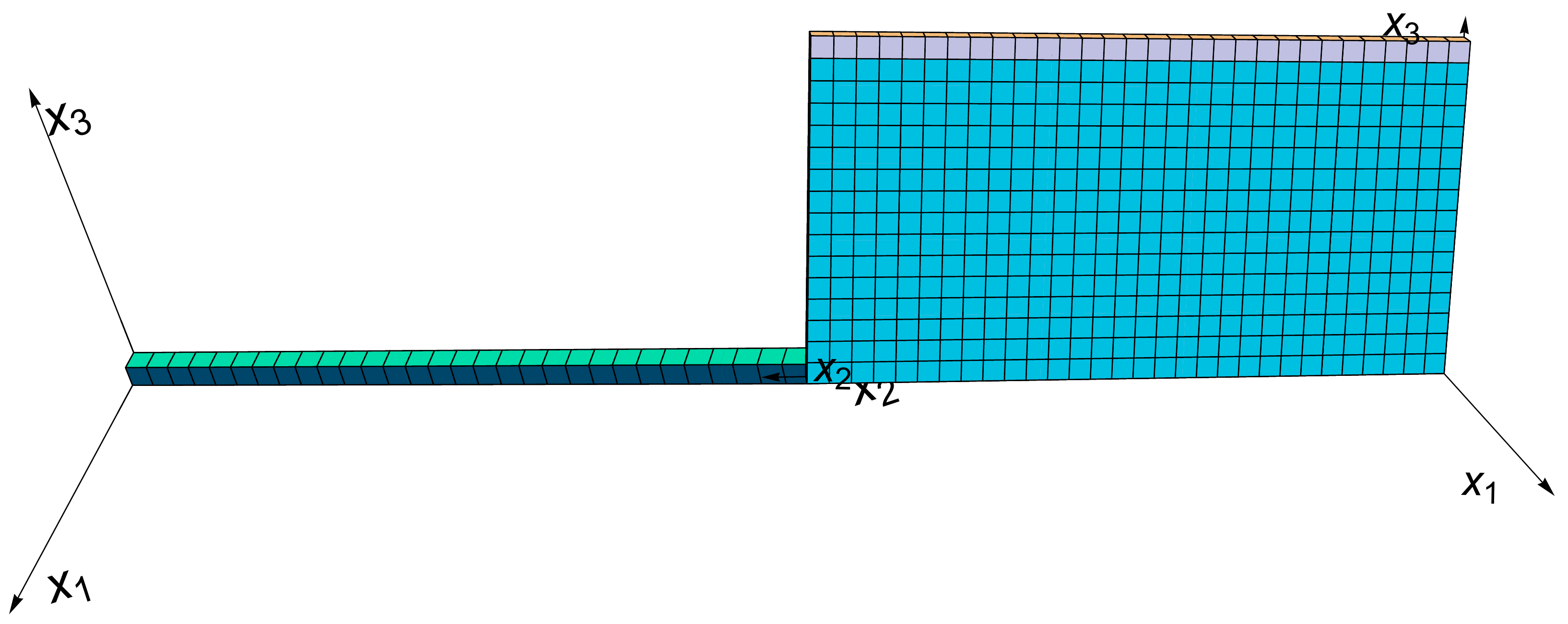}
\caption{The plane partition configuration of the building block $\blacksquare$ in the box-antibox construction. 
Note that the internal leg (along the $x_2\sim \hat{x}_2$ direction) is infinitely long, which makes it possible for the same state to transition from a row of boxes on the left to a high wall on the right.}
\label{fig:blacksquarebab}
\end{center}
\end{figure}

Denote the two configurations in (\ref{bimodule}) as $\blacksquare$ and $\overline{\blacksquare}$ respectively,  the corresponding operators are
\begin{equation}\label{N2BBfermionreview}. 
\begin{aligned}
\blacksquare: \qquad \qquad  x:\, \textrm{creation} \qquad  P:\, \textrm{charge}  \qquad  y:\, \textrm{annihilation}\,, \\
\overline{\blacksquare}: \qquad \qquad \bar{x}:\, \textrm{creation} \qquad  \bar{P}:\, \textrm{charge}  \qquad  \bar{y}:\, \textrm{annihilation} \,,
\end{aligned}
\end{equation}
which is the analogue of (\ref{N2BBleftreview}) that captures the essential part of the $(e,\psi,f)$'s action (given in equation (\ref{Yaction})) on the left plane partitions, and similarly for the $(\hat{e},\hat{\psi},\hat{f})$'s action on the right plane partitions. 
The building block $\blacksquare$ is shown in Figure \ref{fig:blacksquarebab}.
To the left plane partition, it is the ground state of the module $(0, {\square},0)$, i.e.\ a semi-infinitely long row of boxes along the $x_2$ direction. 
To the right one, it is the ground state of $(0,{\overline{\square}},0)$ module, i.e.\ the conjugate of $(0, {\square},0)$; and it is realized as a ``high wall" sitting at $\hat{x}_1=-1$ and  extending along the $\hat{x}_2$ directions, with the height (i.e. along the $\hat{x}_3$ direction) reaching \emph{one box lower} than the ``ceiling".
For a derivation of this ``high wall" configuration and for examples on other ``high wall" states, see section 3 of \cite{Gaberdiel:2018nbs}.
The building block $\overline{\blacksquare}$ is given by the left-right mirror image of the one shown in Figure \ref{fig:blacksquarebab}.

Repeated application of the creation operators $x$ and $\bar{x}$ generate pairs of plane partitions whose asymptotics are correlated in the following way 
\begin{equation}\label{tppvacuum}
[(0, \lambda \otimes \bar{\rho^{t}}, 0)\,\, ,\,\, (0,\bar{\lambda^{t}} \otimes \rho,0)]\,,
\end{equation}
where the bar denotes conjugate and the  $t$ denotes ``transpose". (For detailed explanation see \cite{Gaberdiel:2018nbs}).
Such a pair of plane partitions whose asymptotics along the $x_2$ and $\hat{x}_2$ directions are correlated as in (\ref{tppvacuum}) were called \textit{twin plane partitions} in \cite{Gaberdiel:2018nbs, Li:2019nna}. 
Those in (\ref{tppvacuum}) are in the vacuum module, and a generic representation is labeled by
\begin{equation}\label{tpp}
\bm{\Lambda}: \qquad [(\mu_1, \lambda \otimes \bar{\rho^{t}}, \mu_3)\,\, , \,\, (\hat{\mu}_1,\bar{\lambda^{t}} \otimes \rho,\hat{\mu}_3)]\,.
\end{equation}
Applying the creation operator $e$ and $\hat{e}$ then creates descendants of (\ref{tpp}), represented by additional boxes $\square$'s and hatted boxes $\hat{\square}$'s on top of (\ref{tpp}).

\subsubsection{Determine glued algebra via twin plane partitions}

The glued algebra contains the following three types of operators
\begin{equation}\label{operatorfullset}
\begin{aligned}
\textrm{type }\bm{c}: & \quad \psi, \hat{\psi}, P, \bar{P}\,; \qquad
\textrm{type }\bm{s}: \quad  e, f, \hat{e}, \hat{f}\,; \qquad
\textrm{type }\bm{g}: \quad  x, y,  \bar{x}, \bar{y}\,.
\end{aligned}
\end{equation}
Type $(\bm{c})$ operators are the charge operators whose eigenstates are the set of all twin plane partitions
\begin{equation}\label{eigendefreview}
\begin{aligned}
\psi(u) \, |\bm{\Lambda}\rangle &= \bm{\Psi}_{\bm{\Lambda}}(u) \, |\bm{\Lambda}\rangle\ , \\
\hat{\psi}(u) \, |\bm{\Lambda}\rangle &= \hat{\bm{\Psi}}_{\bm{\Lambda}}(u) \, |\bm{\Lambda}\rangle\ ,
\end{aligned}
\qquad \textrm{and}\qquad
\begin{aligned}
P(u) \, |\bm{\Lambda}\rangle &= \textbf{P}_{\bm{\Lambda}}(u) \, |\bm{\Lambda}\rangle\ , \\
\bar{P}(u) \, |\bm{\Lambda}\rangle &= \bar{\textbf{P}}_{\bm{\Lambda}}(u) \, |\bm{\Lambda}\rangle\ .
\end{aligned}
\end{equation}
Type $(\bm{s})$ operators create/annihilate single boxes, and type  ($\bm{g}$) operators create/annihilate non-trivial states along the internal leg.
The goal is to find the OPEs between all pairs of operators in (\ref{operatorfullset}).

The main constraint in building this glued algebra is that it needs to have a natural representation in terms of the set of twin plane partitions. 
Namely, the OPEs are fixed together with the action of all operators on an arbitrary twin plane partition state:
\begin{equation}\label{Oformreview}
	\mathcal{O}(z) |\bm{\Lambda}\rangle = \sum_{i} \frac{O[\bm{\Lambda} \rightarrow \bm{\Lambda}'_i]}{z-z^*_{i}} |\bm{\Lambda}_{i}'\rangle\,,
\end{equation}
where $i$ runs over all possible final states $\bm{\Lambda}'_i$ that can be created by applying $\mathcal{O}$ on the initial state $\Lambda$. 

Generalizing (\ref{Yaction}), we first write down the ansatz for the action of operators (\ref{operatorfullset}) in the glued algebra.
As in  (\ref{Yaction}), the eigenvalue function $\bm{\Psi}_{\bm{\Lambda}}(z)$ (resp.\ $ \hat{\bm{\Psi}}_{\bm{\Lambda}}(z))$ controls the action of single-box operators $e$ and $f$ (resp.\ $\hat{e}$ and $\hat{f}$) on twin plane partitions:
\begin{equation}\label{ppartpsi}
\begin{aligned}
\psi(z)|\bm{\Lambda} \rangle & = \bm{\Psi}_{\bm{\Lambda}}(z)|\bm{\Lambda} \rangle  \ ,\\
e(z) | \bm{\Lambda} \rangle & =  \sum_{ \square\in {\rm Add}(\bm{\Lambda})}\frac{\Big[ -  \frac{1}{\sigma_3} {\rm Res}_{w = h(\square)} \bm{\Psi}_{\bm{\Lambda}}(w) \Big]^{\frac{1}{2}}}{ z - h(\square) } 
| \bm{\Lambda} + \square \rangle \ , \\
f(z) | \bm{\Lambda} \rangle & =  \sum_{ \square \in {\rm Rem}(\bm{\Lambda})}\frac{\Big[ +  \frac{1}{\sigma_3} {\rm Res}_{w = h(\square)} \bm{\Psi}_{\bm{\Lambda}}(w) \Big]^{\frac{1}{2}}}{ z - h(\square) } | \bm{\Lambda} - \square \rangle \ ;
\end{aligned}
\end{equation}
and the hatted version. 

Comparing (\ref{N2BBleftreview}) and (\ref{N2BBfermionreview}), we have the ansatz for the gluing operators
\begin{equation}\label{xansatzreview}
\begin{aligned}
P(z) \, |\bm{\Lambda}\rangle &= \textbf{P}_{\bm{\Lambda}}(z) \, |\bm{\Lambda}\rangle\ , \\
x(z)|\bm{\Lambda}\rangle &=\sum_{\blacksquare\in \textrm{Add}(\lambda)}\frac{
\Big[ \textrm{Res}_{w=p_{+}({\blacksquare})}  \textbf{P}_{\bm{\Lambda}}(w) \Big]^{\frac{1}{2}} 
}{z-p_+(\blacksquare)}|[\bm{\Lambda}+\blacksquare]\rangle+ \sum_{\overline{\blacksquare}\in \textrm{Rem}(\hat{\rho})}\frac{
\Big[ \textrm{Res}_{w=p_{-}(\overline{\blacksquare})} \textbf{P}_{\bm{\Lambda}}(w) \Big]^{\frac{1}{2}}
}{z-p_-(\overline{\blacksquare})}|[\bm{\Lambda}-\overline{\blacksquare}]\rangle  \ , \\
y(z)|\bm{\Lambda}\rangle &= \!\!\! \sum_{\blacksquare\in \textrm{Rem}({\lambda})}
\frac{ \Big[ \textrm{Res}_{w=p_{+}({\blacksquare})} \, \textbf{P}_{\bm{\Lambda}}(w) \Big]^{\frac{1}{2}}
}{z-p_+(\blacksquare)} \, |[\bm{\Lambda}-\blacksquare]\rangle
+ \!\!\! \sum_{\overline{\blacksquare}\in \textrm{Add}(\hat{\rho})}\frac{\Big[ \textrm{Res}_{w=p_{-}(\overline{\blacksquare})} \, \textbf{P}_{\bm{\Lambda}}(w) \Big]^{\frac{1}{2}}}{z-p_-(\overline{\blacksquare})}|[\bm{\Lambda}+\overline{\blacksquare}]\rangle  \ ;
\end{aligned}
\end{equation}
where $\lambda$ and $\hat{\rho}$ were defined in (\ref{tpp}).
And similarly for $(\bar{x},\bar{P},\bar{y})$.

Note that $x$ can not only add $\blacksquare$ in $\lambda$ but also remove $\overline{\blacksquare}$ in $\hat{\rho}$.
This is based on the fact that the tensor product of the fundamental and the anti-fundamental contains a singlet and an adjoint. 
In terms of the plane partitions, the adjoint is just the configuration with the fundamental (i.e.\ a single long row of boxes) and the anti-fundamental (a high wall of width-one) coexisting (which is possible since the high wall is behind the $x_1=0$ plane).
The singlet is the vacuum, namely,  if we place the long row (i.e.\ the fundamental representation) on top of the high wall (i.e.\ the anti-fundamental representation), the height of the wall reaches maximum and there is no more place to add or remove any boxes --- effectively we get the vacuum. 
And similarly for the other three operators, $y$, $\bar{x}$, and $\bar{y}$.
The difference between e.g.\ the $\blacksquare$-adding actions of $x$ and $\bar{y}$ are both in their poles, i.e.\ $p_{+}(\blacksquare)$ v.s.\ $\bar{p}_{-}(\blacksquare)$, and in their coefficients. For details see \cite{Gaberdiel:2018nbs}.

We see that once we determine the four charge functions $(\bm{\Psi}_{\bm{\Lambda}}(z) , \hat{\bm{\Psi}}_{\bm{\Lambda}}(z))$ and $(\bm{\textbf{P}}_{\bm{\Lambda}}(z) , \hat{\bm{\textbf{P}}}_{\bm{\Lambda}}(z))$ as functions of the twin plane partition $\bm{\Lambda}$, we obtain the full algebra; or vice versa: knowing the algebra would allow us to fix the action of all the operators on the set of twin plane partitions. 
In practice, these two aspects are fixed together, see details in \cite{Gaberdiel:2018nbs, Li:2019nna}.

\subsection{New VOAs from gluing multiple affine Yangians of $\mathfrak{gl}_1$}

We have just seen how constraints from twin plane partitions allow us to fix the algebra that is isomorphic to (the UEA) of $\mathfrak{u}(1)\oplus \mathcal{W}^{\mathcal{N}=2}_{\infty}$ algebra.
We now explain how the same procedure, slightly generalized,  can be used to construct new affine Yangian algebras that are isomorphic to new VOAs with higher spin currents.

Recall that there are four important ingredients in the construction reviewed above:
\begin{enumerate}
\label{list3}
\item Two copies of  affine Yangians of $\mathfrak{gl}_1$ --- $\mathcal{Y}$ and $\hat{\mathcal{Y}}$.
\item The gluing operators transform, w.r.t.\ $\mathcal{Y}$ and $\hat{\mathcal{Y}}$, as bimodules
\begin{equation}\label{bimodule2}  
[(0, {\square},0)\,\, ,\,\,  (0,{\overline{\square}},0)  ]=:[\square, \overline{\square}]
\quad \textrm{and}\quad 
 [(0, \overline{\square},0) \,\, ,\,\,  (0,{\square},0)]=:[\overline{\square}, {\square}] \,.
\end{equation}
\item The parameters of $\mathcal{Y}$ and $\hat{\mathcal{Y}}$ are related by
\begin{equation}\label{hip0}
h_i=\hat{h}_i \,.
\end{equation}
\item The conformal dimension of the gluing operators is $\frac{3}{2}$, i.e.\ that of the supercharge $G^{\pm}$. 
In terms of $\psi_0$ and $\hat{\psi}_0$, this translates into 
\begin{equation}\label{hirhohalf}
h_1\, h_3\, \psi_0+\hat{h}_1\, \hat{h}_3\, \hat{\psi}_0=-1\,.
\end{equation}
\end{enumerate}

In \cite{Li:2019nna}, we asked the question of, assuming we keep (1) and (2) unchanged, can we still modify (3) and (4)? 
And if yes, what are the most general parameters in (3) and (4)? 
Finally, does a parameter change in (3) or (4) cause any significant effect on the glued algebra?

The answers are the following. 
First,  both  (3) and (4) can be modified, resulting in a two-parameter generalization of the isomorphism (\ref{N2decompY}), i.e.\ the corresponding VOAs are a two-parameter generalization of the $\mathcal{N}=2$ supersymmetric $\mathcal{W}_{\infty}$ algebra. 
Second, given how the gluing operators transform (as (\ref{bimodule2})), the constraints from twin plane partitions allow us to solve for all possible solutions for the parameters of $\mathcal{Y}$ and $\hat{\mathcal{Y}}$.
Finally, the change in (\ref{hirhohalf}) only shifts the conformal dimension of the gluing operators; whereas the change in  (\ref{hip0}) switches the statistics of the gluing operators (i.e. from fermionic to bosonic) and hence results in more drastic changes of the algebra.

\subsubsection{Constraints on parameters from twin plane partitions}

A priori, the parameters of $\mathcal{Y}$ and $\hat{\mathcal{Y}}$ are
\begin{equation}\label{YYhatparameters}
\begin{cases}
\mathcal{Y}: \qquad h_i \, (\textrm{with}\, \sum_i h_i=0) \quad \textrm{and} \quad \psi_0\\
\hat{\mathcal{Y}}: \qquad \hat{h}_i \, (\textrm{with}\, \sum_i \hat{h}_i=0) \quad \textrm{and} \quad \hat{\psi}_0\,,
\end{cases}
\end{equation}
subject to the scaling automorphism (\ref{automode01}) and its hatted version.
Now let's first consider the constraint on (\ref{YYhatparameters}) from twin plane partitions. 
The main constraints come from the fact that a gluing operator (i.e.\ one from $\{x,y, \bar{x},\bar{y}\}$) affects both left and right plane partitions, hence its pole can be determined from both sides. 
The requirement is that the values from the two sides need to coincide:\footnote{
Note that for both sides }
\begin{equation}\label{zleftright}
z^{*}_{\textrm{left}}=z^{*}_{\textrm{right}}\,,
\end{equation}
since they are from the action of the same gluing operator. 
Applying this argument on different twin plane partition states generates various constraints on the parameters in (\ref{YYhatparameters}).

For the algebra with gluing operators transforming as (\ref{bimodule2}), the crucial constraints are from the fact that the gluing operator $x$ creates a configuration that is in the representation $(0,{\square},0)$ w.r.t.\ $\mathcal{Y}$ and $(0, \overline{\square},0)$ w.r.t.\ $\hat{\mathcal{Y}}$.
We now list all the constraints on (\ref{YYhatparameters}) from twin plane partitions.

\begin{enumerate}
\item From the left and right plane partitions separately, i.e.\  for $\mathcal{Y}$ and $\hat{\mathcal{Y}}$ separately, we have 
\begin{flalign}\label{projection1}
\text{(a1)}
&&
   & \sum^3_{i=1} h_i =0 
	\qquad \textrm{and} \qquad 
	\sum^3_{i=1} \hat{h}_i =0 \,.
	&
\end{flalign}
\item In twin plane partitions, the two corners share a common $x_2$ direction, hence $h_2$ and $\hat{h}_2$ should be directly related. 
Consider the action of $x(z)$ on the initial state of $|\square\rangle$. 
The $(\bm{\Psi}_{\bm{\Lambda}}(z) , \hat{\bm{\Psi}}_{\bm{\Lambda}}(z))$ charge functions of the final state $x(z)|\square\rangle$ are
\begin{equation}\label{xonbox}
x(z)|\square\rangle:\qquad \begin{cases}
\begin{aligned}
\bm{\Psi}(u): &\qquad \psi_0(u) \,\varphi_3(u)\, \varphi_2(u-z^{*}_{\textrm{left}})  \\
\hat{\bm{\Psi}}(u): &\qquad \hat{\psi}_0(u)\, \hat{\varphi}^{-1}_2(-(u-z^*_{\textrm{right}})-\hat{\sigma}_3 \hat{\psi}_0)  \ .
\end{aligned}
\end{cases}
\end{equation}
For both sides, we need to find the pole $z^{*}$ such that the corresponding $\bm{\Psi}$ charge function corresponds to a sensible plane partition. 
The results are
\begin{equation}
z^{*}_{\textrm{left}}=h_2 \qquad \textrm{and} \qquad z^{*}_{\textrm{right}}=\hat{h}_2 \,,
\end{equation}
which gives 
\begin{equation}\label{xonboxresult}
|\blacksquare+\hat{\square}_{\textrm{top}}\rangle: \quad\begin{cases}
\begin{aligned}
\bm{\Psi}(u): &\quad \psi_0(u) \,  \varphi_2(u)  \\
\hat{\bm{\Psi}}(u): &\quad \hat{\psi}_0(u)\, \hat{\varphi}^{-1}_2(-u-\hat{\sigma}_3 \hat{\psi}_0)\, \hat{\varphi}_3(u-\hat{h}_2+\hat{\sigma}_3 \hat{\psi}_0)   \ ,
\end{aligned}
\end{cases}
\end{equation}
where $\varphi_2$ and $\hat{\varphi}^{-1}_2$ are from the contribution of $\blacksquare$, and $\hat{\varphi}_3$ accounts for the contribution from $\hat{\square}_{\textrm{top}}$. 
Then the constraint (\ref{zleftright}) gives
\begin{flalign}\label{A2}
\text{(a2)}
&&
    & h_2=\hat{h}_2\,.
&
\end{flalign}

\item Now consider the action of $x(z)$ on the initial state $|\blacksquare+\textrm{bud}(\blacksquare)_{1} \rangle$ in order to add a $\blacksquare$ at the position $(x_1,x_3)= (1,0)$, i.e.\ to produce the final state  $|\blacksquare\blacksquare_1\rangle$.
Here the $\textrm{bud}(\blacksquare)_1$, i.e.\ the minimal bud for $\blacksquare$ at $(x_1,x_3)= (1,0)$, denotes the $s_1$ additional $\square$'s lined up along the $x_2$ direction from $x_2=0$ and in the $(x_1,x_3)=(1,0)$ position where the new $\blacksquare$ is about to be added. (For the derivation of the bud condition, see section 6.1 of \cite{Li:2019nna}.)
The pole from the left side is
\begin{equation}
z^{*}_{\textrm{left}}=h_1+s_1 h_2\,,
\end{equation}
where $h_1$ comes from the position of the $\blacksquare$ at $(x_1,x_3)= (1,0)$ and $s_1 h_2$ comes from the displacement of the pole by the presence of the $s_1$ additional boxes along the $x_2$ direction. 

On the other hand, while the action of $x(z)$ creates a $\square$
in the asymptotic Young diagram along $x_2$ direction for the left plane partition, it creates a $\overline \square$ in the asymptotic Young diagram along the $\hat{x}_2$ direction for the right plane partition. 
In \cite{Gaberdiel:2018nbs, Li:2019nna}, the transverse coordinates of the $\overline\square$'s are negative.\footnote{ 
For the reasoning behind this convention see section 3 of  \cite{Gaberdiel:2018nbs}.}
Therefore, the pole from the consideration of the right plane partition is
\begin{equation}
z^{*}_{\textrm{right}}=-\hat{h}_3\,.
\end{equation}
The argument (\ref{zleftright}) then gives 
\begin{equation}\label{h1hh3}
h_1+s_1\, h_2=-\hat{h}_3\,.
\end{equation}
Repeating the argument on creating a second $\blacksquare$ at the coordinate $(x_1,x_3)=(0,1)$, we have
\begin{equation}\label{h3hh1}
h_3+s_3\, h_2=-\hat{h}_1 \,.
\end{equation}
In summary, the condition on creating a second $\blacksquare$ gives
\begin{flalign}\label{A3}
\text{(a3)}
&&
    & h_1+s_1 \, h_2=-\hat{h}_3 \qquad \textrm{and} \qquad h_3+s_3 \, h_2=-\hat{h}_1 \,,
&
\end{flalign}
with $s_{1,3}\in\mathbb{Z}$ and $s_1+s_3=2$ (due to constraint (\ref{projection1})).

\item In \cite{Li:2019nna}, we further constrained that twin plane partitions do not have buds of negative length, i.e.\ buds that stick out of the left or right wall. 
This gives 
\begin{flalign}\label{A4}
\text{(a4)}
&&
     & s_1  \geq 0 
	\qquad \textrm{and} \qquad 
	s_3 =2-s_1 \geq 0 \ ,
&
\end{flalign}
where we have used $s_1+s_3=2$.
\end{enumerate}

The four constraints above give the following three sets of solutions of (\ref{YYhatparameters}).
First, they all have
\begin{equation}\label{h2}
	\hat h_2 = h_2\,.
\end{equation}
Then the three solution of (\ref{A4}) gives 
\begin{eqnarray}\label{3solutions}
(s_1,s_3)=(2,0):& \qquad\,\, \hat{h}_1=-h_3 \qquad& \hat h_2 = h_2 \qquad\hat{h}_3=h_3-h_2 \,,\nonumber\\
(s_1,s_3)=(1,1):& \qquad\,\, \hat{h}_1=h_1 \qquad&\hat h_2 = h_2\qquad \, \hat{h}_3=h_3 \,,\\
(s_1,s_3)=(0,2):& \qquad\qquad \hat{h}_1=h_1-h_2 \qquad &\hat h_2 = h_2 \qquad \hat{h}_3=-h_1 \,, \nonumber
\end{eqnarray}
which can also be written as
\begin{equation}\label{hhbp}
\hat{h}_1=h_1-p\, h_2 \,, \qquad \hat{h}_2=h_2 \,, \qquad \hat{h}_3=h_3+p \ h_2\,.
\end{equation}
where $p\equiv \frac{s_3-s_1}{2}$, thus $p=-1,0,1$ for the three cases in (\ref{3solutions}).
In \cite{Li:2019nna}, it was shown that the three solutions (\ref{3solutions}) correspond to toric Calabi-Yau threefolds $\mathcal{O}(-1-p)\oplus \mathcal{O}(-1+p)\rightarrow \mathbb{P}^1$ with $p=-1,0,1$.
In particular, the four constraints (a1)-(a4) are in one-to-one correspondence with the four constraints for a toric Calabi-Yau threefold with two vertices in the toric diagram, for details see \cite{Li:2019nna}.
It was then shown, by a character analysis, that the gluing operators are bosonic when $p=0$ and fermionic when $p=\pm 1$ \cite{Li:2019nna}.

\subsubsection{Constraints on parameters from conformal dimensions of gluing operators}

As we have seen, the relations between $h_i$ and $\hat{h}_i$ are related to the self-statistics of the gluing operators. 
We now explain that the equation involving $\psi_0$ and $\hat{\psi}_0$ are related to the conformal dimension of the gluing operators.

From the spin-two part of the map (\ref{Wepsif}) we have 
\begin{equation}
L_0=\frac{1}{2}(\psi_2+\hat{\psi}_2)\,,
\end{equation}
for the derivation see  \cite{Gaberdiel:2017dbk}.
The conformal dimension of the gluing operators are identical to that of the first state $|\blacksquare\rangle$ created by $x$ out of the vacuum:
\begin{equation}
|\blacksquare\rangle:\qquad \begin{cases}
\begin{aligned}
\bm{\Psi}(u): &\qquad \psi_0(u)\, \varphi_2(u)  \\
\hat{\bm{\Psi}}(u): &\qquad \hat{\psi}_0(u)\, \hat{\varphi}^{-1}_2(-u-\hat{\sigma}_3 \hat{\psi}_0)  \ ,
\end{aligned}
\end{cases}
\end{equation}
which gives 
\begin{equation}
h_{\blacksquare}=1-\frac{1}{2}(h_1 h_3 \psi_0+\hat{h}_1\hat{h}_3\hat{\psi}_0) =: 1+\rho\,,
\end{equation}
where we have defined a new parameter
\begin{equation}\label{rhodefreview}
\rho\equiv -\frac{1}{2}(h_1 h_3 \psi_0+\hat{h}_1\hat{h}_3\hat{\psi}_0) \in \frac{1}{2} \mathbb{Z}_0\,,
\end{equation}
which has to be an integer or a half-integer.

For the algebra to correspond to the $\mathcal{N}=2$ supersymmetric $\mathcal{W}_{\infty}$ algebra, $\rho=\frac{1}{2}$, i.e.\ the gluing operators' conformal dimension is $\frac{3}{2}$ --- that of the supercharges $G^{\pm}$.
For a generic algebra from the gluing construction that satisfies the condition (1) and (2) on page 13, their parameters $\psi_0$ and $\hat{\psi}_0$ only need to be related via (\ref{rhodefreview}).

\section{$(\square,\square)$ gluing operators and perturbative twin plane partitions}
\label{sec:3tpp}

In section~2, we have seen that the gluing construction 
for the $\mathcal{N}=2$ supersymmetric $\mathcal{W}_{\infty}$
satisfies four basic conditions (see page 13).
Relaxing the last two conditions results in a two-parameter generalization of the $\mathcal{N}=2$ $\mathcal{W}_{\infty}$ algebra. 
In this section, we relax the conditions further. 
In particular, we choose the gluing operators to transform as bimodules
\begin{equation}\label{boxboxdef}
[(0, {\square},0)\,\, ,\,\,  (0,{{\square}},0)  ]=:[\square, {\square}]\,,
\end{equation}
w.r.t.\ the two affine Yangians of $\mathfrak{gl}_1$.
The resulting algebra is denoted schematically as
\begin{equation}\label{GC2bb}
\mathcal{Y}(q)\,\oplus\, \tilde{\mathcal{Y}}(\tilde{q})\, \oplus\, [\square, \square]\,,
\end{equation}
where we use $\tilde{\mathcal{Y}}$ to denote the right Yangian, in order to distinguish from the box-antibox construction reviewed in section 2. 
Note that since the relations (\ref{3solutions}) between the  parameters of the two affine Yangians of $\mathfrak{gl}_1$ depend on the choices of bimodules, we would have to solve them for the current choice (\ref{GC2bb}).

\subsection{Operators in $[\square, \square]$  gluing construction}

\begin{figure}[h!]
\begin{center}
\includegraphics[width=0.7\textwidth]{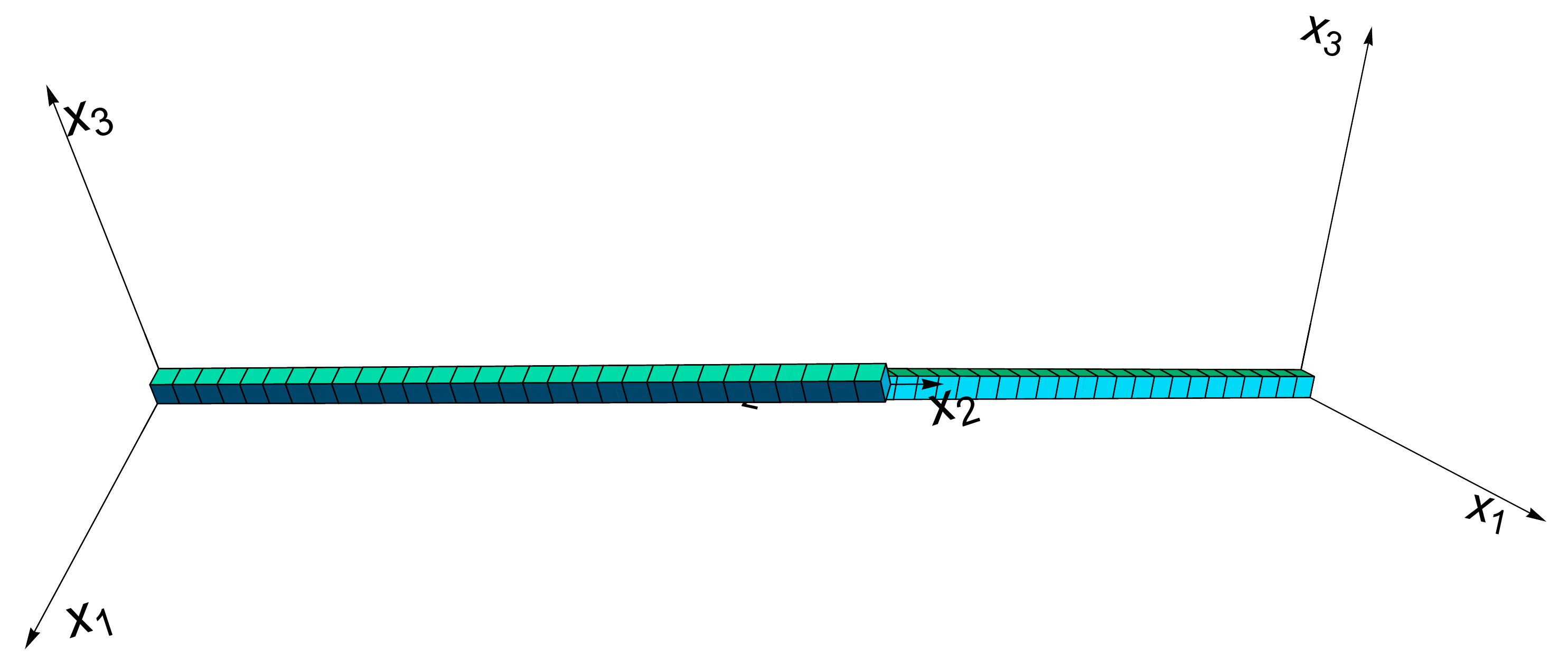}
\caption{The plane partition configuration of the building block $\blacksquare$ in the $[\square,\square]$ gluing construction. 
Note that the long rows of boxes in the left and right plane partitions are connected seamlessly. }
\label{fig:blacksquarebb}
\end{center}
\end{figure}

The algebra in (\ref{GC2bb}) consists of two affine Yangians of $
\mathfrak{gl}_1$, $\mathcal{Y}$ given by (\ref{bosonicdef}) and $\tilde{\mathcal{Y}}$ by the tilde version of (\ref{bosonicdef}).
Correspondingly, they act on the left and right plane partitions, respectively: 
\begin{equation}\label{N2BBleft}
\begin{aligned}
&{{\square}}: \qquad \qquad e:\, \textrm{creation} \qquad  \psi:\, \textrm{charge}  \qquad  f:\, \textrm{annihilation}\,,\\
&\tilde{\square}: \qquad \qquad \tilde{e}:\, \textrm{creation} \qquad  \tilde{\psi}:\, \textrm{charge}  \qquad  \tilde{f}:\, \textrm{annihilation} \,.
\end{aligned}
\end{equation}
Similar to the box-antibox construction, the building block for the configurations along the internal leg connecting the two plane partitions is\footnote{
Note that we recycle the symbol $\blacksquare$ --- which was used in \cite{Gaberdiel:2018nbs, Li:2019nna} to denote the $[\square,\bar{\square}]$ building block (shown in Figure \ref{fig:blacksquarebab}) --- for lack of a better choice; however in this paper (except for section \ref{sec:review}) it denotes the $[\square,\square]$ building block (shown in Figure \ref{fig:blacksquarebb}).}
\begin{equation}\label{BB}
\blacksquare\equiv [{\square},\square]\,,
\end{equation}
which was defined in (\ref{boxboxdef}).
It consists of boxes $\square$'s in positions 
\begin{equation}\label{boxboxleft}
\begin{aligned}
& x_1=0\,,\qquad x_2=0,1,2,\cdots,\infty\,, \qquad x_3=0 
\end{aligned}
\end{equation}
of the left plane partition and tilde boxes $\tilde{\square}$'s in positions  
\begin{equation}\label{boxboxright}
\begin{aligned}
&\tilde{x}_1=0\,,\qquad \tilde{x}_2=0,1,2,\cdots,\infty\,, \qquad \tilde{x}_3=0 
\end{aligned}
\end{equation}
of the right plane partition.
The building block $\blacksquare$ is shown in Figure \ref{fig:blacksquarebb}. 
Compared to its counterpart in the box-antibox construction, shown in Figure \ref{fig:blacksquarebab}, this plane partition configuration is more elegant, since the left and right plane partitions can be connected seamlessly.
This is one of the main advantages of the current construction over the box-antibox construction in \cite{Gaberdiel:2018nbs, Li:2019nna}.

Similar to (\ref{N2BBleft}), we define the corresponding operators
\begin{equation}\label{N2BBfermion}
\begin{aligned}
\blacksquare: \qquad \qquad  \x:\, \textrm{creation} \qquad  Q:\, \textrm{charge}  \qquad  \y:\, \textrm{annihilation} \,.\end{aligned}
\end{equation}
In summary, the algebra from the gluing construction (\ref{GC2bb}) contains the following three types of operators
\begin{equation}\label{operatorfullsetBB}
\begin{aligned}
\textrm{type }\bm{c}:  \quad \psi, \tilde{\psi}, Q \,; \quad\qquad
\textrm{type }\bm{s}: \quad  e, f, \tilde{e}, \tilde{f}\,; \quad\qquad
\textrm{type }\bm{g}: \quad  r, s \,. 
\end{aligned}
\end{equation}
Type $\bm{c}$ operators are the charge operators whose eigenstates are the set of all relevant twin plane partitions (to be defined in subsection 3.2), type $\bm{s}$ operators create/annihilate single boxes, and type  $\bm{g}$ operators create/annihilate non-trivial states along internal legs.
The goal is to find the OPEs between all pairs of the operators in (\ref{operatorfullsetBB}).

\subsection{Perturbative twin plane partitions}
\label{sec:ptpp}

As in the box-antibox construction of \cite{Gaberdiel:2018nbs, Li:2019nna}, the algebra can be completely determined by the constraints from its action on the relevant plane partition type representations.
For the $[\square, \square]$ construction, they are pairs of plane partitions connected by the internal leg, whose states are built by repeated actions of $r$, hence their asymptotics along the $x_2$ and $\tilde{x}_2$ directions are correlated by 
\begin{equation}\label{TPP1nonV}
\Lambda: \qquad [(\mu_1, \lambda, \mu_3)  \,\, \,, \,\, (\tilde{\mu}_1,\lambda,\tilde{\mu}_3)]\,,
\end{equation}
or
\begin{equation}\label{TPP2nonV}
\Lambda: \qquad [(\mu_1, \lambda, \mu_3)\,\, \,, \,\, (\tilde{\mu}_1,\lambda^t,\tilde{\mu}_3)]\,.
\end{equation}
As we will show later by a  character analysis, these two choices, i.e.\ whether the symmetric direction of the left Young diagram --- namely, the asymptotics of the left plane partition along the $x_2$ direction --- is correlated with the symmetric or the anti-symmetric direction of the right Young diagram, are determined by whether the gluing operators are bosonic or fermionic.\footnote{
Note that when there is only one affine Yangian of $\mathfrak{gl}_1$ (and one plane partition) in the system, the two directions in the $x_1-x_3$ plane (since we are focusing on the asymptotic Young diagram along the $x_2$ direction) are related by a $\mathbb{Z}_2$ duality symmetry --- the residue symmetry of the $\mathcal{S}_3$ permutation, therefore the designation of the $x_{1,3}$ direction as the ``symmetric" or ``anti-symmetric" direction of the asymptotic Young diagram is merely a convention. 
However, since in this paper we have two plane partitions connected by an internal leg, the relative orientation of the two asymptotic Young diagrams of the left and right plane partitions (that are given by the same internal leg configuration) matters, e.g.\ in the character computations later (see (\ref{chiBtpp}) and (\ref{chiFtpp})).
Namely the residue duality symmetry is $\mathbb{Z}_2$ instead of $\mathbb{Z}_2\times \mathbb{Z}_2$.}

Risking confusion with representations in the box-antibox construction of \cite{Gaberdiel:2018nbs, Li:2019nna}, we will name these plane partitions ``perturbative twin plane partitions", for want of a better name.\footnote{
To distinguish the representations in the box-antibox construction and those here, given in  (\ref{TPP1nonV}) and (\ref{TPP2nonV}), we might want to rename the former to ``twain plane partition", since those in the present construction are more symmetric (between the left and right corners).}
The ``perturbative" is to emphasize that there is no conjugate representations involved --- recall that the conjugate representation is characterized by walls of height $\sim N$ (for $\mathcal{W}_{1+N}$ algebra).
Furthermore, when there is no risk of confusion, we will often drop  the ``perturbative" in the remainder of the paper.

Note that from the $\mathcal{W}$ algebra's point of view, a representation and its conjugate are on equal footing --- neither is more ``non-perturbative" than the other. 
However for the affine Yangian, since the representations are described in terms of plane partitions, the fundamental and anti-fundamental representations (in terms of the plane partitions) are very different: the former is given in terms of a long row of single boxes whereas the latter in terms of a wall of infinite height and width-one (as $N\rightarrow \infty$) --- it is in this sense that we call the anti-fundamental representation, and the conjugate representations in general, non-perturbative.

This is also related to the fact that in the convention of \cite{Gaberdiel:2017dbk} and this paper, the $x_3$ direction is the direction that is not visible directly by the coset representations, see eq.\ (\ref{WtoPP}). 
As a consequence, in the map (\ref{h123}) and (\ref{psi0}) between the affine Yangian parameter $(h_1,h_2, h_3, \psi_0)$ and the $\mathcal{W}_{\infty}$ algebra parameters $(N,k)$,  the $h_3$ scales differently from $h_1$ and $h_2$ in the 't Hooft limit of $N\rightarrow \infty$ and $k\rightarrow \infty$ with $\frac{N}{N+k}$ fixed, namely,  $h_3\sim \frac{1}{N} h_{1,2}$. 
(Or in terms of the $\lambda_i=-\frac{\sigma_3\psi_0}{h_i}$ parameters, $\lambda_3\sim N \lambda_{1,2}$.)
Note that it is precisely along the $x_3$ direction that the ``high wall" corresponding to the anti-fundamental representation is erected.

\subsubsection{Set of twin plane partitions and coordinate system}
\label{sec:setofTPP}

To study the constraints on the algebra, it is enough to focus on the vacuum representation of the glued algebra.
A vacuum twin plane partition configuration is given by a pair of plane partitions with asymptotics
\begin{equation}\label{TPP1}
[(0, \lambda, 0)\,\,\,,\,\, (0,\lambda,0)]\,,
\end{equation}
or
\begin{equation}\label{TPP2}
[(0, \lambda, 0) \,\,\,,\,\, (0,\lambda^t,0)]\,,
\end{equation}
depending on whether the gluing operators are bosonic or fermionic. 
The descendants of (\ref{TPP1}) and (\ref{TPP2}) have additional boxes on the left and right corners. 
Namely, a generic twin plane partition representation in the vacuum module has the following building blocks:
\begin{itemize}
\item A bi-representation $[\lambda, {\lambda}]$ or $[\lambda, {\lambda}^t]$,  generated by $\x$.
\item A collection $\mathcal{E}$ of individual $\square$'s in the left plane partition, generated by $e$.
\item A collection $\tilde{\mathcal{E}}$ of individual $\tilde{\square}$'s in the right plane partition, generated by $\tilde e$.
\end{itemize}

The coordinate systems for $\square$ and $\tilde{\square}$ are simply
\begin{equation}\label{xxtildebound}
x_i(\square)=0,1,2,\cdots,\infty\qquad \textrm{and}\qquad \tilde{x}_i(\tilde{\square})=0,1,2,\cdots,\infty\,.
\end{equation}
The coordinate functions of the $\square$ and $\tilde{\square}$ are
\begin{equation}\label{coorfunctionfloor}
h(\square)  \equiv \sum^{3}_{i=1}x_i(\square)\,h_i 
\qquad \textrm{and}\qquad 
\tilde{h}(\tilde{\square})  \equiv \sum^3_{i=1}\tilde{x}_i(\tilde{\square})\,\tilde h_i  \ .
\end{equation}
The restriction on the coordinates (\ref{xxtildebound}) means that we do not allow any box outside the left and right walls. 
As in the box-antibox construction, this bound imposes strong constraints on the algebra's action on the twin plane partitions, in particular, on the allowed final states. 
\bigskip

The coordinate systems for $\blacksquare$ are simpler than in \cite{Li:2019nna}, since they are symmetric w.r.t.\ the two sides. In particular, without loss of generality, we can choose their $(x_1,x_3)$ 
and $(\tilde{x}_1,\tilde{x}_3)$ 
coordinates to be correlated by\footnote{
The other natural choice with $x_1(\blacksquare)=\tilde{x}_1(\blacksquare)$ and $x_3(\blacksquare)=\tilde{x}_3(\blacksquare)$ is equivalent to (\ref{coordcorre}) by a mere relabeling with $\tilde{x}_1\leftrightarrow \tilde{x}_3$ and $\tilde{h}_1\leftrightarrow \tilde{h}_3$.}
\begin{equation}\label{coordcorre}
\begin{aligned}
& x_1(\blacksquare)=\tilde{x}_3(\blacksquare) \qquad \textrm{and} \qquad x_3(\blacksquare)=\tilde{x}_1(\blacksquare)\,,
\end{aligned}
\end{equation}
hence we would only need the $(x_1,x_3)$ coordinates for $\blacksquare$; and they take the value of 
\begin{equation}
\begin{aligned}
&\blacksquare :\qquad x_1(\blacksquare)\,, x_3(\blacksquare)=0,1,2, \cdots,\infty\,. 
\end{aligned}
\end{equation}
A $\blacksquare$ in $\lambda$ has a pair of coordinate functions since it is visible from both corners
\begin{equation}\label{gcharge}
g(\blacksquare)=h_1\, x_1(\blacksquare)+h_3 \, x_3(\blacksquare)\qquad \textrm{and} \qquad \tilde{g}(\blacksquare)=\tilde{h}_1 \, \tilde{x}_1(\blacksquare)+\tilde{h}_3 \,\tilde{x}_3(\blacksquare)\,.
\end{equation}

\subsubsection{Charge functions of perturbative twin plane partitions}

The main constraint in building the glued algebra is that it needs to have a natural representation in terms of the perturbative twin plane partition. 
Namely, the OPEs are fixed together with the action of all operators on an arbitrary twin plane partition state:
\begin{equation}\label{Oform}
	\mathcal{O}(z) |\Lambda\rangle = \sum_{i} \frac{O[{\Lambda} \rightarrow \Lambda'_i]}{z-z^*_{i}} |{\Lambda}_{i}'\rangle\,.
\end{equation}
The pole structure $z^{*}$ is relatively easy to determine, based on the fact that the resulting state $ |\Lambda_{i}'\rangle$
also needs to be a valid twin plane partition. 
However, the coefficient $O[\Lambda \rightarrow \Lambda'_i]$ can only be determined together with the OPEs. 

In \cite{Li:2019nna}, we found a procedure of fixing all OPEs together with the action (\ref{Oform}), for the gluing operator that transforms as $[\square, \overline{\square}]$ and  $[\overline{\square}, \square]$.
In this paper, we will follow this procedure and adapt it for the current case, with the gluing operators that transforms as $[\square, \square]$.

We first write down the ansatz for all the operators (\ref{operatorfullsetBB}) on an arbitrary twin plane partition state.
First of all, a  twin plane partition is the eigenstate of the charge operators $({\psi}(z) , \tilde{{\psi}}(z))$ and ${Q}(z)$:
 \begin{equation}\label{eigendef}
\begin{aligned}
\psi(z) \, |{\Lambda}\rangle &= \bm{\Psi}_{{\Lambda}}(z) \, |{\Lambda}\rangle\ , \\
\tilde{\psi}(z) \, |{\Lambda}\rangle &= \tilde{\bm{\Psi}}_{{\Lambda}}(z) \, |{\Lambda}\rangle\ ,
\end{aligned}
\qquad \textrm{and}\qquad
\begin{aligned}
Q(z) \, |{\Lambda}\rangle &= \textbf{Q}_{{\Lambda}}(z) \, |{\Lambda}\rangle\ . 
\end{aligned}
\end{equation}

The charge function $\bm{\Psi}_{{\Lambda}}(z)$ (resp.\ $ \tilde{\bm{\Psi}}_{{\Lambda}}(z))$ controls the action of single-box operators $e$ and $f$ (resp.\ $\tilde{e}$ and $\tilde{f}$) on twin plane partitions. 
Namely, the action of $\{e,\psi,f\}$ on an arbitrary twin plane partition $\Lambda$ is given by the analogue of (\ref{ppartpsi}), where $\bm{\Lambda}$ is replaced by the perturbative twin plane partition $\Lambda$,  and the action of $(\tilde{e},\tilde{\psi},\tilde{f})$ is given by the tilde version.
The charge functions $\bm{\Psi}_{{\Lambda}}(z)$ and $ \tilde{\bm{\Psi}}_{{\Lambda}}(z)$ will be fixed later, see section \ref{sec:psicharge}. 
They need to reproduce (\ref{bosonicdef}) and its tilde version.

Similarly, the charge function $\bm{\P}_{{\Lambda}}(z)$ controls the action of gluing operators $\x$ and $\y$ on twin plane partitions by
\begin{equation}\label{xansatz}
\begin{aligned}
Q(z) \, |{\Lambda}\rangle &= \textbf{Q}_{{\Lambda}}(z) \, |{\Lambda}\rangle\ , \\
r(z)|{\Lambda}\rangle &=\sum_{\blacksquare\in \textrm{Add}({\Lambda})}\frac{
\Big[ \textrm{Res}_{w=p_{+}({\blacksquare})}  \textbf{Q}_{{\Lambda}}(w) \Big]^{\frac{1}{2}} 
}{z-p_+(\blacksquare)}|[{\Lambda}+\blacksquare]\rangle\,, \\
s(z)|{\Lambda}\rangle &= \!\!\! \sum_{\blacksquare\in \textrm{Rem}({\Lambda})}
\frac{ \Big[ \textrm{Res}_{w=p_{-}({\blacksquare})} \, \textbf{Q}_{{\Lambda}}(w) \Big]^{\frac{1}{2}}
}{z-p_-(\blacksquare)} \, |[{\Lambda}-\blacksquare]\rangle\,.
\end{aligned}
\end{equation}

In principle, if we can determine the charge functions $(\bm{\Psi}_{{\Lambda}}(z) , \tilde{\bm{\Psi}}_{{\Lambda}}(z))$ and 
$\bm{\P}_{{\Lambda}}(z)$ for arbitrary twin plane partition $|{\Lambda}\rangle$ in the vacuum module (\ref{TPP1}) or (\ref{TPP2}), we can derive all the OPEs from there. 
However, as we will see below, some OPEs (especially those between single-box operators and gluing operators) are more easily determined (at least partially) directly using the constraints from twin plane partitions. 
These OPEs can in turn fix part of the charge functions. 

\subsection{Evaluating $(\mathbf{\Psi}, \mathbf{\tilde{\Psi}})$ charge functions}
\label{sec:psicharge}

Before we start fixing the entire algebra, some crucial aspects, such as the parameters of the algebra and the conformal dimension of the gluing operators, can already be fixed once we know the $(\bm{\Psi} , \tilde{\bm{\Psi}})$ charge function for a given twin plane partition and therefore OPEs between $(\psi, \tilde{\psi})$ and all other operators.
We will now determine the $(\bm{\Psi} , \tilde{\bm{\Psi}})$ charge function and leave the $\textbf{Q}$ charge function for the next section.

Same as in the box-antibox construction, the $(\bm{\Psi}_{{\Lambda}}(z), \tilde{\bm{\Psi}}_{{\Lambda}}(z))$ charge functions have the following decomposition
\begin{equation}\label{XLambda}
\bm{\Psi}_{{\Lambda}}(u) = \psi_0(u)\, \Biggl\{
\prod_{\blacksquare \in {\lambda}}  \psi_{\,\blacksquare}(u) \,
\prod_{ {\square} \in \mathcal{E}} \psi_{\,{\square}}(u) \, 
  \Biggr\} \quad \textrm{and}\quad
\tilde{\bm{\Psi}}_{{\Lambda}}(u) = \tilde{\psi}_0(u)\, \Biggl\{
\prod_{\blacksquare \in {\lambda}}  \tilde{\psi}_{\,\blacksquare}(u) \,
\prod_{\tilde{{\square}} \in \tilde{\mathcal{E}} } \tilde{\psi}_{\tilde{{\square}}}(u)  \Biggr\} 
\end{equation}
where $\lambda$, $\mathcal{E}$, and $\tilde{\mathcal{E}}$ are defined below (\ref{TPP2}).
Note that the $\square$'s on the left side are invisible to $\tilde{\bm{\Psi}}_{\Lambda}(u)$ while the $\tilde{\square}$'s on the right side are invisible to $\bm{\Psi}_{\Lambda}(u)$.
We can immediately fix all the factors in (\ref{XLambda}).

First of all, since it needs to reproduce the bosonic OPEs (\ref{bosonicdef}) and its tilde version, each $\square$ (resp.\ $\tilde{\square}$) again contributes to the $\bm{\Psi}_{{\Lambda}}(z)$ (resp.\ $\tilde{\bm{\Psi}}_{{\Lambda}}(z)$) charge function by
\begin{equation}\label{psichargebox}
\psi_{\square}(u)=\varphi_3 (u-h(\square)) \qquad \textrm{and} \qquad \tilde{\psi}_{\tilde{\square}}(u)=\tilde{\varphi}_3 (u-\tilde{h}(\tilde{\square}))\,.
\end{equation}
Secondly, since to the left plane partition, the $\blacksquare$ is composed of infinite number of $\square$'s at the coordinate 
\begin{equation}
x_1=x_1(\blacksquare)\,,\qquad x_2=0,1,2,\dots,\infty\,, \qquad 
x_3=x_3(\blacksquare)\,,
\end{equation}
as shown in Figure \ref{fig:blacksquarebb}, the contribution of $\blacksquare$ to the $\bm{\Psi}_{\Lambda}(u)$ charge function comes from all these $\square$'s:
\begin{equation}
\psi_{\blacksquare}(u)=\prod^{\infty}_{n=0}\varphi_3(u-n h_2-g(\blacksquare))=\varphi_2(u-g(\blacksquare))\,,
\end{equation}
where
\begin{equation}\label{psiudef0}
\varphi_2(u) \equiv\frac{u(u +h_2)}{(u -h_1)(u -h_3)}   \,.
\end{equation}
Similarly for the right plane partition, the $\blacksquare$ is composed of infinite number of $\tilde{\square}$'s at the coordinate
\begin{equation}
\tilde{x}_1=\tilde{x}_1(\blacksquare) = x_3 (\blacksquare)\,,\qquad \tilde{x}_2=0,1,2,\dots,\infty\,, \qquad 
\tilde{x}_3=\tilde{x}_3(\blacksquare)=x_1(\blacksquare)\,,
\end{equation}
as shown in Figure \ref{fig:blacksquarebb}.
Hence each $\blacksquare$ contributes to the charge function $(\bm{\Psi}_{{\Lambda}}(z) , \tilde{\bm{\Psi}}_{{\Lambda}}(z))$ by 
\begin{equation}\label{chargebbox}
\psi_{\blacksquare}(u)=\varphi_2(u-g(\blacksquare))  \qquad \textrm{and}\qquad  \tilde{\psi}_{\blacksquare}(u)=\tilde{\varphi}_2(u-\tilde{g}(\blacksquare)) \,.
\end{equation}
These are summarized in the $(\bm{\Psi}_{{\Lambda}}(u), \tilde{\bm{\Psi}}_{{\Lambda}}(u))$ part of Table \ref{tab2}.

\subsection{OPEs between $(\psi, \tilde{\psi})$ and all other operators}
\label{sec:psiOPEs}

Since to fix some basic properties of the glued algebra, such as the relation between parameters of $\mathcal{Y}$ and $\tilde{\mathcal{Y}}$, we already need to know OPEs between $(\psi, \tilde{\psi})$ and all other operators, we will fix them in this subsection.
All other OPEs will be left for section 4.

\begin{figure}[h!]
	\centering
	\includegraphics[trim=2cm 14cm 4cm 4cm, width=0.9\textwidth]{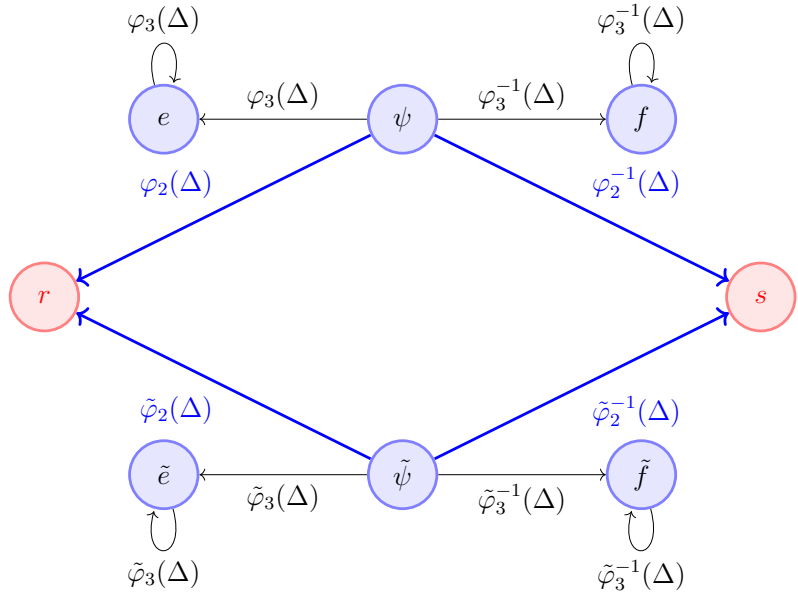}
	\caption{All operators in the $[\square,\square]$ gluing construction, together with their OPEs with the charge operators $\psi$ and $\tilde{\psi}$.}
	\label{OPEeverybody2bb}
\end{figure}

First of all, the OPEs of $(\psi , \tilde{\psi})$ and single-box operators $\bm{s}$ are already given by (\ref{bosonicdef}) and its tilde version.  
They can also be derived from the $(\bm{\Psi}_{{\Lambda}}(z) , \tilde{\bm{\Psi}}_{{\Lambda}}(z))$ charge functions of $\square$  and $\tilde{\square}$ in (\ref{psichargebox}). 
(See the top and bottom parts of Figure \ref{OPEeverybody2bb}.)
\smallskip

Using the same logic, the $(\bm{\Psi}_{{\Lambda}}(z) , \tilde{\bm{\Psi}}_{{\Lambda}}(z))$ charge function of $\blacksquare$ in (\ref{chargebbox})  determines the OPE relations between the charge operator $(\psi , \tilde{\psi})$ and all the gluing operators $\bm{g}$:
\begin{equation}\label{psirs}
\begin{aligned}
&\psi(z) \, \x(w)  \sim  \varphi_2(\Delta) \, \x(w) \,\psi(z)  \,, \qquad\qquad\,\,\,
\tilde{\psi}(z) \, \x(w)  \sim  \tilde\varphi_2(\Delta) \, r(w)\, \tilde{\psi}(z)\,,  \\
&\psi(z) \, \y(w)  \sim  \varphi_2^{-1}(\Delta) \, \y(w) \,\psi(z)  \,, \qquad\qquad
\tilde{\psi}(z) \, \y(w)  \sim \tilde \varphi_2^{-1}(\Delta) \,\, \y(w) \,\tilde{\psi}(z) \,. 
\end{aligned}
\end{equation}
See the thick blue arrows in Fig.~\ref{OPEeverybody2bb}. 

Actually, there is an alternative, more hand-wavy but essentially the same, method to obtain some of the OPEs above.
Since $\x$ corresponds to the twin plane partition configuration consisting of boxes at (\ref{boxboxleft}) and tilde boxes at (\ref{boxboxright}) (as shown in Figure \ref{fig:blacksquarebb}), one can think of $\x$ as composed of infinitely many $e$ and $\tilde{e}$ operators, acting on different coordinates along the $x_2$ and $\tilde{x}_2$ axes:
\begin{equation}\label{xintermsofeehat}
r(u)=\prod^{\infty}_{n=0} e(u+n h_2) \prod^{\infty}_{n=0} \tilde{e}(u+n \tilde{h}_2)\,.
\end{equation}
Passing the $\psi$ operator over all the $e$ operators in (\ref{xintermsofeehat}) immediately gives
\begin{equation}
\psi(z) \, r(w)\sim \left[\prod^{\infty}_{n=0} \varphi_3(\Delta-n h_2)\right]  r(w) \, \psi(z) =\varphi_2(\Delta)\, r(w)\,  \psi(z)\,,
\end{equation}
and similarly for $\tilde{\psi }\,\x$ OPE:
\begin{equation}
\tilde{\psi}(z) \, r(w)\sim \left[\prod^{\infty}_{n=0} \tilde{\varphi}_3(\Delta-n \tilde{h}_2)\right]  r(w) \tilde{\psi}(z) =\tilde{\varphi}_2(\Delta)\, r(w) \, \tilde{\psi}(z)\,.
\end{equation}

\subsection{Pole structure of algebra's action on twin plane partitions}
\label{sec:pole}

As in the box-antibox construction, once we fix the allowed set of twin plane partitions (see section~\ref{sec:setofTPP}) and their $(\bm{\Psi},\tilde{\bm{\Psi}})$ charge functions (see section~\ref{sec:psicharge}), one can immediately fix the action of single-box operators $\bm{s}=\{e,f,\tilde{e},\tilde{f}\}$ in (\ref{Oform}), i.e.\ both the allowed final states  $|{\Lambda}'_i\rangle$ (determined by poles $z^{*}_i$) and  the coefficients $\bm{s}[{\Lambda} \rightarrow {\Lambda}'_i]$ that appear in (\ref{Oform}).
This part is completely parallel to the box-antibox construction --- even to the case of affine Yangian of $\mathfrak{gl}_1$, since the $\square$ generated by $\{e,\psi, f\}$ is only visible to the $\bm{\Psi}_{\Lambda}(u)$ charge function whereas the  $\tilde{\square}$ generated by $\{\tilde{e},\tilde{\psi}, \tilde{f}\}$ is only visible to $\tilde{\bm{\Psi}}_{\Lambda}(u)$.

Moreover, as we explain now,  the results from section~\ref{sec:psicharge} and \ref{sec:psiOPEs} also determine, for any initial state  $|{\Lambda}\rangle$ and gluing operator $\bm{g}$, all allowed final states  $|{\Lambda}'_i\rangle$ (and therefore the associated poles $z^{*}_i$). 
(Note that it is harder to fix the coefficients $\bm{g}[{\Lambda} \rightarrow {\Lambda}'_i]$.)

Consider an arbitrary initial state $|\Lambda\rangle$ and an operator $\mathcal{O}$ in the process (\ref{Oform}). (To be generic, we consider any $\mathcal{O}$ in the list (\ref{operatorfullsetBB}), although the case of $\mathcal{O}=r$ or $s$ is the most interesting one.)
The goal is to fix all the allowed final states $|\Lambda'\rangle$ and their associated poles $w^*$.
It takes three steps.
\begin{enumerate}
\item Evaluate the $(\bm{\Psi},\tilde{\bm{\Psi}})$ charge function of the initial state, denoted by
\begin{equation}
\textrm{initial state } |\Lambda\rangle:\qquad \begin{cases}
\bm{\Psi}(u): &  \qquad \bm{\Psi}_{\Lambda}(u)\\
\tilde{\bm{\Psi}}(u): &\qquad \tilde{\bm{\Psi}}_{\Lambda}(u) \ .
\end{cases} 
\end{equation}
\item Assume that the pole of the $\mathcal{O}(z)$ action in (\ref{Oform}) is $z^{*}$, and evaluate the $(\bm{\Psi},\tilde{\bm{\Psi}})$ charge function of the putative final state, using the result of OPEs between $(\psi, \tilde{\psi})$ and all other operators computed in section~\ref{sec:psiOPEs}:
\begin{equation}\label{finalstate}
\textrm{final state } |\Lambda'\rangle:\qquad \begin{cases}
\bm{\Psi}(u): &  \qquad \bm{\Psi}_{\Lambda'}(u)=\psi[\mathcal{O}](u-z^*)\cdot\bm{\Psi}_{\Lambda}(u)\\
\tilde{\bm{\Psi}}(u): &\qquad \tilde{\bm{\Psi}}_{\Lambda'}(u)=\tilde{\psi}[\mathcal{O}](u-z^*)\cdot\tilde{\bm{\Psi}}_{\Lambda}(u) \ ,
\end{cases} 
\end{equation}
where $\psi[\mathcal{O}](u)$ and $\tilde{\psi}[\mathcal{O}](u)$ are defined as
\begin{equation}\label{operatorcharge}
\begin{aligned}
&\psi[e](u)=\varphi_3(u)\,, \quad \psi[f](u)=\varphi^{-1}_3(u)\,,\quad 
\tilde{\psi}[e](u)=1\,, \quad\qquad \tilde{\psi}[f](u)=1\,,\\
&\psi[\tilde{e}](u)=1 \,,\quad \qquad \psi[\tilde{f}](u)=1\,, \qquad\quad\ \tilde{\psi}[\tilde{e}](u)=\tilde{\varphi}_3(u) \,,\quad \tilde{\psi}[\tilde{f}](u)=\tilde{\varphi}^{-1}_3(u)\,,\\
&\psi[r](u)=\varphi_2(u) \,,\quad \psi[s](u)=\varphi^{-1}_2(u) \,,\quad 
\tilde{\psi}[r](u)=\tilde{\varphi}_2(u) \,,\quad \tilde{\psi}[s](u)=\tilde{\varphi}^{-1}_2(u)\,.
\end{aligned}
\end{equation}
(To see this, simply apply the OPEs between $(\psi,\tilde{\psi})$ and $\mathcal{O}$ computed in section~\ref{sec:psiOPEs} on the initial state $|\Lambda\rangle$.)
\item Examine the two lines in (\ref{finalstate}) separately. Demanding $\bm{\Psi}_{\Lambda'}(u)$ to correspond to a sensible left plane partition fixes a set of allowed poles 
\begin{equation}
z^{*}\in \{z^*_{\textrm{left}}\}\,.
\end{equation} 
Repeating for $\tilde{\bm{\Psi}}_{\Lambda'}(u)$ fixes 
\begin{equation}
z^{*}\in \{z^*_{\textrm{right}}\}\,.
\end{equation} 
The final states are those with
\begin{equation}\label{leftrightpole}
z^{*}=z^{*}_{\textrm{left}}=z^{*}_{\textrm{right}}\,.
\end{equation}

\end{enumerate}
This method would allow us to fix the allowed pole $p_+(\blacksquare)$ in the ansatz (\ref{xansatz}).
However, in selecting the poles that satisfy (\ref{leftrightpole}), we will need to know the relation between the parameters $\{h_i\}$ of $\mathcal{Y}$ and $\{\tilde{h}_i\}$ of $\tilde{\mathcal{Y}}$. 
Therefore the computation will take the following three steps:
\begin{enumerate}
\item Analyze the action of $r(z)$ adding the second $\blacksquare$ and determine the ``bud condition" for the second $\blacksquare$.
\item Based on the ``bud condition" for the second $\blacksquare$, determine the relation between the parameters $\{h_i\}$ of $\mathcal{Y}$ and $\{\tilde{h}_i\}$ of $\tilde{\mathcal{Y}}$.
\item Use this relation to analyze the ``bud condition" for generic $\blacksquare$ and fix the allowed pole $p_+(\blacksquare)$ in the ansatz (\ref{xansatz}).
\end{enumerate}

\subsection{Constraints on parameters from perturbative twin plane partitions}

Recall that in the box-antibox construction, the relations (\ref{3solutions})  between parameters of $\mathcal{Y}$ and $\hat{\mathcal{Y}}$ 
were solved using constraints from twin plane partitions, in particular, by an analysis of bud conditions. 
We now repeat this exercise for the $[\square,{\square}]$ construction.

\subsubsection{Bud condition for first two $\blacksquare$'s}
\label{sec:bud12}

We first derive the bud condition for the $[\square,{\square}] $ construction.
As we will see, the crucial differences from the one in the $[\square,\overline{\square}] \oplus [\overline{\square},{\square}]$ construction are that (1) now the bud has to have precisely the minimal length, and (2) it is now possible to distribute the boxes in the bud between the left and right plane partitions.

Let's start from the vacuum. 
Since
\begin{equation}\label{ronvacuum}
r(z) |\emptyset\rangle=\frac{1}{z}|\blacksquare\rangle \,,
\end{equation}
there is no need for additional boxes: the minimal bud has length zero. 
Now we show that it is also not possible to have any bud longer than this.

Take $|\square\rangle$ for example.
Similar to (\ref{xonbox}), we now have 
\begin{equation}
r(z)|\square\rangle:\qquad \begin{cases}
\begin{aligned}
\bm{\Psi}(u): &\qquad \psi_0(u) \,\varphi_3(u)\, \varphi_2(u-z^{*})  \\
\tilde{\bm{\Psi}}(u): &\qquad \tilde{\psi}_0(u)\, \tilde{\varphi}_2(u-z^*)  \ .
\end{aligned}
\end{cases}
\end{equation}
Now $z^{*}$ has to be chosen such that $\bm{\Psi}(u)$ corresponds to a valid plane partition on the left and $\tilde{\bm{\Psi}}(u)$ to a plane partition on the right.
The left plane partition demands 
\begin{equation}\label{zsolutionleft}
z^*=z^{*}_{\textrm{left}}=h_2\,.
\end{equation}
However, the right plane partition demands 
\begin{equation}\label{zsolutionright}
z^*=z^{*}_{\textrm{right}}=0\,.
\end{equation}
Since (\ref{zsolutionleft}) and (\ref{zsolutionright}) cannot hold at the same time,\footnote{
The solution $h_2=0$ is not allowed because it would make the resulting algebra degenerate. 
At the level of plane partitions, all $\square$'s with the same $(x_1,x_3)$ coordinates but different $x_2$ coordinates would contribute the same amount to the $\bm{\Psi}$ charge function, etc.} 
we conclude that
\begin{equation}\label{ronbox0}
\boxed{r(z)|\square\rangle=0}\,.
\end{equation}
This is in contrast with the corresponding process in the $[\square,\overline{\square}] \oplus [\overline{\square},{\square}]$ construction:
\begin{equation}\label{xonboxtop}
x(z)|\square\rangle=\frac{1}{z-h_2} |\blacksquare+\hat{\square}_{\textrm{ top}}\rangle\,,
\end{equation}
see around eq (\ref{xonboxresult}).
We have concluded that for the first $\blacksquare$, the bud condition is 
\begin{equation}
\ell=0\,,
\end{equation} 
as opposed to $\ell\geq 0$ in the $[\square,\overline{\square}] \oplus [\overline{\square},{\square}]$ construction.
\bigskip

We now move to $r(z)$'s action that creates the next $\blacksquare$. 
The starting point is the state created in (\ref{ronvacuum}), i.e.\ the $\blacksquare$ with position $(x_1,x_3)=(0,0)$.
The goal is to create a second $\blacksquare$ either at the position 
\begin{equation}
(x_1,x_3)=(1,0) \qquad \textrm{or} \qquad (x_1,x_3)=(0,1)\,,
\end{equation}
denoted by $\blacksquare_1$ and $\blacksquare_3$, respectively. 
Let's focus on $\blacksquare_1$ first. 
Similar to the $[\square,\overline{\square}] \oplus [\overline{\square},{\square}]$ construction, one might need additional $\square$'s acting as ``bud" for the second $\blacksquare$, i.e.\ at the position
\begin{equation}\label{budleft}
(x_1,x_3)=(1,0) \qquad \textrm{and} \qquad x_2=0,1,\cdots, s_1-1\,.
\end{equation}

With $|\blacksquare+s_1 \square_2\rangle$ as the initial state, the charge functions for the final state are 
\begin{equation}\label{rforbs1}
r(z)|\blacksquare+s_1 \square_2\rangle:\qquad \begin{cases}
\begin{aligned}
\bm{\Psi}(u): &\qquad \psi_0(u) \,\left(\prod^{s_1-1}_{j=0}\varphi_3(u-j h_2)\right) \, \varphi_2(u-z^{*})  \\
\tilde{\bm{\Psi}}(u): &\qquad \tilde{\psi}_0(u)\, \tilde{\varphi}_2(u-z^*)  \ .
\end{aligned}
\end{cases}
\end{equation}
Then demanding that $\bm{\Psi}(u)$  and $\tilde{\bm{\Psi}}(u)$ correspond to sensible left and right plane partitions, respectively, we have
\begin{equation}
z^*=z^{*}_{\textrm{left}}=h_1+s_1 h_2 \qquad \textrm{and} \qquad 
z^*=z^{*}_{\textrm{right}}=\tilde{h}_3\,.
\end{equation}
Again, since both poles are from the same $r(z)$ action, they have to coincide and we have
\begin{equation}\label{h1th3}
h_1+s_1\, h_2=\tilde{h}_3\,.
\end{equation}
Next, we repeat this exercise for the creation of $\blacksquare_3$ and obtain 
\begin{equation}\label{h3th1}
h_3+s_3\, h_2=\tilde{h}_1\,.
\end{equation}
We pause to mention that (\ref{h1th3}) and (\ref{h3th1}) are the analogues of (\ref{h1hh3}) and (\ref{h3hh1}) in the $[\square,\overline{\square}] \oplus [\overline{\square},{\square}]$ construction, respectively.
\bigskip

To proceed, we need a relation between $h_2$ and $\tilde{h}_2$. 
Recall that for the $[\square,\overline{\square}] \oplus [\overline{\square},{\square}]$ construction, the derivation of (\ref{A2}) was based on the process (\ref{xonboxtop}).
However, since now the process (\ref{xonboxtop}) is replaced by (\ref{ronbox0}), we cannot use the old derivation.

This problem is solved by noticing that different from the $[\square,\overline{\square}] \oplus [\overline{\square},{\square}]$ construction, the gluing operator $r(u)$ transforms as  
\begin{equation}
[(0, {\square},0)\,\, ,\,\,  (0,{{\square}},0)  ]
\end{equation} w.r.t.\ the two affine Yangians of $\mathfrak{gl}_1$, and therefore looks like a long row of boxes from both the left and right plane partitions.
As a consequence, there exists another set of bud conditions from the right plane partition. 
Take $\blacksquare_{\tilde{1}}$ for example, the bud can also be composed of 
$\tilde{\square}$'s, sitting at the right plane partition at
\begin{equation}
(\tilde{x}_1,\tilde{x}_3)=(0,1) \qquad \textrm{and} \qquad \tilde{x}_2=0,1,\cdots, \tilde{s}_1-1\,,
\end{equation}
mirroring (\ref{budleft}). 
And similarly for $\blacksquare_{\tilde{3}}$.

Repeating the derivation for (\ref{h1th3}) and (\ref{h3th1}) for the right plane partition, we have
\begin{equation}\label{th1h3}
\tilde{h}_1+\tilde{s}_1\, \tilde{h}_2=h_3 
\end{equation}
and
\begin{equation}\label{th3h1}
\tilde{h}_3+\tilde{s}_3\, \tilde{h}_2=h_1\,,
\end{equation}
where $\tilde{s}_1$ (resp.\ $\tilde{s}_3$) is the length of the bud from the right plane partition for $\blacksquare_{\tilde{1}}$ (reps.\ $\blacksquare_{\tilde{3}}$).

Now the $\blacksquare_{1}$ from the left plane partition actually corresponds to exactly the same row of boxes as the 
$\blacksquare_{\tilde{3}}$ from the right plane partition, and similarly for the pair with $1\leftrightarrow 3$:
\begin{equation}
\blacksquare_{1}=\blacksquare_{\tilde{3}} \qquad \textrm{and} \qquad \blacksquare_{3}=\blacksquare_{\tilde{1}}\,.
\end{equation}
Namely the conditions (\ref{h1th3}) and (\ref{th3h1}) actually are the bud conditions for exactly the same configuration. 
Therefore the number of additional boxes needed from the left should be the same as the one from the right, which gives  
\begin{equation}\label{s1ts3}
s_1=\tilde{s}_3 \,, 
\end{equation}
and similarly for (\ref{h3th1}) and (\ref{th1h3}), which gives 
\begin{equation}\label{s3ts1}
s_3=\tilde{s}_1\,.
\end{equation}

Taking the four equations (\ref{h1th3}), (\ref{h3th1}),  (\ref{th1h3}), and (\ref{th3h1}) together with (\ref{s1ts3}) and (\ref{s3ts1}), we have
\begin{equation}\label{h2th2}
\boxed{h_2=-\tilde{h}_2}
\end{equation}
which is the analogue of (\ref{A2}) in the $[\square,\overline{\square}] \oplus [\overline{\square},{\square}]$ construction.

\subsubsection{Relations between parameters of $\mathcal{Y}$ and $\tilde{\mathcal{Y}}$}

We are now ready to solve for the relation between the parameters of $\mathcal{Y}$ and $\tilde{\mathcal{Y}}$.
A priori, the parameters of $\mathcal{Y}$ and $\tilde{\mathcal{Y}}$ are
\begin{equation}\label{YYtildeparameters}
\begin{cases}
&\mathcal{Y}: \qquad h_i \, (\textrm{with}\, \sum_i h_i=0) \quad \textrm{and} \quad \psi_0\\
&\tilde{\mathcal{Y}}: \qquad \tilde{h}_i \, (\textrm{with}\, \sum_i \tilde{h}_i=0) \quad \textrm{and} \quad \tilde{\psi}_0\,,
\end{cases}
\end{equation}
subject to the scaling automorphism (\ref{automode01}) and its tilde version.
Now we can list the four constraints from the twin plane partitions.

\begin{enumerate}
\item From the left and right plane partition separately, i.e.\ for $\mathcal{Y}$ and $\tilde{\mathcal{Y}}$ separately, we have 
\begin{flalign}\label{B1}
\text{(b1)}
&&
    & \sum^3_{i=1} h_i =0 
	\qquad \textrm{and} \qquad 
	\sum^3_{i=1} \tilde{h}_i =0 \,.
&
\end{flalign}
\item In twin plane partitions, the two corners share a common $x_2$ direction, hence $h_2$ and $\tilde{h}_2$ should be directly related. 
In particular, a $\blacksquare$ looks like a long row of boxes from both the left and right plane partitions.
The discussion on the bud conditions earlier in section \ref{sec:bud12} then gives 
\begin{flalign}\label{B2}
\text{(b2)}
&&
    & h_2=-\tilde{h}_2\,.
&
\end{flalign}

\item The bud condition for adding the second $\blacksquare$, at the position $(x_1,x_3)=(1,0)$ and $(x_3,x_1)=(0,1)$, gives 
\begin{flalign}\label{B3}
\text{(b3)}
&&
    & h_1+s_1 \, h_2=\tilde{h}_3 \qquad \textrm{and} \qquad h_3+s_3 \, h_2=\tilde{h}_1 \,,
&
\end{flalign}
with $s_{1,3}\in\mathbb{Z}$ and $s_1+s_3=2$ (due to constraint (\ref{B1})).

\item Finally, we further demand that twin plane partitions do not have buds of negative length, namely the buds cannot stick out of the left or right wall. (See the constraint (\ref{xxtildebound}) and the discussion below.)
This gives 
\begin{flalign}\label{B4}
\text{(b4)}
&&
    & s_1 \geq 0 
	\qquad \textrm{and} \qquad 
	s_3 =2-s_1 \geq 0 \ ,
&
\end{flalign}
where we used $s_1+s_3=2$.
\end{enumerate}

The four constraints above give the following three sets of solutions of (\ref{YYtildeparameters}).
First, they all have  $	\tilde h_2 = -h_2$.
Then the three solution of (\ref{B4}) gives 
\begin{eqnarray}\label{3solutions2}
(s_1,s_3)=(2,0):& \qquad \tilde{h}_1=h_3 \qquad &\tilde{h}_2=-h_2\qquad \tilde{h}_3=-h_3+h_2 \,,\nonumber\\
(s_1,s_3)=(1,1):& \qquad \tilde{h}_1=-h_1 \qquad &\tilde{h}_2=-h_2\qquad  \tilde{h}_3=-h_3\,,\\
(s_1,s_3)=(0,2):& \qquad \tilde{h}_1=-h_1+h_2 \qquad &\tilde{h}_2=-h_2\qquad\tilde{h}_3=h_1 \,. \nonumber
\end{eqnarray}

Defining
\begin{equation}\label{pdef}
p\equiv \frac{s_3-s_1}{2}\,,
\end{equation}
the three cases in (\ref{3solutions2}) can be written as
\begin{equation}\label{hthp}
\tilde{h}_1=-h_1+p\, h_2 \,, \qquad \tilde{h}_2=-h_2 \,, \qquad \tilde{h}_3=-h_3-p \ h_2\,.
\end{equation}
The constraint (\ref{B4}) translates into
\begin{equation}\label{pconstraint}
|p|\leq 1 \,.
\end{equation}
And $p=-1,0,1$ correspond to the three cases in (\ref{3solutions}), respectively.
Finally, the three solutions (\ref{3solutions2}) are to be compared with the three solutions (\ref{3solutions}) for the $[\square,\overline{\square}] \oplus [\overline{\square},{\square}]$ construction. 
One can see that if we fix $h_i$, the parameters of the right affine Yangians are related by $\tilde{h}_i=-\hat{h}_i$.
We will explain in section 6 how to understand this relation.

\subsubsection{Relations to Toric Calabi-Yau}

In \cite{Li:2019nna}, a close relation between the geometry of the twin plane partition and the toric Calabi-Yau threefold in the $T^2\times \mathbb{R}$ fibration was noted. 
Remarkably, the four constraints (\ref{projection1}), (\ref{A2}), (\ref{A3}), and (\ref{A4}) match one-to-one to four constraints in defining the toric CY$_3$ with two vertices.

In the current construction, the four constraint (\ref{B1})-(\ref{B4}) from twin plane partitions can also match one-to-one to the four constraints from toric CY$_3$, once we flip the sign of the $(p,q)$ charge of the second vertex --- this correspond to the sign flip 
\begin{equation}
\tilde{h}_i=-\hat{h}_i\,.
\end{equation}
More specifically, consider the two-vertex diagram, in which the $(p,q)$ charges of the first and second vertex are
\begin{equation}\label{pqweb2}
V_i=(p,q)_i \qquad \textrm{and} \qquad \tilde{V}_i=(\tilde{p},\tilde{q})_i \qquad \textrm{with}\quad i=1,2,3\,,
\end{equation}
where all the charges are integers. 
First of all, the Calabi-Yau condition demands
\begin{flalign}\label{T1}
\text{(t1)}
&&
    & \sum^3_{i=1} V_{i}=0 \qquad \textrm{and}\qquad \sum^{3}_{i=1}\tilde{V}_i=0\,.
&
\end{flalign}
Choosing their second directions to be the joined leg between the two vertices and all directions pointing outwards, we get the second constraint
\begin{flalign}\label{T2}
\text{(t2)}
&&
    & V_2=-\tilde{V}_2\,.
&
\end{flalign}
Thirdly, the smoothness condition means $V_1\wedge V_2=\tilde{V}_1\wedge\tilde{V}_2$, which can be translated to
\begin{flalign}\label{T3}
\text{(t3)}
&&
    & V_1+c_1 V_2=\tilde{V}_3 \qquad \textrm{and} \qquad V_3+(2-c_1) V_2=\tilde{V}_1\,,
&
\end{flalign}
where $c_1\in \mathbb{Z}$. 
Finally, demanding that the four external directions in (\ref{pqweb2}) do not intersect gives the constraint
\begin{flalign}\label{T4}
\text{(t4)}
&&
    & |p|\leq 1 \qquad \textrm{where}\qquad p\equiv1-c_1 \,.
&
\end{flalign}
We see that the four constraint (\ref{B1})-(\ref{B4}) from twin plane partitions can also match one-to-one to the four constraints (\ref{T1})-(\ref{T4}) from the toric Calabi-Yau geometry, in essentially the same way as in \cite{Li:2019nna}.
The three solutions (\ref{3solutions2}) correspond to toric Calabi-Yau threefolds $\mathcal{O}(-1-p)\oplus \mathcal{O}(-1+p)\rightarrow \mathbb{P}^1$ with $p=-1,0,1$.

\begin{figure}[h!]
\begin{center}
\includegraphics[width=1\textwidth]{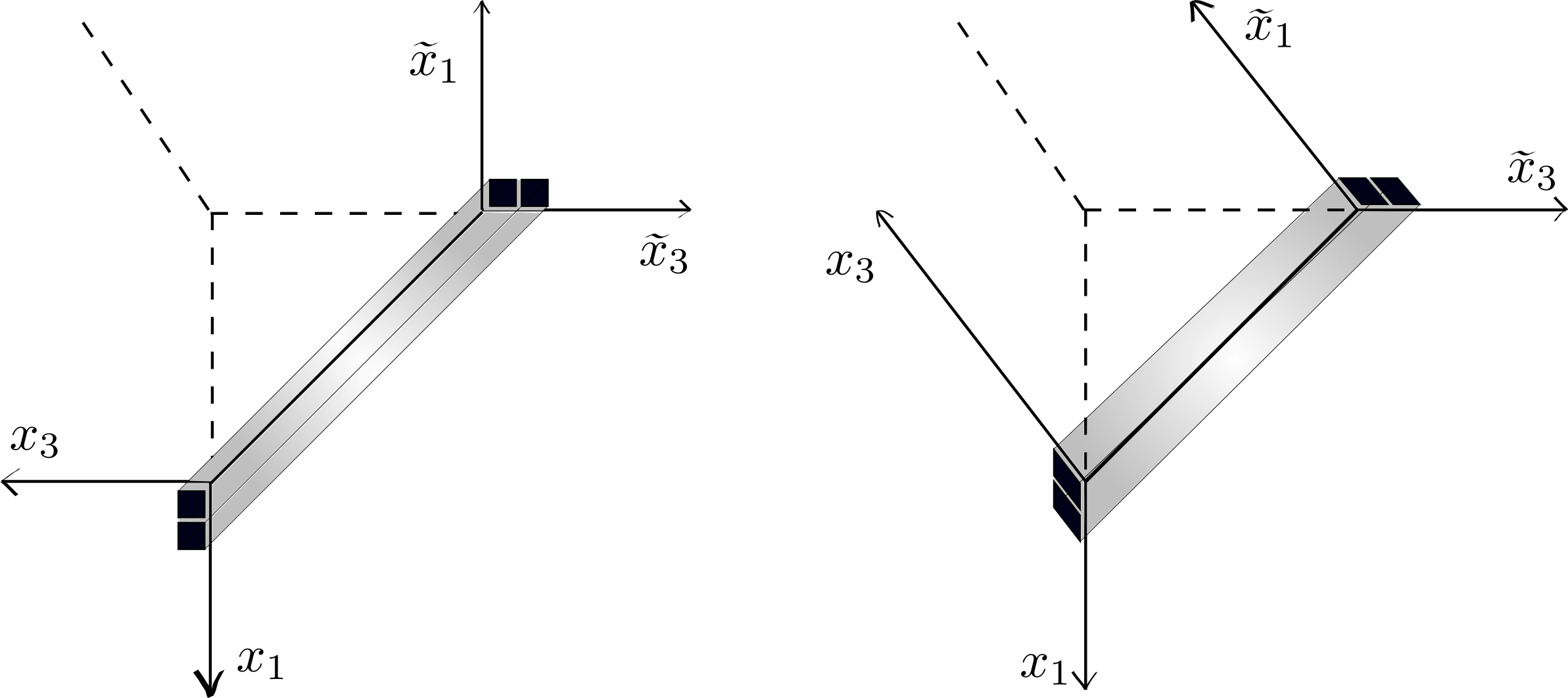}
\caption{The toric diagrams for $\mathcal{O}(-1)\oplus \mathcal{O}(-1)\rightarrow \mathbb{P}^1$ and $\mathcal{O}(-2)\rightarrow \mathbb{P}^1$, corresponding to $p=0$ and $p=-1$, respectively.
The twin plane partition $|\blacksquare\blacksquare_1\rangle= |\blacksquare\blacksquare_{\tilde{3}}\rangle$ is projected onto the diagram. 
}
\label{fig:room}
\end{center}
\end{figure}

In Figure \ref{fig:room} we show these toric diagrams for $p=0$ and $p=-1$, together with a projection of the twin plane partition $|\blacksquare\blacksquare_1\rangle= |\blacksquare\blacksquare_{\tilde{3}}\rangle$ onto the diagram.\footnote{
In \cite{Li:2019nna}, a similar picture was shown (Figure 13) to illustrate the interpretation of the shape of the room, i.e.\ where twin plane partitions reside, as the $T^3$ fibration of the toric Calabi-Yau threefolds. Note that in that picture, the projection of the rows of boxes was not to be understood as the projection of the twin plane partition itself, but merely as a schematic explanation of how the left Young diagram and the conjugate of the right Young diagram is correlated.}
\subsubsection{Bud condition for generic $\blacksquare$'s}
\label{sec:budgeneric}

With the result on the relation between the parameters $\{h_i\}$ of $\mathcal{Y}$ and $\{\tilde{h}_i\}$ of $\tilde{\mathcal{Y}}$, we are now ready to generalize the derivation for the bud condition for the second $\blacksquare$ (see section~\ref{sec:bud12}) to a generic $\blacksquare$ and obtain its pole $p_+(\blacksquare)$.
An $r(z)$ action contributes to the $(\bm{\Psi},\tilde{\bm{\Psi}})$ charge function 
\begin{equation}\label{rcharge}
r(z) :\qquad \begin{cases}
\begin{aligned}
\bm{\Psi}(u): &\qquad  \varphi_2(u-z^{*})  \\
\tilde{\bm{\Psi}}(u): &\qquad  \tilde{\varphi}_2(u-z^*)  \ .
\end{aligned}
\end{cases}
\end{equation}
On the other hand, a  $\blacksquare$ at a generic position 
\begin{equation}\label{xxtilde13}
x_1(\blacksquare)=\tilde{x}_3(\blacksquare) \qquad \textrm{and} \qquad x_3(\blacksquare)=\tilde{x}_1(\blacksquare) 
\end{equation} 
contributes to the $(\bm{\Psi},\tilde{\bm{\Psi}})$ charge function 
\begin{equation}\label{bbcharge}
\blacksquare :\qquad \begin{cases}
\begin{aligned}
\bm{\Psi}(u): &\qquad  \varphi_2(u-g(\blacksquare))  \\
\tilde{\bm{\Psi}}(u): &\qquad  \tilde{\varphi}_2(u-\tilde{g}(\blacksquare) ) \ ,
\end{aligned}
\end{cases}
\end{equation}
where $g(\blacksquare)$ and $\tilde{g}(\blacksquare)$ are the coordinate functions defined in (\ref{gcharge}).
For (\ref{rcharge}) to have a chance to reach (\ref{bbcharge}), a priori we might need a bud at either or both ends of the $\blacksquare$:
\begin{equation}
(n_{\textrm{left}},n_{\textrm{right}})-\textrm{bud}  :\quad \begin{cases}
&x_1=x_1(\blacksquare)\,,\quad x_2=0,1,\cdots, n_{\textrm{left}}-1\,,\quad\,\,\, x_{3}=x_3(\blacksquare)\,, \\
&\tilde{x}_1=x_3(\blacksquare)\,,\quad \tilde{x}_2=0,1,\cdots, n_{\textrm{right}}-1\,,\quad \tilde{x}_{3}=x_1(\blacksquare) \ ,
\end{cases}
\end{equation}
which contributes to the $(\bm{\Psi},\tilde{\bm{\Psi}})$ charge function 
\begin{equation}\label{budcharge}
(n_{\textrm{left}},n_{\textrm{right}})-\textrm{bud}  :\quad \begin{cases}
\begin{aligned}
\bm{\Psi}(u): &\quad \prod^{n_{\textrm{left}}-1}_{j=0}  \varphi_3(u-g(\blacksquare)-j h_2) \qquad (\textrm{bud of length } n_{\textrm{left}} )  \\
\tilde{\bm{\Psi}}(u): &\quad   \prod^{n_{\textrm{right}}-1}_{j=0}  \tilde{\varphi}_3(u-\tilde{g}(\blacksquare)-j \tilde{h}_2)  \quad\,\,\, (\textrm{bud of length } n_{\textrm{right}})\ .
\end{aligned}
\end{cases}
\end{equation}

Combining the charges from (\ref{rcharge}) and (\ref{budcharge}), we see that to match (\ref{bbcharge}), the $\bm{\Psi}$ and $\tilde{\bm{\Psi}}$ charge functions demand
\begin{equation}\label{poleleft}
z^*_{\textrm{left}}=g(\blacksquare)+n_{\textrm{left}}\, h_2 \qquad \textrm{and}\qquad z^*_{\textrm{right}}=\tilde{g}(\blacksquare)+n_{\textrm{right}}\, \tilde{h}_2\,,
\end{equation}
respectively, 
where we have used the following identities
\begin{equation}\label{2identities}
\varphi_3(u)=\frac{\varphi_2(u)}{\varphi_2(u-h_2)} \qquad \textrm{and} \qquad \varphi_2(-u)=\varphi_2(u-h_2) \,,
\end{equation}
together with their tilde versions. 
Demanding the two poles in (\ref{poleleft}) to coincide, we obtain the total length of the bud:
\begin{equation}\label{budlength}
B(\blacksquare)\equiv n_{\textrm{left}}+n_{\textrm{right}}=x_1(\blacksquare)(1-p)+x_3(\blacksquare)(1+p) \,,
\end{equation}
where we have used the coordinate functions in (\ref{gcharge}), the relation (\ref{hthp}) between parameters of $\mathcal{Y}$ and $\tilde{\mathcal{Y}}$, and the relation (\ref{xxtilde13}).
There is a very useful equation relating $B(\blacksquare)$ with the two coordinate functions of $\blacksquare$ defined in (\ref{gcharge}):
\begin{equation}
\tilde{g}(\blacksquare)-g(\blacksquare)=B(\blacksquare)\, h_2\,.
\end{equation}

We recognize that the length of the bud (\ref{budlength}) for a generic $\blacksquare$ is identical to the minimal length for the corresponding bud in the box-antibox construction. 
However, the main differences are that in the $[\square,\square]$ construction here, (1) the length of the bud cannot exceed (\ref{budlength}), and (2) it is possible to distribute the $B(\blacksquare)$ number of boxes into $n_{\textrm{left}}$ number of $\square$'s for the left plane partition and $n_{\textrm{right}}$ number of $\tilde{\square}$'s for the right plane partition.  
Accordingly, the pole for the action of $r(z)$ in adding $\blacksquare$ (see the second equation in the ansatz (\ref{xansatz})) can take
\begin{equation}\label{ppleft}
p_{+}(\blacksquare)=z^*_{\textrm{left}}=g(\blacksquare)+n_{\textrm{left}}\, h_2 \qquad \textrm{with}\qquad n_{\textrm{left}}=0,1,\cdots, B(\blacksquare)\,,
\end{equation}
depending on the bud configuration in the initial state $|\Lambda\rangle$.
A priori, there are $B(\blacksquare)+1$ possible bud configurations for $\blacksquare$.
Finally, the pole (\ref{ppleft}) can also be written as 
\begin{equation}\label{ppright}
p_{+}(\blacksquare)=z^*_{\textrm{right}}=\tilde{g}(\blacksquare)+n_{\textrm{right}}\, \tilde{h}_2 \qquad \textrm{with}\qquad n_{\textrm{right}}=0,1,\cdots, B(\blacksquare)\,.
\end{equation}

\subsection{Constraints on conformal dimension of gluing operators}
\label{sec:rhoparameter}

Similar to the $[\square,\overline{\square}] \oplus [\overline{\square},{\square}]$ construction, the equation involving $\psi_0$ and $\tilde{\psi}_0$ is related to the conformal dimension of the gluing operators.

The conformal dimension of the gluing operators $\{r,s\}$ is identical to the conformal weight $h_{\blacksquare}$ of the first state $|\blacksquare\rangle$ created by $r$ out of the vacuum:
\begin{equation}\label{boxstate}
|\blacksquare\rangle:\qquad \begin{cases}
\begin{aligned}
\bm{\Psi}(u): &\qquad \psi_0(u)\, \varphi_2(u)  \\
\tilde{\bm{\Psi}}(u): &\qquad \tilde{\psi}_0(u)\, \tilde{\varphi}_2(u)  \ .
\end{aligned}
\end{cases}
\end{equation}
The conformal weight of the state (\ref{boxstate}) is computed by 
\begin{equation}
L_0=\frac{1}{2}(\psi_2+\tilde{\psi}_2)\,,
\end{equation}
which gives 
\begin{equation}\label{bboxweight}
h_{\blacksquare}\equiv 1-\frac{1}{2}(h_1 h_3 \psi_0+\tilde{h}_1\tilde{h}_3\tilde{\psi}_0) =: 1+\rho\,, 
\end{equation}
where the parameter $\rho$ is defined as
\begin{equation}\label{rhodef}
\rho\equiv -\frac{1}{2}(h_1 h_3 \psi_0+\tilde{h}_1\tilde{h}_3\tilde{\psi}_0) \in \frac{1}{2} \mathbb{Z}_0\,,
\end{equation}
which has to be an integer or a half integer, since the conformal dimension $h_{\blacksquare}\in\frac{1}{2}\mathbb{Z}_0$.
Later we will show that, quite remarkably, this constraint $\rho \in\frac{1}{2} \mathbb{Z}_0$ can also come from twin plane partitions.

\subsection{Character of the algebra}

Depending on whether the gluing operators are bosonic or fermionic, the character of the glued algebra is
\begin{equation}
\chi^{\textrm{Boson}}_0(q,y) = 
	\prod_{n=1}^{\infty} 
	\frac{1}{(1-q^n)^{2n}  \(1-y q^{n+\rho}\)^{n} } \,,\label{chiB}
\end{equation}
	or
\begin{equation}
	\chi^{\textrm{Fermion}}_0(q,y) = \prod_{n=1}^{\infty} \frac{(1+y\, q^{n+\rho})^{n}}{(1-q^n)^{2n}} \,.\label{chiF}
\end{equation}

To show that they indeed reproduce the counting of the twin plane partitions, use the expansion
\begin{equation}\label{eq:char-id-bosonic}
\prod_{n=1}^{\infty} (1-y\, q^{n+\rho})^{-n} = \sum_{\lambda} y^{|\lambda|}\, \chi_{\lambda}^{({\rm w})}(q) \cdot \tilde{\chi}_{\lambda}^{({\rm w})}(q) 
\end{equation}
for the bosonic case and
\begin{equation}\label{eq:char-id-fermionic}
\prod_{n=1}^{\infty} (1+y\, q^{n+\rho})^{n} = \sum_{\lambda} y^{|\lambda|}\, \chi_{\lambda}^{({\rm w})}(q) \cdot \tilde{\chi}_{\lambda^t}^{({\rm w})}(q) 
\end{equation}
for the fermionic case, where the sum is over all possible Young diagrams $\lambda$, $|\lambda|$ is the number of boxes in $\lambda$, and $\chi_{\lambda}^{(\textrm{w})}(q)$ (resp.\ $\tilde{\chi}_{\lambda}^{(\textrm{w})}(q)$) is the wedge character --- i.e.\ the full character divided by the vacuum character ---  of the representation $\lambda$ of the left (resp.\ right) affine Yangian of $\mathfrak{gl}_1$.
Note that the main difference between the bosonic and fermionic cases is that for the fermionic one, the representations in the two factors are related by a transpose.

Plugging the two expansions (\ref{eq:char-id-bosonic}) and (\ref{eq:char-id-fermionic}) into the character (\ref{chiB}) and (\ref{chiF}), respectively, we have
\begin{equation}
\chi^{\textrm{Boson}}_0(q,y) = \sum_{\lambda} y^{|\lambda|} \, \chi_{\lambda}(q) \cdot \tilde{\chi}_{\lambda}(q)\label{chiBtpp}
\end{equation}
	or
\begin{equation}
	\chi^{\textrm{Fermion}}_0(q,y) = \sum_{\lambda} y^{|\lambda|}\,  \chi_{\lambda}(q) \cdot \tilde{\chi}_{\lambda^t}(q) \,,
 \label{chiFtpp}
\end{equation}
where we have used
\begin{equation}
\chi_{\lambda}(q)\equiv \chi_{\lambda}^{({\rm w})}(q) \cdot \chi_{\textrm{vac}}(q) \,,\qquad \textrm{where}\quad \chi_{\textrm{vac}}(q)=\prod_{n=1}^{\infty} 
	\frac{1}{(1-q^n)^{n} } \,.
\end{equation}

Let's first look at the bosonic case (\ref{chiB}).
When the gluing operators are bosonic, for each state $\lambda$ along the internal leg, it is viewed by the left and right plane partitions as $\lambda$ and $\lambda$, namely, the asymptotics of the left and right plane partitions are $(0,\lambda,0)$ and $(0,\lambda,0)$, respectively.
Thus the characters of the left and right plane partition are $\chi_{\lambda}(q)$ and $\tilde{\chi}_{\lambda}(q)$, respectively.
Summing over all possible internal states $\lambda$, each weighted by $y^{|\lambda|}$, we obtain the full vacuum character of the glued algebra given by (\ref{chiBtpp}).
The proof for the fermionic case --- namely (\ref{chiF}) is equivalent to (\ref{chiFtpp}) --- is parallel.

Finally, the character formulae (\ref{chiBtpp}) and (\ref{chiFtpp}) suggest that the symmetric direction of the left Young diagram is correlated with the symmetric (resp.\ anti-symmetric) direction of the right Young diagram if the gluing operators are bosonic (resp.\ fermionic).

\subsection{Self-statistics of gluing operators}

Finally, we explain what determines whether the gluing operators are bosonic or fermionic. 
To connect the character formulae derived above to the stacking of $\blacksquare$ by repeated action of the creation operator $r(z)$, we expand (\ref{eq:char-id-bosonic}) for the bosonic case
\begin{equation}\label{eq:bosonic-char-buds}
\begin{split}
	 & \prod_{n=1}^{\infty} (1-y\, q^{n+\rho})^{-n}
	 \\
	 = & 1 
	+ y \chi_{\blacksquare}^{({\rm w})}(q) \cdot \tilde{\chi}_{ \blacksquare}^{({\rm w})}(q) 
	+ y^2 
		\left(\chi_{\blacksquare\blacksquare}^{({\rm w})}(q) \cdot \tilde{\chi}_{{{\blacksquare\blacksquare}}}^{({\rm w})}(q) 
		+
		\chi_{{\substack{\blacksquare\\\blacksquare}}}^{({\rm w})}(q) \cdot \tilde{\chi}_{{\substack{\blacksquare\\\blacksquare}}}^{({\rm w})}(q) 
		\right)+ O(y^3)
	\\
	= & 1 
	+ y \left( q^{ h_{\blacksquare}} + \dots \right) 
	+ y^2 
	\left( 
		 q^{2h_{\blacksquare}} + q^{2h_{\blacksquare}+2} + \dots 
	\right)
	+ O(y^3)\,,
\end{split}
\end{equation}
and (\ref{eq:char-id-fermionic}) for the fermionic case
\begin{equation}\label{eq:fermionic-char-buds}
\begin{split}
	& \prod_{n=1}^{\infty} (1+y\, q^{n+\rho})^{n} \\
	 = & 1 
	+ y \chi_{\blacksquare}^{({\rm w})}(q) \cdot \tilde{\chi}_{\blacksquare}^{({\rm w})}(q) 
	+ y^2 
		\left(\chi_{\blacksquare\blacksquare}^{({\rm w})}(q) \cdot \tilde{\chi}_{{\substack{\blacksquare\\\blacksquare}}}^{({\rm w})}(q) 
		+
		\chi_{{\substack{\blacksquare\\\blacksquare}}}^{({\rm w})}(q) \cdot\tilde{ \chi}_{{\blacksquare\blacksquare}}^{({\rm w})}(q) 
		\right)
		+ O(y^3)
	\\
	= & 1 
	+ y \left( q^{h_{\blacksquare}} + \dots \right) 
	+ y^2 
	\left( 
		 2 q^{2h_{\blacksquare}+1} + \dots 
	\right)
	+ O(y^3)\,,
\end{split}
\end{equation}
where $h_{\blacksquare}=1+\rho$ is the conformal dimension of the first $\blacksquare$ and also that of the gluing operator $r(z)$; and we have used the explicit expressions for the wedge characters for the first few representations. 
Comparing the two expansions (\ref{eq:bosonic-char-buds}) and (\ref{eq:fermionic-char-buds}), we see that the first difference appears at the order $y^2$, namely, when one adds the second $\blacksquare$. 

For the bosonic case, there are two states, with conformal dimensions
\begin{equation}\label{2statesB}
h_{\blacksquare}+h_{\blacksquare} \qquad \textrm{and}\qquad h_{\blacksquare}+(h_{\blacksquare} +2 h_{\square})\,,
\end{equation}
where we have used $h_{\square}=1$.
This matches exactly with the bud condition for $(s_1,s_3)=(2,0)$ or $(s_1,s_3)=(0,2)$, or equivalently $p=\pm1$.
Namely, the first state in (\ref{2statesB}) corresponds to stacking the second $\blacksquare$ along the direction where the bud length is zero, whereas the second state corresponds to the direction where the bud length is two.

On the other hand, for the fermionic case, the two states at the order $y^2$ both have conformal dimension
\begin{equation}\label{2statesF}
 h_{\blacksquare}+(h_{\blacksquare} + h_{\square})\,.
\end{equation}
This corresponds to the case of $(s_1,s_3)=(1,1)$ (i.e.\ $p=0$), namely, stacking the second $\blacksquare$ along either direction requires a bud of length one.

Therefore we conclude that the relations (\ref{3solutions2}) between $h_i$ and $\tilde{h}_i$ are related to the self-statistics of the gluing operators by 
\begin{equation}\label{stat3cases}
\begin{aligned}
(s_1,s_3)=(2,0):& \qquad \textrm{bosonic} \quad\qquad (p=-1)\,,\\
(s_1,s_3)=(1,1):& \qquad \textrm{fermionic} \qquad (p=0)\,,\\
(s_1,s_3)=(0,2):& \qquad\textrm{bosonic} \quad \qquad (p=1)\,.
\end{aligned}
\end{equation}
Note that this is also consistent with the fact that on the vacuum, $r(z) r(w)$ vanishes when $p=0$ (because $r(z)$ needs a bud of length-one to act), suggesting a fermionic self-statistics; whereas when $p=\pm1$, $r(z) r(w)$ on the vacuum is non-vanishing (because the bud length for $r(z)$'s action is $s_3$=0 for $p=-1$ and $s_1=0$ for $p=1$), suggesting bosonic self-statistics.

\section{Glued algebra from twin plane partitions}\label{sec:algebra}

In this section, we derive all the OPEs of the glued algebra based on its action on the twin plane partitions.
We will follow the procedure in \cite{Li:2019nna}. 
However, as we will see, in certain crucial steps, e.g.\ the OPEs between single-box generators and the gluing operators and the OPEs among gluing operators, the detailed mechanisms are rather different from the corresponding ones in the box-antibox construction of  \cite{Gaberdiel:2018nbs,Li:2019nna}.

\subsection{OPEs between single-box generators and gluing generators: incomplete}
\label{sec:OPEef-xy:incom}
The OPEs between $(\psi, \tilde{\psi})$ and all other operators were already fixed in section~\ref{sec:psiOPEs}.
We now use the result from section~\ref{sec:pole} to constrain the OPEs between single-box generators $\{e,f,\tilde{e},\tilde{f}\}$ and gluing operators $\{r,s\}$.

It is easy to see that, out of the $8$ OPE relations, there are only two independent functions
\begin{equation}\label{4fns}
G(\Delta) \qquad \textrm{and}\qquad H(\Delta) \,, \end{equation}
corresponding to 
\begin{equation}\label{4OPE}
\begin{aligned}
e(z) \, \x(w) &\sim G(\Delta) \, \x(w) \, e(z) \qquad \textrm{and}\qquad f(z) \, \x(w) \sim H(\Delta) \, \x(w) \, f(z) \,. 
\end{aligned}
\end{equation}
The inverse of the two OPEs (\ref{4OPE}) are obtained by
\begin{equation}
(e,f,\x)\rightarrow (f,e,\y)\quad \Longrightarrow \quad(G,H)\rightarrow(G^{-1},H^{-1})\,.
\end{equation} 
The tilde versions of the $4$ OPEs above are obtained by 
\begin{equation}
(e,f)\rightarrow (\tilde{e},\tilde{f})\quad \Longrightarrow \quad(G,H)\rightarrow(\tilde{G},\tilde{H})\,.
\end{equation} 
in which $(\tilde{G},\tilde{H})$ is obtained by replacing all $h_i$ in $(G,H)$ by $\tilde{h}_i$.
Namely, we only need to fix the two OPEs in (\ref{4OPE}).

As we shall see below, the result of section \ref{sec:pole} only allows us to fix the two functions in (\ref{4fns}) up to two constants. 
They will be determined in the next subsection, together with the single boxes' contributions to the $\bm{\P}_{{\Lambda}}(z)$ charge functions.

\subsubsection{$e\cdot \x$ OPE and $\tilde{e}\cdot \x$ OPE}

Using the schematic expression (\ref{xintermsofeehat}), one immediately has 
\begin{equation}
\begin{aligned}
e(z)\, r(w)&\sim \left[\prod^{\infty}_{n=0} \varphi_3(\Delta-n h_2)\right]  r(w) \,e(z) =\varphi_2(\Delta)\, r(w) \,e(z)\,,\\
\tilde{e}(z) \, r(w)&\sim \left[\prod^{\infty}_{n=0} \tilde{\varphi}_3(\Delta-n \tilde{h}_2)\right]  r(w) \, \tilde{e}(z) =\tilde{\varphi}_2(\Delta)\, r(w) \, \tilde{e}(z)\,,
\end{aligned}
\end{equation}
namely
\begin{equation}
G(\Delta)=\varphi_2(\Delta)\qquad \textrm{and}\qquad \tilde{G}(\Delta)=\tilde{\varphi}_2(\Delta)\,.
\end{equation}

For a more rigorous derivation, we use the method in \cite{Gaberdiel:2018nbs, Li:2019nna}.
The basic idea is to apply the ansatz 
\begin{equation}\label{exOPE}
e(z) \, \x(w) \sim G(\Delta) \, \x(w) \, e(z)
\end{equation}
 on various initial states ${\Lambda}$.
For some ${\Lambda}$, the different order of applying $e$ and $\x$, i.e.\ $e \x$ and $\x e$, produce different sets of final states. 
Then demanding (\ref{exOPE}) to hold on such initial states constrains the function $G(\Delta)$, since it must contain factors in the numerator and denominator to cancel those final states that only appear on one side of the equation.
As we will see, although the logic is the same as in \cite{Gaberdiel:2018nbs, Li:2019nna}, the details are rather different.

First, apply (\ref{exOPE}) on the vacuum $|\emptyset\rangle$.
The l.h.s.\ gives
\begin{equation}\label{exvacuum}
	e(z) r(w) |\emptyset\rangle  
	 \sim e(z) \frac{1}{w} |\blacksquare \rangle
	 \sim \frac{1}{w} \frac{(\#)}{z-h_1} |\blacksquare  + {\square}_1\rangle
	+ \frac{1}{w} \frac{(\#)}{z-h_3} |\blacksquare  + {\square}_3\rangle
\end{equation}
and the r.h.s., without the factor $G(\Delta)$, is
\begin{equation}\label{xevacuum}
	r(w)e(z) |\emptyset\rangle  
	 \sim r(w)
	\frac{1}{z} |{\square}\rangle  
	\sim 0  \ ,
\end{equation}
due to eq.\ (\ref{ronbox0}).
Since the l.h.s.\ (\ref{exvacuum}) contains two final states whereas the r.h.s.\ (\ref{xevacuum}) contains nothing, the $G(\Delta)$ must have two factors in the denominator to cancel the two states in the l.h.s.\ (\ref{exvacuum}):
\begin{equation}\label{Gdenomenator}
G(\Delta)\supset \frac{1}{(\Delta-h_1)(\Delta-h_3)}\,.
\end{equation}

In the box-antibox construction, (\ref{xevacuum}) contains a state $|\blacksquare+\hat{\square}_{\textrm{top}}\rangle$ (due to the process (\ref{xonboxtop})), which would have given the $(\Delta-h_2)$ factor in the numerator of $G$ \cite{Gaberdiel:2018nbs, Li:2019nna}. 
But in the current construction, since
\begin{equation}
r(w)\, |\square \rangle =0\,,
\end{equation}
to obtain the factors in the numerator of $G(\Delta)$, we need a different initial state from $|\square \rangle$.
To generate the constraint on the numerator of $G(\Delta)$, we need an initial state which, when acted on by $\x(w) e(z)$, produces some final states that cannot be created by $e(z)\x(w)$.
The simplest one is 
\begin{equation}
|{\Lambda}\rangle=\begin{cases}
\begin{aligned}
&|\blacksquare+\square_1 \rangle \qquad &p=-1\\
&|\blacksquare \rangle \qquad &p=0\\
&|\blacksquare+\square_3 \rangle \qquad &p=1\,. 
\end{aligned}
\end{cases}
\end{equation}

Take the $p=-1$ case for example. 
Apply (\ref{exOPE}) on $|\blacksquare+\square_1 \rangle$.
The l.h.s.\ gives
\begin{equation}\label{exnext}
\begin{aligned}
	&e(z) r(w) |\blacksquare+\square_1 \rangle 
	 \sim e(z) \frac{(\#)}{w-h_3} |\blacksquare\blacksquare_3 +\square_1\rangle\\
	& \sim \frac{(\#)}{w-h_3}[ \frac{(\#)}{z-2h_3} |\blacksquare\blacksquare_3  + {\square}_1+\square_3\rangle
	+ \frac{(\#)}{z-h_1-h_3} |\blacksquare\blacksquare_3  + {\square}_1+\square_{1,3}\rangle\\
	&\qquad\qquad  + \frac{(\#)}{z-h_1-h_2} |\blacksquare\blacksquare_3  + \textrm{bud}_1\rangle+ \frac{(\#)}{z-2h_1} |\blacksquare\blacksquare_3  + {\square}{\square}_1\rangle
	] \,,
\end{aligned}
\end{equation}
and the r.h.s., without the factor $G(\Delta)$, is
\begin{equation}\label{xenext}
\begin{aligned}
	&r(w)e(z) |\blacksquare+\square_1 \rangle \\
	&
	 \sim r(w)\left[ \frac{(\#)}{z-h_3} |\blacksquare  + {\square}_1+\square_3\rangle
	 +\frac{(\#)}{z-h_1-h_2} |\blacksquare  + \textrm{bud}_1\rangle
	 +\frac{(\#)}{z-2h_1} |\blacksquare  + {\square}{\square}_1\rangle
	 \right] \\
&\sim \frac{(\#)}{z-h_1-h_2} \left(\frac{(\#)}{w-h_1-2h_2}  |\blacksquare\blacksquare_1\rangle +\frac{(\#)}{w-h_3}  |\blacksquare\blacksquare_3+\textrm{bud}_1\rangle\right)\\
&\quad+ \frac{(\#)}{z-2h_1}\frac{(\#)}{w-h_3} |\blacksquare\blacksquare_3  + {\square}{\square}_1\rangle\,.
\end{aligned}
\end{equation}
Compare the l.h.s.\ (\ref{exnext}) and the r.h.s.\ (\ref{xenext}).
The l.h.s.\ (\ref{exnext}) contains four final states and the  r.h.s.\ (\ref{xenext}) contains three.
Two states, $|\blacksquare\blacksquare_3  + {\square}{\square}_1\rangle$ and $|\blacksquare\blacksquare_3+\textrm{bud}_1\rangle$, appear on both sides.
The two factors in the denominator of $G(\Delta)$ fixed in (\ref{Gdenomenator}) cancel the two extra states in the l.h.s.\ (\ref{exnext}).
To cancel the extra state $ |\blacksquare\blacksquare_1\rangle $ in (\ref{xenext}), the numerator of $G(\Delta)$ must contain a factor $\Delta+h_2$.
Taking this together with (\ref{Gdenomenator}), and the fact that $G$ is quadratic, we have
\be\label{eq:e-x-OPE-coeff-incomplete}
	G(\Delta) = \frac{ (\Delta+a)(\Delta +   h_2) }{(\Delta - h_1)(\Delta - h_3)}\ ,
\ee
where $a$ is a constant to be fixed later.
The analysis for $\tilde{G}$ is parallel.

\subsubsection{$f\cdot \x$ OPE and $\tilde{f}\cdot \x$ OPE}

Closely related to the $e\cdot \x$ OPE is the $f\cdot \x$ OPE:
\begin{equation}\label{frOPE}
f(z) \, \x(w) \sim H(\Delta) \, \x(w) \, f(z)\,.
\end{equation}
Apply the ansatz (\ref{frOPE}) on the initial state $|\square\rangle$.
The l.h.s.\ annihilates $|\square\rangle$, whereas the r.h.s.\ gives
\begin{equation}
\x(w) \, f(z) |\square\rangle \sim \x(w) \, \frac{1}{z} |\emptyset\rangle \sim\frac{1}{w}  \, \frac{1}{z} |\blacksquare\rangle \,.
\end{equation}
Since the state $|\blacksquare\rangle$ only appears on the r.h.s., the numerator of $H(\Delta)$ must contain a factor of $\Delta$ to kill this state.
On the other hand, we also know that $H$ is linear, and $G$ and $H$ satisfy the following relation \cite{Gaberdiel:2018nbs}:
\begin{equation}
G(\Delta) H(\Delta)=\varphi_2(\Delta)\,.
\end{equation}
Therefore
\begin{equation}\label{Hincomplete}
	H(\Delta) = \frac{ \Delta }{\Delta+a }\ .
\end{equation}
The computation for $\tilde{H}$ is similar.

\subsection{OPEs between $Q$ and single-box operators}

In the previous subsection, we fixed all OPEs between single-box generators $\{e,f,\tilde{e},\tilde{f}\}$ and gluing operators $\{r,s\}$ up to a constant $a$. 
Now we will show that this constant can be fixed together with the OPEs between $Q$ and single-box operators $\{e,f,\tilde{e},\tilde{f}\}$, since they are both related to the single boxes' contribution to the $\textbf{Q}$ charge function. 

\subsubsection{Interlude: evaluating $\textbf{Q}$ charge function}

An important part in the construction of the glued algebra is to determine the charge function $\bm{\P}_{{\Lambda}}(z)$.
Similar to the $(\bm{\Psi}_{{\Lambda}}(z) , \tilde{\bm{\Psi}}_{{\Lambda}}(z))$ charge function (\ref{XLambda}), the $\bm{\P}_{{\Lambda}}(z)$ charge function has the decomposition
\label{step71} 
\begin{equation}\label{PPbarcomponent}
\begin{aligned}
\textbf{Q}_{{\Lambda}}(u) &= Q_0(u)\, \Biggl\{
\prod_{\blacksquare \in {\lambda}}  Q_{\,\blacksquare}(u) \,
\prod_{ {\square} \in \mathcal{E}} Q_{\,{\square}}(u) \, 
\prod_{\tilde{{\square}} \in \tilde{\mathcal{E}} } Q_{\tilde{{\square}}}(u)  \Biggr\} \ .
\end{aligned}
\end{equation}
The only difference is that while $\bm{\Psi}$ only sees $\square$ (and $\tilde{\bm{\Psi}}$ only $\tilde{\square}$), $\textbf{Q}$ sees both the $\square$ and $\tilde{\square}$.

It takes the following three steps to fix all the components in (\ref{PPbarcomponent}):
\begin{enumerate}
\item Fix single boxes' contribution to $\textbf{Q}_{{\Lambda}}(z)$
\item Fix the contribution of the first $\blacksquare$ to $\textbf{Q}_{{\Lambda}}(z)$
\item Fix all higher $\blacksquare$'s contribution to $\textbf{Q}_{{\Lambda}}(z)$
\end{enumerate}

\subsubsection{Single boxes' contribution to $\textbf{Q}$ charge function }

The single boxes' contributions to the $\textbf{Q}_{{\Lambda}}(z)$ charge function, i.e. $Q_{\,{\square}}(u)$ and $Q_{\,{\tilde{\square}}}(u)$, are related to the following two OPEs between single-box creators and gluing operators
\begin{equation}\label{Qbox}
\begin{aligned}
Q_{\,{\square}}(u): \qquad &e(z) \, \x(w) \sim G(\Delta) \, \x(w) \, e(z) \\
\textrm{and}\quad Q_{\,{\tilde{\square}}}(u):\qquad &\tilde{e}(z) \, \x(w) \sim \tilde{G}(\Delta) \, \x(w) \, \tilde{e}(z)\,,
\end{aligned}
\end{equation}
respectively.
We only need to study the first one, since the second one is its tilde version.

To understand the first equation in (\ref{Qbox}), apply the $e\cdot\x$ OPE on a generic twin plane partition $|{\Lambda}\rangle$, and we obtain
\begin{equation}
\begin{aligned}
	&
	{
	\Big[ -  \frac{1}{\sigma_3} {\rm Res}_{u = h(\square)} \bm{\Psi}_{{\Lambda}+\blacksquare}(u) \Big]^{\frac{1}{2}}
	}
	{
		\Big[ \textrm{Res}_{u=p_{+}({\blacksquare})} \, \textbf{Q}_{{\Lambda}}(u) \Big]^{\frac{1}{2}} 
	}
	\\
	& \qquad =
	G( h(\square) -p_+(\blacksquare)) \ 
	{
		\Big[ \textrm{Res}_{u=p_{+}({\blacksquare})} \, \textbf{Q}_{{\Lambda} + \square}(u) \Big]^{\frac{1}{2}} 
	}
	{
		\Big[ -  \frac{1}{\sigma_3} {\rm Res}_{u = h(\square)} \bm{\Psi}_{{\Lambda}}(u) \Big]^{\frac{1}{2}}
	} \ , 
\end{aligned}
\end{equation}
which gives
\begin{equation}\label{Pfromex}
\begin{aligned}
	Q^{1/2}_{\square}(p_+(\blacksquare) ) 
		& = G^{-1}( h( \square )  - p_+(\blacksquare) ) \ \Big[  \psi_{\blacksquare}(h( {\square} ) ) \Big]^{1/2} \,.
\end{aligned}
\end{equation}
Namely, $Q_{\square}(u)$ and $G(\Delta)$ contain the same information. 
Similarly for the tilde version.

Recall that $G(\Delta)$ was fixed by the pole structures of the $e$ and $x$ actions up to one constant $a$ in (\ref{eq:e-x-OPE-coeff-incomplete}). 
Taking the result of $\psi_{\blacksquare}$ in (\ref{chargebbox}), we see the most natural solution for (\ref{Pfromex}) is 
\begin{equation}
a=0
\end{equation}
in  (\ref{eq:e-x-OPE-coeff-incomplete}), which would give
\begin{equation}\label{Pbox1}
Q_{{\square}}(u)=\varphi^{-1}_2(-u+h({\square}))
\qquad \textrm{and}\qquad
G(\Delta)=\varphi_2(\Delta)\,.
\end{equation}
Similarly for the tilde version
\begin{equation}\label{Pbox1t}
Q_{\tilde{\square}}(u)=\tilde{\varphi}^{-1}_2(-u+\tilde{h}(\tilde{\square}))
\qquad \textrm{and}\qquad
\tilde{G}(\Delta)=\tilde{\varphi}_2(\Delta)\,.
\end{equation}
This also fixes the $H$ and $\tilde{H}$:
\begin{equation}\label{HHt}
\begin{aligned}
H(\Delta)=\frac{\Delta}{\Delta} \qquad \textrm{and}\qquad
\tilde{H}(\Delta)=\frac{\Delta}{\Delta}\,.
\end{aligned}
\end{equation}

\subsubsection{OPEs between $Q$ and single-box operators}
\label{sec:OPEQsingle}

\begin{figure}[h!]
	\centering
\includegraphics[trim=2cm 14cm 4cm 4cm, width=0.9\textwidth]{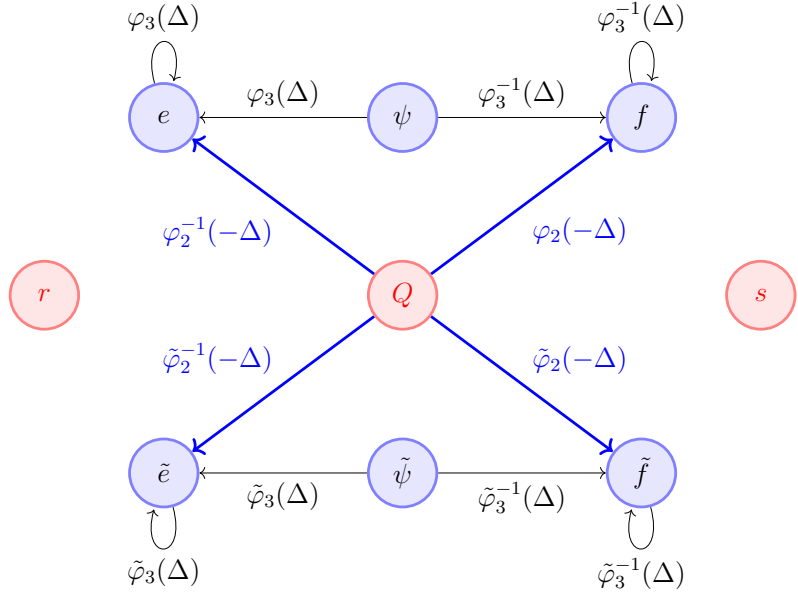}
\caption{OPEs between the charge operator $Q$ and the four single box generators $\{\e,f,\tilde{e},\tilde{f}\}$.}
	\label{OPEfermionicPebb}
\end{figure}

The results on the single boxes' contributions to $\textbf{Q}_{{\Lambda}}(z)$ immediately give us the OPEs between $Q$ and the four single-box generators $\{\e,f,\tilde{e},\tilde{f}\}$:
\begin{equation}\label{Qsingle}
\begin{aligned}
&Q(z) \,e(w) \sim \varphi^{-1}_2(-\Delta) \, e(w)\,Q(z)\,, \qquad \qquad Q(z) \,f(w) \sim \varphi_2(-\Delta) \, f(w)\, Q(z)\,, \\
&Q(z) \,\tilde{e}(w) \sim \tilde{\varphi}^{-1}_2(-\Delta) \, \tilde{e}(w) \,Q(z)\,, \qquad \qquad Q(z) \,\tilde{f}(w) \sim \tilde{\varphi}_2(-\Delta) \, \tilde{f}(w)\,Q(z)\,, 
\end{aligned}
\end{equation}
shown by thick blue lines in Figure~\ref{OPEfermionicPebb}.

The derivation is parallel to the one for $\psi e$ OPE in section~\ref{sec:psiOPEs}. 
Take the $Q\cdot e$ OPE for example. 
Applying the ansatz
\begin{equation}\label{Pe}
Q(z)e(w)\sim K(z-w) \,e(w) Q(z)
\end{equation} 
 on an arbitrary twin plane partition $|\Lambda \rangle$, and using the action of $Q$ in (\ref{xansatz}), we get
\begin{equation}\label{QeOPEstep}
\begin{aligned}
 &\qquad\qquad\sum_{\square \in \textrm{Add}(\Lambda)} \frac{E(\Lambda\rightarrow \Lambda+\square)}{w-h(\square)} \, \textbf{Q}_{\Lambda+\square}(z) \, |\Lambda+\square\rangle \\
=&
 \sum_{\square \in \textrm{Add}(\Lambda)} K(z-w) \,\frac{E(\Lambda\rightarrow \Lambda+\square)}{w-h(\square)} \,\textbf{Q}_{\Lambda}(z)\, |\Lambda+\square\rangle\,,
\end{aligned}
\end{equation}
which gives
\begin{equation}
K(z-h(\square))=\frac{\textbf{Q}_{\Lambda+\square}(z)}{\textbf{Q}_{\Lambda}(z)}=Q_{\square}(z)=\varphi^{-1}_2(-z+h(\square))\,,
\end{equation}
where in the last step we have used the result $Q_{\square}(z)$ (\ref{Pbox1}). 
This fixes the  OPE (\ref{Pe}) to be
\begin{equation}
Q(z) \,e(w) \sim \varphi^{-1}_2(-\Delta) \,e(w)\,Q(z)\,.
\end{equation}
The other three cases are completely parallel.

\subsubsection{OPEs between single-box generators and gluing generators: final}

\begin{figure}[h!]
	\centering
\includegraphics[trim=2cm 14cm 4cm 4cm, width=0.9\textwidth]{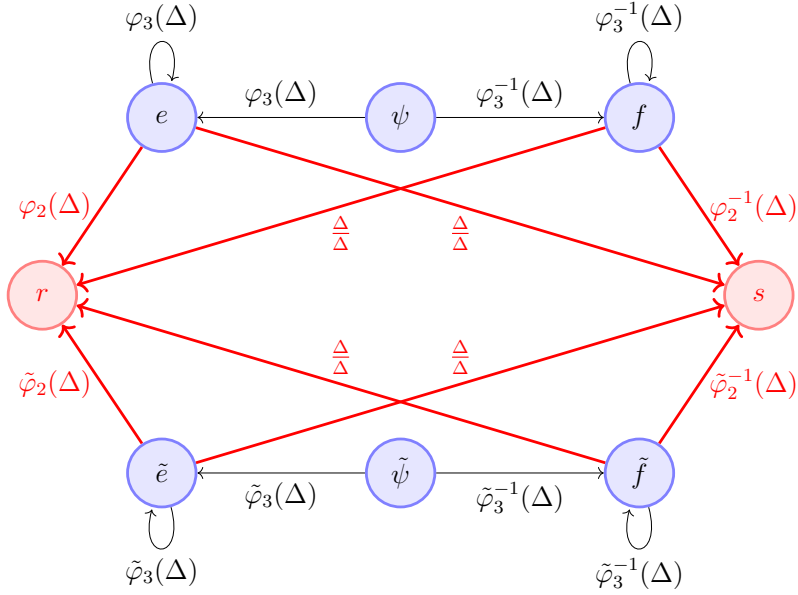}
\caption{OPEs between the single-box generators $\{e,f, \tilde{e},\tilde{f}\}$ and the gluing operators $\{r,s\}$. 
	}
	\label{OPEbosonicxy-conjbbfinal}
\end{figure}

With (\ref{Pbox1}), (\ref{Pbox1t}), and (\ref{HHt}), we can now write down the OPEs between $\{e,f,\tilde{e},\tilde{f}\}$ and the creation operator $\x$:
\be\label{singlegluingOPE1}
\begin{aligned}
e(z) \, \x(w) & \sim \varphi_2(\Delta)\, \x(w) \, e(z) \ ,  
 \  & \qquad
f(z) \, \x(w) & \sim \frac{\Delta}{\Delta} \, \x(w) \, f(z) \ , \\
\tilde{e}(z) \, \x(w) & \sim \tilde{\varphi}_2(\Delta) \, \x(w) \, \tilde{e}(z) \ , 
 \  &\qquad
\tilde{f}(z) \, \x(w) & \sim \frac{\Delta}{\Delta} \, \x(w) \, \tilde{f}(z) \ ;
\end{aligned}
\ee
and those for the corresponding annihilation operator $s$: 
\be\label{singlegluingOPE2}
\begin{aligned}
e(z) \, \y(w) & \sim \frac{\Delta}{\Delta} \, \y(w) \, e(z) \ ,  
 \  &\qquad
f(z) \, \y(w) & \sim \varphi^{-1}_2(\Delta) \, \y(w) \, f(z) \ , \\
\tilde{e}(z) \, \y(w) & \sim \frac{\Delta}{\Delta} \, \y(w) \, \tilde{e}(z) \ , 
 \  &\qquad
\tilde{f}(z) \, \y(w) & \sim \tilde{\varphi}^{-1}_2(\Delta)  \, \y(w) \, \tilde{f}(z) \ .
\end{aligned}
\ee
See the thick red arrows in Fig.~\ref{OPEbosonicxy-conjbbfinal}.

\subsection{OPEs among $\{r,Q,s\}$}

In this subsection, we fix the OPEs among $\{r, Q, s\}$.
This will also allow us to fix  $\blacksquare$'s contribution to the $\textbf{Q}$ charge function.

Given the similarity between the generators $(e,\psi, f)$ for $\square$ in (\ref{N2BBleft}) and the generators $(r,Q,s)$ for $\blacksquare$ in (\ref{N2BBfermion}), and in particular, the similarity between the action of $(e,\psi,f)$ on twin plane partitions (\ref{ppartpsi}) and the action of $(r,Q,s)$ (\ref{xansatz}),
we can write down the ansatz for the OPEs among $(r,Q,s)$ in a similar form to those among $(e,\psi,f)$ in (\ref{bosonicdef}):
\begin{equation}\label{Pxy}
\begin{aligned}
\begin{aligned}
Q(z)\, \x(w)&\sim S_p(\Delta) \ \x(w)\ Q(z) \,, \\
Q(z)\, \y(w)&\sim S^{-1}_p(\Delta)\ \y(w)\ Q(z)\,, 
\end{aligned}
&\qquad
\begin{aligned}
\x(z)\ \x(w) &\sim \epsilon_p\, S_p(\Delta) \ \x(w)\ \x(z)\,, \\
\y(z)\ \y(w) &\sim  \epsilon_p\, S^{-1}_p(\Delta)\ \y(w)\ \y(z)\,, \\
\end{aligned}\\
[\x(z), \y(w)]_{ \epsilon_p\,} & \sim \frac{Q(z)-Q(w)}{z-w}\,,
\end{aligned}
\end{equation}
where $S_p(\Delta)$ plays the role similar to the $\varphi_3(u)$ function for the $\square$: it reflects the change to the $\textbf{Q}$ charge function after we apply an $\x$ operator; and 
\begin{equation}\label{eq:epsilon-p}
	\epsilon_p \equiv \begin{cases}
		-1 \qquad &  (p  = 0) \\
		1 \qquad &  (p  = \pm1) 
	\end{cases}
\end{equation}
reflects the self-statistics of the gluing operators.
Note that the ansatz (\ref{Pxy}) is compatible with the ansatz (\ref{xansatz}) for the actions of $\{r,Q,s\}$ on twin plane partitions.

\subsubsection{Relation between $S_p$ and $\blacksquare$'s contribution to $\textbf{Q}$ charge function}

In the ansatz (\ref{Pxy}), $S_p$ plays the role similar to the $\varphi_3$ function for the OPEs among $\{e,\psi,f\}$ in (\ref{bosonicdef}).
Recall that $\varphi_3(z-h(\square))$ gives the contribution of $\square$ to the $\bm{\Psi}$ charge function, see (\ref{psichargebox}). 
In other words, it records the change to the $\bm{\Psi}$ charge function after we apply the $e$ operator once:
\begin{equation}
\frac{\bm{\Psi}_{{\Lambda}+\square}(z)}{\bm{\Psi}_{{\Lambda}}(z) }=\varphi_3(z-h(\square))\,,
\end{equation}
where $h(\square)$, as the coordinate function of the $\square$ added, is also the pole of the $e(z)$'s action.

For the $\blacksquare$,  applying 
\begin{equation}\label{QrOPE}
Q(z)\ r(w)\sim S_p(\Delta)\  r(w)\ Q(z)
\end{equation} in (\ref{Pxy}) on an arbitrary twin plane partition $|\Lambda\rangle$ and then using (\ref{xansatz}), we see that 
$S_p$ records the change to the $\textbf{Q}$ charge function after the application of $r$ operator on $|\Lambda\rangle$:
\begin{equation}\label{rchangeQ}
\frac{\textbf{Q}_{[{\Lambda}+\blacksquare]}(z)}{\textbf{Q}_{{\Lambda}}(z) }=S_p(z-p_{+}(\blacksquare))\,,
\end{equation}
where $|[{\Lambda}+\blacksquare]\rangle$ denotes the final configuration after we apply $r$ on $\Lambda$, which is not always $|{\Lambda}+\blacksquare\rangle$, in particular, the bud has disappeared into the $\blacksquare$ just created.
And $p_+(\blacksquare)$ is the pole for adding $\blacksquare$ in (\ref{xansatz}), given in (\ref{ppleft}) or equivalently in (\ref{ppright}).
Note that unlike the case for $\square$, the pole $p_+(\blacksquare)$ is not always equal to the coordinate function $g(\blacksquare)$ of $\blacksquare$ (\ref{gcharge}).

The relation (\ref{rchangeQ}) allows us to fix the contribution of $\blacksquare$ to the $\textbf{Q}$ charge function recursively.
First, consider the vacuum $|{\Lambda}\rangle=|\emptyset\rangle$.
The pole for adding the first $\blacksquare$ is
\begin{equation}
p_+(\blacksquare)=0 \qquad \textrm{for} \qquad g(\blacksquare)=0\,,
\end{equation} 
since $n_{\textrm{left}}=B(\blacksquare)=0$ for the first $\blacksquare$.
The relation (\ref{rchangeQ}) specializes to
\begin{equation}
\frac{\textbf{Q}_{\blacksquare}(u)}{\textbf{Q}_{0}(u) }=S_p(u) \qquad \textrm{for} \qquad g(\blacksquare)=0\,.
\end{equation}
Using the decomposition of $\textbf{Q}$ in (\ref{PPbarcomponent}), in particular,
\begin{equation}
\textbf{Q}_{0}(u)=Q_0(u) \qquad \textrm{and}\qquad \textbf{Q}_{\blacksquare}(u)=Q_0(z)Q_{\blacksquare}(u)\,,
\end{equation}
we immediately see that the first $\blacksquare$ created from the vacuum contributes to the $\P$ charge function by 
\begin{equation}\label{Pblacksquare0}
Q_{\blacksquare}(u)=S_p(u) \qquad \textrm{with} \qquad g(\blacksquare)=0\,.
\end{equation}
To proceed, we need to fix the function $S_p(\Delta)$.

\subsubsection{Fixing $S_p$}

The strategy for fixing the $S_p$ function is to apply the relation (\ref{rchangeQ}) on various initial states $|\Lambda\rangle$ and obtain constraints by comparing different resulting equations. 

Recall that for the $[\square,\overline{\square}] \oplus [\overline{\square},{\square}]$ construction, a crucial step in deriving $S_p$ was to use the process (\ref{xonboxtop}).
Since now in the $[\square,\square]$ construction, the process (\ref{xonboxtop}) is replaced by (\ref{ronbox0}), we cannot use the old derivation. Instead, we use the fact that in the $[\square,\square]$ construction, the $\blacksquare$ looks like a long row of boxes from both sides, and the bud for growing higher $\blacksquare$'s can sit at both corners. 

In particular, consider the process that generates the state $|\blacksquare\blacksquare_1\rangle$ starting from $|\blacksquare\rangle$.
Since the require bud can sit at both corners, we can consider two initial states,\footnote{
In fact, we can also consider the initial state where the bud is broken into two pieces, one on the left and one on the right. 
However, this does not give any additional constraint when fixing $S_p$.
} 
namely $|\blacksquare+\textrm{bud}(\blacksquare_1)\rangle$ and $|\blacksquare+\textrm{bud}(\blacksquare_{\tilde{3}})\rangle$, on which applying $r$ can generate  the state $|\blacksquare\blacksquare_1\rangle$. 
Let's now consider them in turn.  

\begin{enumerate}
\item Apply $\x$ on
$|{\Lambda}_1 \rangle\equiv |\blacksquare+\textrm{bud}(\blacksquare_1)\rangle$,
where
\begin{equation}
\textrm{bud}(\blacksquare_1)=\{ \square_k \,|\,k=0,\cdots, B(\blacksquare_1)-1\} =\{\square_{(1,k,0)} \,|\,k=0,\cdots, (1-p)-1\}\,,
\end{equation}
where $\square_{k}=\square_{(1,k,0)}$ denotes the $k^{\textrm{th}}$ $\square$ in the left bud, at the position  $(x_1,x_2,x_3)=(1,k,0)$.
The pole of the action is 
\begin{equation}\label{pole11}
p_+(\blacksquare_1)=h_1+(1-p)h_2\,.
\end{equation}

Since we have already derived the contribution to the $\textbf{Q}$ charge function from the first $\blacksquare$ (\ref{Pblacksquare0}) and from individual $\square$'s (\ref{Pbox1}), we can immediately write down the charge function of the initial state:
\begin{equation}\label{Qcharge1}
\textbf{Q}_{{\Lambda}_1}(u)=Q_0(u)\, S_p(u)\prod^{(1-p)-1}_{k=0}\varphi^{-1}_2(-u+h_1+k h_2) \,.
\end{equation}
Applying (\ref{rchangeQ}) with the position of the pole (\ref{pole11}), we get the charge function for the  final state
\begin{equation}
\textbf{Q}_{[{\Lambda}_1+\blacksquare]}(u)=S_p(u-h_1-(1-p)h_2)\ \textbf{Q}_{{\Lambda}_1}(u)\,.
\end{equation}

\item Apply $\x$ on $|{\Lambda}_{\tilde{3}} \rangle\equiv |\blacksquare+\textrm{bud}(\blacksquare_{\tilde{3}})\rangle$, where
\begin{equation}
\textrm{bud}(\blacksquare_{\tilde{3}})=\{\tilde{\square}_{k} \,|\,k=0,\cdots, B(\blacksquare_{\tilde{3}})-1\} =\{\tilde{\square}_{(0,k,1)}\,|\,k=0,\cdots, (1-p)-1\}\,,
\end{equation}
where $\tilde{\square}_{k}=\tilde{\square}_{(0,k,1)}$ denotes the $k^{\textrm{th}}$ $\tilde{\square}$ in the right bud, at the position  $(\tilde{x}_1,\tilde{x}_2,\tilde{x}_3)=(0,k,1)$.
The charge function of the initial state is
\begin{equation}\label{Qcharge3t}
\textbf{Q}_{{\Lambda}_{\tilde{3}}}(u)=Q_0(u)\, S_p(u)\prod^{(1-p)-1}_{k=0}\tilde{\varphi}^{-1}_2(-u+\tilde{h}_3+k \tilde{h}_2) \,,
\end{equation}
and the one for the final state is
\begin{equation}
\textbf{Q}_{[{\Lambda}_{\tilde{3}}+\blacksquare]}(u)=S_p(u-\tilde{h}_3-(1-p)\tilde{h}_2)\ \textbf{Q}_{{\Lambda}_{\tilde{3}}}(u)\,.
\end{equation}
\end{enumerate}

Now, since the two final states are the same, we have
\begin{equation}
\textbf{Q}_{[{\Lambda}_1+\blacksquare]}(u)=\textbf{Q}_{[{\Lambda}_{\tilde{3}}+\blacksquare]}(u)\,,
\end{equation}
namely
\begin{equation}\label{Q1Q3t}
\textbf{Q}_{{\Lambda}_1}(u)\, S_p(u-h_1-(1-p)h_2)=\textbf{Q}_{{\Lambda}_{\tilde{3}}}(u)\, S_p(u-\tilde{h}_3-(1-p)\tilde{h}_2)\,.
\end{equation}
Plugging the $\textbf{Q}$ charge functions (\ref{Qcharge1}) and (\ref{Qcharge3t}) into the relation (\ref{Q1Q3t}) then gives the  constraint
\begin{equation}\label{Sp1}
\frac{S_p(u-h_1-(1-p)h_2)}{S_p(u-\tilde{h}_3-(1-p)\tilde{h}_2)}=\prod^{(1-p)-1}_{k=0}\frac{\varphi_2(-u+h_1+k h_2)}{\tilde{\varphi}_2(-u+\tilde{h}_3+k \tilde{h}_2)}\,.
\end{equation}

Now we can repeat this exercise for the final state $|\blacksquare\blacksquare_3\rangle$, and obtain a constraint mirroring (\ref{Sp1}):
\begin{equation}\label{Sp3}
\frac{S_p(u-h_3-(1+p)h_2)}{S_p(u-\tilde{h}_1-(1+p)\tilde{h}_2)}=\prod^{(1+p)-1}_{k=0}\frac{\varphi_2(-u+h_3+k h_2)}{\tilde{\varphi}_2(-u+\tilde{h}_1+k \tilde{h}_2)}\,.
\end{equation}

\begin{figure}[h!]
	\centering
\includegraphics[trim=7cm 14cm 3cm 4cm, width=0.75\textwidth]{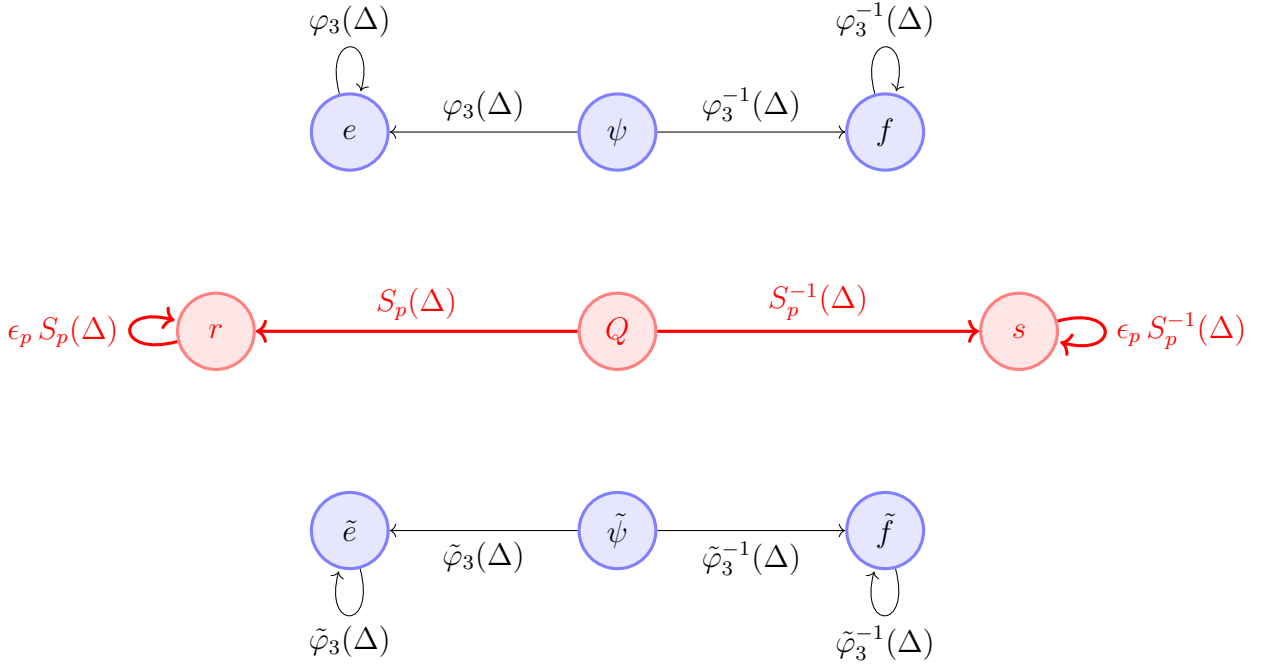}
\caption{OPEs among gluing operators. 
	}
	\label{OPEfermionicxybb}
\end{figure}

Simplifying (\ref{Sp1}) and (\ref{Sp3}), we arrive at
\begin{equation}\label{SpC1}
\frac{S_{p}(u-h_2)}{S_p(u)}=\frac{\varphi_2(u-h_2)}{\tilde{\varphi}_2(u)} \qquad \textrm{for}\quad p=0,\pm1\,.
\end{equation}
In addition, examining the $r\cdot r$ OPE in the ansatz (\ref{Pxy}), we see that $S_p$ needs to satisfy
\begin{equation}\label{SpC2}
S_p^{-1}(u)=S_p(-u) \,.
\end{equation}
Combining the two constraints (\ref{SpC1}) and (\ref{SpC2}), we can solve $S_p$: 
\begin{equation}\label{SpSolution}
	{S_p(u) = \frac{u+\delta_p}{u-\delta_p} }
	\qquad\textrm{with}\qquad
	\delta_p = \begin{cases}
	h_1 \qquad & \ (p=1) \\
	 0 \qquad & \ (p=0) \\
	h_3 \qquad & \ (p=-1) \\
	\end{cases} \,.
\end{equation}
Thus we have fixed the OPE among $\{r,Q,s\}$ in (\ref{Pxy}), where $S_p$ is given by (\ref{SpSolution}). 
They are shown by thick red lines in Figure~\ref{OPEfermionicxybb}.
The result of $S_p$ also tells us the contribution of the first $\blacksquare$ to the $\textbf{Q}$ charge function in (\ref{Pblacksquare0}).

\subsubsection{$\blacksquare$'s contribution to $\textbf{Q}$ charge function}
\label{sec:blackbforQ}

It remains to compute the contribution of higher $\blacksquare$ to the $\textbf{Q}$ charge function. 
Consider the process of creating the $\blacksquare$ at position $(x_1,x_3)$ on top of an initial state $|\Lambda+\textrm{bud}(\blacksquare)\rangle$: 
\begin{equation}
r(z)|\Lambda+\textrm{bud}(\blacksquare)\rangle\rightarrow\frac{\#}{z-p_+(\blacksquare)}|\Lambda+\blacksquare\rangle\,,
\end{equation}
where without loss of generality we choose the bud configuration such that all boxes are on the left side:
\begin{equation}
\textrm{bud}(\blacksquare)=\{\tilde{\square}_{(x_1, k, x_3)} \,|\,k=0,\cdots, B(\blacksquare)-1\} \,,
\end{equation}
whose contribution to the $\textbf{Q}$ charge function is
\begin{equation}
Q_{\textrm{bud}(\blacksquare)}(z)= \prod^{B(\blacksquare)-1}_{k=0}\varphi^{-1}_2(-z+g(\blacksquare)+k h_2) \,,
\end{equation}
where $g(\blacksquare)$ is the $\blacksquare$'s coordinate function defined in (\ref{gcharge}). 
And the pole $p_{+}(\blacksquare)$ for this initial configuration is
\begin{equation}\label{poleleftforQ}
p_{+}(\blacksquare)=g(\blacksquare)+B(\blacksquare)\, h_2\,.
\end{equation}

Applying (\ref{QrOPE}) on the initial state $|\Lambda+\textrm{bud}(\blacksquare)\rangle$, we have
\begin{equation}
\textbf{Q}_{\Lambda+\blacksquare}(z)=S_p(z-p_+(\blacksquare))\ \textbf{Q}_{\Lambda+\textrm{bud}(\blacksquare)}(z)\,.
\end{equation}
Now using the decomposition of the $\textbf{Q}$ charge function (\ref{PPbarcomponent}) and canceling the $\textbf{Q}_{\Lambda}$ on both sides, we arrive at 
\begin{equation}\label{PblacksquareGene}
Q_{\blacksquare}(z)= S_p(z-p_{+}(\blacksquare)) \prod^{B(\blacksquare)-1}_{k=0}\varphi^{-1}_2(-z+g(\blacksquare)+k h_2) \,.
\end{equation}
Using (\ref{poleleftforQ}) and the recursion relation (\ref{SpC1}), the function (\ref{PblacksquareGene}) can also be written as
\begin{equation}\label{PblacksquareGene1}
Q_{\blacksquare}(z)= S_p(z-g(\blacksquare)) \prod^{B(\blacksquare)-1}_{k=0}\tilde{\varphi}^{-1}_2(-z+\tilde{g}(\blacksquare)+k\, \tilde{h}_2) \,,
\end{equation}
which is what we would have obtained directly if we had started with the initial condition where all the boxes in the bud are in the right plane partition.
It is reassuring to see that the two expressions (\ref{PblacksquareGene}) and (\ref{PblacksquareGene1}) are indeed equivalent.

\section{Extension of algebra with $[\overline{\square},\overline{\square}]$ gluing operators}
\label{sec:extended}

In section 3 and 4, we constructed the algebra with gluing operators transforming as $[\square, {\square}]$ w.r.t.\ the two affine Yangians of $\mathfrak{gl}_1$, using their action on the representation space of a pair of plane partitions.
This pair of plane partitions is the complement of a crystal with two corners and their connecting edge partially melted, generalizing the melting crystal picture for the $\mathcal{W}_{1+\infty}$ algebra (in the context of topological strings).

Now we will show that the algebra constructed in section 3 and 4 allows an extension with a pair of additional gluing operators transforming as $[\overline{\square}, \overline{\square}]$ w.r.t.\ the two affine Yangians of $\mathfrak{gl}_1$.
Instead of also extending the representation space (by pairs of plane partitions connected by ``high walls"), we will still use  the space of twin plane partitions defined in section~\ref{sec:ptpp}, which has simpler and more transparent geometric interpretation, and show how the additional operators act on them.
Quite remarkably, the extended algebra can still be derived based on constraints from these actions. 

\subsection{Extension with $[\overline{\square},\overline{\square}]$ gluing operators}

Extending (\ref{GC2bb}) with a pair of creation and annihilation operators that transform as
\begin{equation}\label{bimodule2AB}
 [(0, \overline{\square},0) \,\, ,\,\,  (0,\overline{\square},0)]=:[\overline{\square}, \overline{\square}] 
\end{equation}
w.r.t.\ the two affine Yangians of $\mathfrak{gl}_1$ gives a glued algebra
\begin{equation}\label{GC2}
\mathcal{Y}(q)\,\oplus\, \tilde{\mathcal{Y}}(\tilde{q})\, \oplus\, [\square, \square] \,\oplus\, [\overline{\square}, \overline{\square}]\,.
\end{equation}

Naively, one might follow (\ref{BB}) and define 
$\overline{\blacksquare}\equiv [\overline {\square},\overline{\square}] $ as the individual building block created by the new gluing operator transforming as (\ref{bimodule2AB}) and its corresponding creation, annihilation, and charge operators similar to (\ref{N2BBfermion}):
\begin{equation}\label{N2ABABfermion}
\begin{aligned}
\overline{\blacksquare}: \qquad \qquad \bar{\x}:\, \textrm{creation} \qquad  \bar{Q}:\, \textrm{charge}  \qquad  \bar{\y}:\, \textrm{annihilation}. \\
\end{aligned}
\end{equation}
However, as we show in appendix A, the plane partition configuration that corresponds to the building block $\overline{\blacksquare}$ is a ``high wall" connecting the left and right plane partitions. 
Since one of the motivations for the current paper is to have a more elegant set of plane partition representations than the box-antibox construction, which requires gluing of ``long rows" with ``high walls", we restrict ourselves to the nicer perturbative twin plane partition representation constructed in section~\ref{sec:ptpp}, namely, we will not create high walls made of $\overline{\blacksquare}$'s. 
The additional operators will only act on the original twin plane partitions.

Similar to (\ref{N2BBfermion}), we again have a triplet of operators $\{\bar{r},\bar{Q},\bar{s}\}$ that transform as $ [\overline {\square},\overline{\square}] $ w.r.t.\ the two affine Yangians of $\mathfrak{gl}_1$.
To write down the ansatz for their action on the twin plane partition, the intuition is that the $\blacksquare$ and $\overline{\blacksquare}$ should annihilate each other, leaving
a few $\square$ and $\tilde{\square}$ behind, since both $\blacksquare$ and $\overline{\blacksquare}$ have positive conformal dimensions.
Therefore, eq.\ (\ref{N2ABABfermion}) should map to
\begin{equation}\label{N2ABABfermionnew}
\begin{aligned}
{\blacksquare}: \qquad \qquad \bar{\x}:\, \textrm{remove} \qquad  \bar{Q}:\, \textrm{charge}  \qquad  \bar{\y}:\, \textrm{add}
\end{aligned}
\end{equation}
and their action on $|\Lambda\rangle$ take the form
\begin{equation}\label{rbaransatz}
\begin{aligned}
\bar{Q}(u) \, |{\Lambda}\rangle &= \bar{\textbf{Q}}_{{\Lambda}}(u) \, |{\Lambda}\rangle\, ,\\
\bar{r}(w)|{\Lambda}\rangle &= \sum_{\blacksquare\in \textrm{Rem}(\Lambda)} \frac{
 \Big[ \textrm{Res}_{u=\bar{p}_{-}({\blacksquare})} \, \bar{\textbf{Q}}_{{\Lambda}}(u) \Big]^{\frac{1}{2}} 
}{w-\bar{p}_-({\blacksquare})}| [{\Lambda}-{\blacksquare}]\rangle  \ , \\
\bar{s}(w)|{\Lambda}\rangle &= \sum_{\blacksquare\in \textrm{Add}(\Lambda)} 
\frac{\Big[ \textrm{Res}_{u=\bar{p}_{+}({\blacksquare})} \, \bar{\textbf{Q}}_{{\Lambda}}(u) \Big]^{\frac{1}{2}}
}{w-\bar{p}_{+}({\blacksquare})}|[{\Lambda}+{\blacksquare}]\rangle  \ , 
\end{aligned}
\end{equation}
where similar to (\ref{PPbarcomponent}), the $\bar{\textbf{Q}}$ charge function has the decomposition 
\begin{equation}\label{PPbarcomponentABAB}
\begin{aligned}
\bar{\textbf{Q}}_{\Lambda}(u) &= \bar{Q}_0(u)\, \Biggl\{
\prod_{\blacksquare \in \lambda}  \bar{Q}_{\,\blacksquare}(u) \,
\prod_{ {\square } \in \mathcal{E}} \bar{Q}_{\,{\square}}(u) \, 
\prod_{\tilde{{\square}} \in \tilde{\mathcal{E}} } \bar{Q}_{\tilde{{\square}}}(u)  \Biggr\} \ .
\end{aligned}
\end{equation}
Compare (\ref{rbaransatz}) with the action of $\{r,Q,s\}$ $|\Lambda\rangle$ in (\ref{xansatz}), the difference between $\bar{r}$ and $s$ is in both the poles and the coefficients. And similar for the pair $\bar{s}$ and $r$. 
We will explain this in more detail later in section~\ref{sec:poleextended} and \ref{sec:Qbarcharge}.

Adding these to the set of operators (\ref{operatorfullsetBB}) in the unextended algebra (\ref{GC2bb}), we have the full list of operators for the gluing construction (\ref{GC2})
\begin{equation}\label{operatorfullsetABAB}
\begin{aligned}
\textrm{type }\bm{c}: & \quad \psi, \tilde{\psi}, Q, \qquad \bar{Q}\\
\textrm{type }\bm{s}:& \quad  e, f, \tilde{e}, \tilde{f}\\
\textrm{type }\bm{g}:& \quad  \x, \y,\qquad\qquad  \bar{\x}, \bar{\y}\\
\end{aligned}
\end{equation}
where we have separated the extending operators (\ref{N2ABABfermionnew}) from those in the subalgebra (\ref{GC2bb}).
As before the type $\bm{c}$ are the charge operators, type $\bm{s}$ are the operators that create/annihilate single boxes $\square$ and $\tilde{\square}$, and type $\bm{g}$ are the operators that affect building blocks ${\blacksquare}$.

To construct the extended algebra (\ref{GC2}), with the list of all operators in (\ref{operatorfullsetABAB}), we only need to determine the following additional OPEs:
\begin{itemize}
\item OPEs between $\{\psi, \tilde{\psi}\}$ and $\{\bar{r}, \bar{s}
\}$.
\item OPEs between $\{e, f, \tilde{e}, \tilde{f}\}$ and $\{\bar{r}, \bar{Q}, \bar{s}\}$
\item OPEs among $\{\bar{r}, \bar{Q}, \bar{s}\}$
\item OPEs between $\{r,s\}$ and $\{\bar{r},\bar{s}\}$
\end{itemize}
They are determined together with the charge function $\bar{\textbf{Q}}_{\Lambda}(u)$.

\subsection{OPEs between $\{\psi, \tilde{\psi}\}$ and $\{\bar{r}, \bar{s}\}$}
As shown in \cite{Gaberdiel:2017hcn,Gaberdiel:2018nbs}, for (the ground state of) a representation $\Lambda$, its $\bm{\Psi}$ charge function is related to the $\bm{\Psi}$ charge function of the conjugate $\bar{\Lambda}$ by
\begin{equation}
\bm{\Psi}_{\Lambda}(u)=\psi_0(u) \psi_{\Lambda}(u) \qquad \Longleftrightarrow \qquad \bm{\Psi}_{\bar{\Lambda}}(u)=\psi_0(u) \psi^{-1}_{\Lambda}(-u-\sigma_0\psi_0) \,.
\end{equation}
Since $(\bar{r},\bar{s})$ are the conjugate of $(r,s)$, the OPE relations between the charge operators $(\psi , \tilde{\psi})$ and the gluing operators $(\bar{r},\bar{s})$ can be obtained by transforming those between $(\psi , \tilde{\psi})$ and $(r,s)$ in (\ref{psirs}):
\begin{equation}
\begin{aligned}
\qquad\psi(z) \, \bar{\x}(w)  &\sim  \varphi^{-1}_2(-\Delta-\sigma_3\psi_0) \, \bar{\x}(w)\, \psi(z)  \,, \\
\qquad\tilde{\psi}(z) \, \bar{\x}(w)  &\sim  \tilde\varphi^{-1}_2(-\Delta-\tilde \sigma_3\tilde{\psi}_0) \, \bar{r}(w) \,\tilde{\psi}(z) \,,
\end{aligned}
\end{equation}
and
\begin{equation}\label{psiFBy0} 
\begin{aligned}
\psi(z) \, \bar{\y}(w)  \sim  \varphi_2(-\Delta -\sigma_3 {\psi}_0 ) \,\, \bar{\y}(w) \,\psi(z)\,,  \\
\qquad
\tilde{\psi}(z) \, \bar{\y}(w)  \sim  \tilde\varphi_2(-\Delta-\tilde\sigma_3\tilde{\psi}_0) \, \bar{s}(w) \,\tilde{\psi}(z)\,.
\end{aligned}
\end{equation}

\subsection{Pole structure of the extended algebra}
\label{sec:poleextended}

Now we can analyze the poles in the actions of $\{\bar{r},\bar{s}\}$ on the twin plane partition in (\ref{rbaransatz}).
To use the procedure outlined in (\ref{sec:pole}), we just need to supplement eq.\ (\ref{operatorcharge})
with
\begin{equation}\label{operatorchargebar}
\begin{aligned}
&\psi[\bar{r}](u)=\varphi^{-1}_2(-u-\sigma_3\psi_{0}) \,,\qquad \psi[\bar{s}](u)=\varphi_2(-u-\sigma_3\psi_{0}) \,,\quad \\
&
\tilde{\psi}[\bar{r}](u)=\tilde{\varphi}^{-1}_2(-u-\tilde{\sigma}_3\tilde{\psi}_{0}) \,,\qquad \tilde{\psi}[\bar{s}](u)=\tilde{\varphi}_2(-u-\tilde{\sigma}_3\tilde{\psi}_{0})\,. \\
\end{aligned}
\end{equation}

\subsubsection{$\bar{r}$ removing  $\blacksquare$}

Let's first consider how $\bar{r}(z)$ can annihilate an existing $\blacksquare$ with coordinate function $g(\blacksquare)$, see the second equation in (\ref{rbaransatz}).
The contribution of $\blacksquare$ to the $(\bm{\Psi},\tilde{\bm
{\Psi}})$ charge function is
\begin{equation}\label{bboxcharge}
\blacksquare : \qquad \begin{cases}
\begin{aligned}
\bm{\Psi}(u): &\qquad \varphi_2(u-g(\blacksquare)) \\
\tilde{\bm{\Psi}}(u): &\qquad \tilde{\varphi}_2(u-\tilde{g}(\blacksquare))\ ,
\end{aligned}
\end{cases}
\end{equation}
where the coordinate functions $g(\blacksquare)$ and $\tilde{g}(\blacksquare)$ are defined in (\ref{gcharge}).
Applying $\bar{r}(z)$ at pole $z^*$ contributes
\begin{equation}\label{rbarcharge}
\bar{r}(z) : \qquad \begin{cases}
\begin{aligned}
\bm{\Psi}(u): &\qquad \varphi^{-1}_2(-u+z^{*}-\sigma_3\psi_0)  \\
\tilde{\bm{\Psi}}(u): &\qquad  \tilde{\varphi}^{-1}_2(-u+z^*-\tilde{\sigma}_3\tilde{\psi}_0)  \ .
\end{aligned}
\end{cases}
\end{equation}
Combining (\ref{bboxcharge}) and (\ref{rbarcharge}) we have
\begin{equation}\label{rbarchargefinal}
\bar{r}(z)\blacksquare : \qquad \begin{cases}
\begin{aligned}
\bm{\Psi}(u): &\qquad \varphi_2(u-g(\blacksquare))\, \varphi^{-1}_2(-u+z^{*}-\sigma_3\psi_0)  \\
\tilde{\bm{\Psi}}(u): &\qquad  \tilde{\varphi}_2(u-\tilde{g}(\blacksquare))\,  \tilde{\varphi}^{-1}_2(-u+z^*-\tilde{\sigma}_3\tilde{\psi}_0)  \ .
\end{aligned}
\end{cases}
\end{equation}
We see that in (\ref{rbarchargefinal}), for the $\bm{\Psi}$ charge function to represent an eligible twin plane partition configuration, we need
\begin{equation}\label{nleft}
z^*=z^{*}_{\textrm{left}}=g(\blacksquare)+\sigma_3\psi_0+(\bar{n}_{\textrm{left}}-1)\,  h_2\,,
\end{equation}
where $\bar{n}_{\textrm{left}} \in \mathbb{Z}_{0}$ is the number of boxes left at the left plane partition (after the $\blacksquare$ is removed by $\bar{r}(z)$), at the position 
\begin{equation}
x_1=x_1(\blacksquare)  \,,\qquad x_2=0,1,\cdots, \bar{n}_{\textrm{left}}-1 \,,\qquad x_3=x_3(\blacksquare)\,, 
\end{equation}
where we have use the two identities (\ref{2identities}).
Similarly for the $\tilde{\bm{\Psi}}$ charge function in (\ref{rbarchargefinal}), we need 
\begin{equation}\label{nright}
z^*=z^{*}_{\textrm{right}}=\tilde{g}({\blacksquare})+\tilde{\sigma}_3\tilde{\psi}_0+(\bar{n}_{\textrm{right}} -1)\, \tilde{h}_2\,,
\end{equation}
where $\bar{n}_{\textrm{right}} \in \mathbb{Z}_{0}$ is the number of boxes left at the right plane partition, at the position 
\begin{equation}
\tilde{x}_1=\tilde{x}_1(\blacksquare)=x_3(\blacksquare) \,,\qquad \tilde{x}_2=0,1,\cdots, \bar{n}_{\textrm{right}}-1\,,\qquad \tilde{x}_3=\tilde{x}_3(\blacksquare)=x_1(\blacksquare)\,.
\end{equation}

Combining (\ref{nleft}) and (\ref{nright}), and using (\ref{h2th2}) and (\ref{hthp}), we obtain a constraint 
\begin{equation}\label{nleftrightbound}
n_{\textrm{left}}+n_{\textrm{right}}=B(\blacksquare)+2-(h_1h_3\psi_0+\tilde{h}_1\tilde{h}_3\tilde{\psi}_0)=B(\blacksquare)+2h_{\blacksquare}=
B(\blacksquare)+2+2\rho \,,
\end{equation}
where $B(\blacksquare)$ is the length of the bud required for the creation of $\blacksquare$, computed in (\ref{budlength}), and we have used the conformal weight $h_{\blacksquare}$ of the gluing operators in (\ref{bboxweight}) and the definition of the $\rho$ parameter in (\ref{rhodef}).
Since both $\bar{n}_{\textrm{left}}$ and $\bar{n}_{\textrm{right}}$ are non-negative integers, the bound (\ref{nleftrightbound}) produces the list of all possible poles for the ${\blacksquare}$-removing action of $\bar{r}(z)$ in (\ref{rbarchargefinal}) 
\begin{equation}\label{rbarpoleleft}
z^*=\bar{p}_{-}({\blacksquare})=g(\blacksquare)+\sigma_3\psi_0+(\bar{n}_{\textrm{left}}-1)\ h_2 \quad \textrm{with}\quad \bar{n}_{\textrm{left}}=0,1,\cdots, B(\blacksquare)+2+2\rho\,,
\end{equation}
or equivalently
\begin{equation}\label{rbarpoleright}
z^*=\bar{p}_{-}({\blacksquare})=\tilde{g}(\blacksquare)+\tilde{\sigma}_3\tilde{\psi}_0+(\bar{n}_{\textrm{right}}-1)\ \tilde{h}_2 \quad \textrm{with}\quad \bar{n}_{\textrm{right}}=0,1,\cdots, B(\blacksquare)+ 2+2\rho\,,
\end{equation}
for the ansatz (\ref{rbaransatz}).

Finally, since both $\bar{n}_{\textrm{left}}$ and $\bar{n}_{\textrm{right}}$ are non-negative integers, the constraint (\ref{nleftrightbound}) also implies that both $h_{\blacksquare}$ and the parameter $\rho$ have to be non-negative integers or half-integers.
Earlier in section~\ref{sec:rhoparameter}, we arrived at this constraint by invoking the fact that $h_{\blacksquare}$ corresponds to the conformal dimension of the gluing operators.
Here we see that the twin plane partition analysis also produces this constraint, independent from the conformal dimension argument. 

\subsubsection{$\bar{s}$ adding  $\blacksquare$}

The process of $\bar{s}$ adding $\blacksquare$ is just the inverse of the one for $\bar{r}$ removing $\blacksquare$. 
Recall that for $r$ to add a $\blacksquare$, one need a bud of length $B(\blacksquare)$, defined in (\ref{budlength}), composed of $n_{\textrm{left}}$ number of $\square$'s and $n_{\textrm{right}}$ number of $\tilde{\square}$'s with $n_{\textrm{left}}+n_{\textrm{right}}=B(\blacksquare)$.
Now, for $\bar{s}$ to add a $\blacksquare$, we now need
a bud of length
\begin{equation}
\bar{n}_{\textrm{left}}+\bar{n}_{\textrm{right}}=B(\blacksquare)+2+2\rho\,,
\end{equation}
again with $\bar{n}_{\textrm{left}}$ number of $\square$'s and $\bar{n}_{\textrm{right}}$ number of $\tilde{\square}$'s, distributed between the left and right plane partitions. 
The pole is still given by (\ref{rbarpoleleft}), or equivalently (\ref{rbarpoleright}).

\subsection{OPEs between single-box generators and $[\overline{\square},\overline{\square}]$ gluing operators}

The OPEs between the single-box operators $\{e, f, \tilde{e}, \tilde{f}\}$ and the gluing operator $\{r, Q, s\}$ are evaluated in section~\ref{sec:OPEef-xy:incom} and \ref{sec:OPEQsingle}.
The procedure for deriving those for $\{\bar{r}, \bar{Q}, \bar{s}\}$ is parallel and a bit lengthy. 
We leave the details to Appendix~\ref{sec:antiboxOPEs} and only present the result here.

The OPE relations between $\{e, f, \tilde{e}, \tilde{f}\}$ and $\{\bar{r}, \bar{s}\}$ are in Appendix~\ref{sec:OPEsgbInc} to \ref{sec:OPEsgbC}:
\be
\begin{aligned}
e(z) \, \bar{\x}(w) & \sim  \frac{(\Delta+\sigma_3{\psi}_0-h_2)}{(\Delta+\sigma_3 {\psi}_0-h_2)}  \, \bar{\x}(w) \, e(z) \ ,  
& \quad 
f(z) \, \bar{\x}(w) & \sim \varphi_2^{-1}(-\Delta - \sigma_3{\psi}_0 ) \, \bar{\x}(w) \, f(z) \ , \\
\tilde{e}(z) \, \bar{\x}(w) & \sim \frac{(\Delta+\tilde \sigma_3 \tilde{\psi}_0 -  \tilde{h}_2) }{(\Delta +\tilde \sigma_3 \tilde{\psi}_0 -  \tilde{h}_2) }  \, \bar{\x}(w) \, \tilde{e}(z) \ , 
& \quad
\tilde{f}(z) \, \bar{\x}(w) & \sim\tilde \varphi_2^{-1}(-\Delta - \tilde \sigma_3 \tilde{\psi}_0)   \, \bar{\x}(w) \, \tilde{f}(z) \ ;
\end{aligned}
\ee
and
\be
\begin{aligned}
e(z) \, \bar{\y}(w) & \sim  \varphi_2(-\Delta - \sigma_3{\psi}_0 )   \, \bar{\y}(w) \, e(z) \ ,  
& \quad 
f(z) \, \bar{\y}(w) & \sim \frac{(\Delta+\sigma_3{\psi}_0-h_2)}{(\Delta+\sigma_3 {\psi}_0-h_2)} \, \bar{\y}(w) \, f(z) \ , \\
\tilde{e}(z) \, \bar{\y}(w) & \sim \tilde \varphi_2(-\Delta - \tilde \sigma_3 \tilde{\psi}_0)  \, \bar{\y}(w) \, \tilde{e}(z) \ , 
& \quad
\tilde{f}(z) \, \bar{\y}(w) & \sim  \frac{(\Delta+\tilde \sigma_3 \tilde{\psi}_0 -  \tilde{h}_2) }{(\Delta +\tilde \sigma_3 \tilde{\psi}_0 -  \tilde{h}_2) }   \, \bar{\y}(w) \, \tilde{f}(z) \ .
\end{aligned}
\ee
The OPE relations between $\bar{Q}$ and $\{e, f, \tilde{e}, \tilde{f}\}$ are computed in Appendix~\ref{sec:QbeOPE}:
\begin{equation}
\begin{aligned}
&\bar{Q}(z) \,e(w) \sim \varphi^{-1}_2(\Delta-\sigma_3\psi_0) \, e(w)\,\bar{Q}(z) \,,\qquad  \bar{Q}(z) \,f(w) \sim \varphi_2(\Delta-\sigma_3\psi_0) \, f(w)\,\bar{Q}(z)\,,\\
&\bar{Q}(z) \,\tilde{e}(w) \sim \tilde{\varphi}^{-1}_2(\Delta-\tilde\sigma_3\tilde\psi_0) \, \tilde{e}(w)\,\bar{Q}(z)\,, \qquad \bar{Q}(z) \,\tilde{f}(w) \sim \tilde\varphi_2(\Delta-\tilde\sigma_3\tilde\psi_0) \, \tilde{f}(w)\,\bar{Q}(z)\,.
\end{aligned}
\end{equation}

\subsection{Evaluating $\bar{\textbf{Q}}$ charge function}
\label{sec:Qbarcharge}
We are now ready to compute the $\bar{\textbf{Q}}$ charge function of an arbitrary twin plane partition function.

First of all, the vacuum piece is
\begin{equation}
\bar{Q}_0(u)=(1+\sigma_3\frac{{\psi}_0}{u})(1-\tilde{\sigma}_3\frac{\tilde{\psi}_0}{u})\,.
\end{equation}
Secondly, in the process of computing the  OPE relations between $\bar{Q}$ and $\{e, f, \tilde{e}, \tilde{f}\}$, we have also fixed the single boxes' contributions to the $\bar{\textbf{Q}}$ charge function in section~\ref{sec:Qbchargebox}:
\begin{equation}\label{Qbchargebox}
\bar{Q}_{{\square}}(u)=\varphi^{-1}_2(-u+h({\square})+\sigma_3\psi_0-h_2)
\quad \textrm{and}\quad
\bar{Q}_{\tilde{\square}}(u)=\tilde{\varphi}^{-1}_2(-u+\tilde{h}({\tilde{\square}})+\tilde{\sigma}_3\tilde{\psi}_0-\tilde{h}_2)\,.
\end{equation}
Finally, to compute $\blacksquare$'s contribution to the $\bar{
\textbf{Q}}$ charge function, we imitate the calculation for the $\textbf{Q}$ charge function in section~\ref{sec:blackbforQ}.
We consider how $\bar{s}(w)$ adds a $\blacksquare$, starting from the configuration where all the boxes in the bud are on the left side. 
The result is 
\begin{equation}\label{bbforQbar}
\bar{Q}_{\blacksquare}(u)={S}_p\bigl( u- \bar{p}_{+}(\blacksquare) \bigr) \!\!
	\prod^{B(\blacksquare)+1+2\rho}_{k=0} \varphi^{-1}_2(-u+(g(\blacksquare)+\sigma_3\psi_0-h_2)+k h_2 )\,,
\end{equation} 
where 
\begin{equation}\label{polenegforQb}
\bar{p}_{+}(\blacksquare)=g(\blacksquare)+[B(\blacksquare)+2+2\rho]\, h_2=\tilde{g}(\blacksquare)+(2+2\rho)\,h_2 \,.
\end{equation}
Again, using (\ref{rhodef}), (\ref{polenegforQb}), and the recursion relation (\ref{SpC1}), the function (\ref{bbforQbar}) can be rewritten as
\begin{equation}\label{bbforQbar1}
\bar{Q}_{\blacksquare}(u)={S}_p\bigl( u- (g(\blacksquare)+\sigma_3\psi_0-h_2) \bigr) \!\!
	\prod^{B(\blacksquare)+1+2\rho}_{k=0} \tilde{\varphi}^{-1}_2(-u+(\tilde g(\blacksquare)+\tilde{\sigma}_3\tilde{\psi}_0-\tilde{h}_2)+k\, \tilde{h}_2 )\,,
\end{equation} 
which is what we would have obtained directly if we had started with the initial state where all the boxes in the bud are placed in the right plane partition. 
The identity between (\ref{bbforQbar}) and (\ref{bbforQbar1}) confirms that, for the $\bar{r}$ and $\bar{s}$ actions, the boxes in the same bud can be distributed between the left and right plane partitions.
This is similar to the $r$ and $s$ actions, see the discussion around (\ref{PblacksquareGene}) and (\ref{PblacksquareGene1}).
The contribution (\ref{Qbchargebox}) and (\ref{bbforQbar}) are collected in the last column of Table~\ref{tab2}.

\subsection{OPEs among $\{\bar{r}, \bar{Q}, \bar{s} \}$}

The OPEs among gluing operators $\{r,Q,s\}$ are (\ref{Pxy}) where $S_p$ is given by (\ref{SpSolution}). 
Based on this we can write down the ansatz for the relations for the triplet $\{\bar{r},\bar{Q},\bar{s}\}$ as
\begin{equation}\label{fermionicbardef}
\begin{aligned}
\begin{aligned}
\bar{Q}(z)\, \bar{r}(w)&\sim \bar{S}_p(\Delta) \ \bar{r}(w)\ \bar{Q}(z)\,,\\
\bar{Q}(z)\, \bar{s}(w)&\sim \bar{S}^{-1}_p(\Delta)\ \bar{s}(w)\ \bar{Q}(z)\,,
\end{aligned}
&\qquad
\begin{aligned}
\bar{r}(z)\ \bar{r}(w) &\sim \epsilon_p \,\bar{S}_p(\Delta) \ \bar{r}(w)\ \bar{r}(z)\,,\\
 \bar{s}(z)\ \bar{s}(w) &\sim  \epsilon_p \, \bar{S}^{-1}_p(\Delta)\ \bar{s}(w)\ \bar{s}(z)\,,\\
\end{aligned}\\
[\bar{r}(z), \bar{s}(w)]_{ \epsilon_p } & \sim \frac{\bar{Q}(z)-\bar{Q}(w)}{z-w}\,,
\end{aligned}
\end{equation}
where $\epsilon_p$ is given by (\ref{eq:epsilon-p}) and records the self-statistics of the gluing operators. 
To solve for $\bar{S}_p(u)$, one can repeat the procedure in solving for $S_p(u)$, and the result is:
\begin{equation}\label{SpbarSolution}
	{\bar{S}_p(u) = \frac{u-\delta_p}{u+\delta_p} }
	\qquad\textrm{with}\qquad
	\delta_p = \begin{cases}
	h_1 \qquad & \ (p=1) \\
	 0 \qquad & \ (p=0) \\
	h_3 \qquad & \ (p=-1) \\
	\end{cases} \,.
\end{equation}
We see that 
\begin{equation}
\bar{S}_{p}(u)=S^{-1}_p(u)=S_{p}(-u)\,.
\end{equation}

\subsection{OPEs between $\{r,s\}$ and $\{\bar{r},\bar{s}\}$}

Finally, we fix the OPEs between the $[\square,\square]$ gluing operators $\{r,s\}$ and the $[\overline{\square},\overline{\square}]$ gluing operators $\{\bar{r},\bar{s}\}$.

First, let's consider 
\begin{equation}\label{rbarsAnsatz}
r(z)\,\bar{s}(w)\sim A_1(z-w)\, \bar{s}(w)\, r(z)\,.
\end{equation}
Apply the ansatz (\ref{rbarsAnsatz}) on an initial state $\Lambda$ on which $r(z)$ can add a $\blacksquare$ (labeled by $\blacksquare_1$) and $\bar{s}(w)$ can add a $\blacksquare$ (labeled by $\blacksquare_2$).  
Using the action of $r(z)$ in (\ref{xansatz}) and the action of $\bar{s}(w)$ in (\ref{rbaransatz}), we have
\begin{equation}
A_1(p_+(\blacksquare_1)-\bar{p}_{+}(\blacksquare_2))=\frac{\left[\textrm{Res}_{u=p_{+}(\blacksquare_1)}\left(Q_{\blacksquare_2}(u)Q^{-1}_{\textrm{long bud}(\blacksquare_2)}(u)\right)\right]^{\frac{1}{2}} }{\left[\textrm{Res}_{u=\bar{p}_{+}(\blacksquare_2)}\left(\bar{Q}_{\blacksquare_1}(u)\bar{Q}^{-1}_{\textrm{bud}(\blacksquare_1)}(u)\right)\right]^{\frac{1}{2}}}\,,
\end{equation}
where the $\textrm{long bud}(\blacksquare_2)$ consists of $B(\blacksquare_2)+2+2\rho$ boxes and the $\textrm{bud}(\blacksquare_1)$ consists of $B(\blacksquare_1)$ boxes.
Plugging in the contributions of $\square$, $\tilde{\square}$, and $\blacksquare$ to the $\textbf{Q}$ and $\bar{\textbf{Q}}$ charge functions (see Table (\ref{tab2})), we get
\begin{equation}
A(\Delta)=S_p \bigl( \Delta  + \sigma_3  \psi_0 - h_2 \bigr) 
	\prod_{k=0}^{ 2\rho+1} 
		\tilde\varphi_2\bigl(-\Delta   - \tilde{\sigma}_3 \tilde{\psi}_0  -k  \tilde{h}_2 \bigr) \,.
\end{equation}
Note that this computation is insensitive to the self-statistics of the gluing operators, so we need to insert the function $\epsilon_p$ (defined in (\ref{eq:epsilon-p})) by hand.
The final result for the $r\, \bar{s}$ OPE is 
\begin{equation}
	r(z) \, \bar{s}(w) 
	 \sim 
	{\epsilon_p}\, \Biggl[ S_p \bigl( \Delta  + \sigma_3  \psi_0 - h_2 \bigr) 
	\prod_{k=0}^{ 2\rho+1} 
		\tilde\varphi_2\bigl(-\Delta   - \tilde{\sigma}_3 \tilde{\psi}_0  -k  \tilde{h}_2 \bigr)  \Biggr]
	\   \bar{s}(w) \, r(z) \ , 
\end{equation}
with $\epsilon_p$ given by (\ref{eq:epsilon-p}).
Similarly, we have 
\begin{equation}
	\bar{r}(z)  \, s(w) 
	 \sim 
	{\epsilon_p}\, \Biggl[ S_p \bigl( -\Delta  + \sigma_3  \psi_0 - h_2 \bigr) 
	\prod_{k=0}^{ 2\rho+1} 
		\tilde\varphi_2\bigl(\Delta   - \tilde{\sigma}_3 \tilde{\psi}_0  -k  \tilde{h}_2 \bigr)  \Biggr]
	\ s(w) \, \bar{r}(z)  \ .
\end{equation}

Now we look at 
\begin{equation}\label{rbarrAnsatz}
r(z)\,\bar{r}(w)\sim A_2(z-w)\, \bar{r}(w)\, r(z)\,.
\end{equation}
Repeating the procedure used for the $r\, \bar{s}$ OPE above, we get
\begin{equation}\label{A2one}
A_2(\Delta)\approx1\,,
\end{equation}
where ``$\approx$" means that we have cancelled the common factors in the numerator and the denominator of $A_2$.
Similar to the $f\, r$ OPE, this doesn't mean that $r\, \bar{r}$ OPE is trivial. 
It only means that the factors in the numerator and denominator of $A$ cancel. (See the discussion around eq.\ (\ref{exampleab}) and (\ref{exampleab2}).)
Therefore, the OPE cannot be fixed using the residue formulae alone. Instead, one needs to apply the ansatz (\ref{rbarrAnsatz}) on various initial states and obtain constraints when there are different final states appearing on both sides --- the same method used to partially fix the $e\, r$ OPE etc. 
Applying this procedure and also using (\ref{A2one}), we get
\begin{equation}
\begin{aligned}
	\x(z) \, \bar{\x}(w) &\sim {\epsilon_p}\, \prod_{k=0}^{ 2\rho+2} \frac{(\Delta+ \tilde{\sigma}_3 \tilde{\psi}_0-\tilde{h}_2+k\tilde{h}_2 )}{(\Delta+ \tilde{\sigma}_3 \tilde{\psi}_0-\tilde{h}_2+k\tilde{h}_2)} \, \bar{\x}(w) \, \x(z) \ ,\\
	\y(z) \, \bar{\y}(w)  &\sim {\epsilon_p}\, \prod_{k=0}^{ 2\rho+2} \frac{(\Delta+ \tilde{\sigma}_3 \tilde{\psi}_0-\tilde{h}_2+k\tilde{h}_2 )}{(\Delta+ \tilde{\sigma}_3 \tilde{\psi}_0-\tilde{h}_2+k\tilde{h}_2)} \, \bar{\y}(w) \, \y(z)  \,.
\end{aligned}
\end{equation}
We have now fixed all the OPEs in the extended algebra (\ref{GC2}).

\begin{table}[h!]
\centering
\vspace{.10\textwidth}
 \rotatebox{90}{
 \begin{varwidth}{.89\textheight}
\resizebox{1.15\columnwidth}{!}{
\begin{tabular}{ccccc}
 & $\bm{\Psi}_{\Lambda}(u)$ 
 & $\tilde{\bm{\Psi}}_{\Lambda}(u)$&$\textbf{Q}_{\Lambda}(u)$
 & $\bar{\textbf{Q}}_{\Lambda}(u)$ \vspace{4pt}\\ 
 \hline  \\[-10pt]
 \hbox{vac.} & $\psi_0(u)$ & $\tilde{\psi}_0(u)$ & $\Bigl(1 + \frac{\sigma_3 \psi_0}{u}\Bigr) \, \Bigl( 1 - \frac{\tilde \sigma_3 \tilde{\psi}_0}{u} \Bigr)  $ &$
\Bigl(1 + \frac{\tilde \sigma_3 \tilde{\psi}_0}{u}\Bigr) \, \Bigl( 1 - \frac{\sigma_3 {\psi}_0}{u} \Bigr)  $ \vspace{4pt} \\
 \hline  \\[-10pt]
$\square$ 
& $\varphi_3(u-h(\square))$ 
& $1$
& $\varphi^{-1}_2(-u+h(\square))$ 
& $\varphi^{-1}_2(-u+h(\square)+\sigma_3 \psi_0-h_2)$ \vspace{4pt}\\
\hline  \\[-10pt]
$\tilde{\square}$ 
&  $1$ 
& $\tilde \varphi_3(u-\tilde{h}(\tilde\square))$ 
& $\tilde \varphi^{-1}_2(-u+\tilde{h}(\tilde\square))$  
& $\tilde \varphi^{-1}_2(-u+\tilde{h}(\tilde\square)+\tilde \sigma_3 \tilde{\psi}_0-\tilde{h}_2)$ 
\vspace{4pt} \\
\hline  \\[-10pt]
$\blacksquare$ 
& $\varphi_2(u-g(\blacksquare))$ 
& $\tilde{\varphi}_2(u-\tilde{g}(\blacksquare))$ 
& ${\displaystyle  S_p\bigl(u - g(\blacksquare) \bigr)\!\!  \prod^{B(\blacksquare)-1}_{k=0}\tilde{\varphi}^{-1}_2(-u+\tilde{g}(\blacksquare)+k \tilde{h}_2)}$
& ${\displaystyle {S}_p\bigl( u- (g(\blacksquare)+\sigma_3\psi_0-h_2) \bigr) \!\!
	\prod^{B(\blacksquare)+1+2\rho}_{k=0} \tilde{\varphi}^{-1}_2(-u+(\tilde g(\blacksquare)+\tilde{\sigma}_3\tilde{\psi}_0-\tilde{h}_2)+k\, \tilde{h}_2 )}$ \vspace{4pt} \\
\vspace{4pt} \\ 
\hline
\end{tabular}
}
\caption{The eigenvalues of the different factors.
}\label{tab2}
\end{varwidth}
}
\end{table}

\section{Summary and discussion}

\subsection{Summary}

In this paper we have constructed a two-parameter family of affine Yangian algebras by gluing two affine Yangians of $\mathfrak{gl}_1$ with additional operators that transform as $[\square,\square]$ w.r.t.\ the two affine Yangians of $\mathfrak{gl}_1$. 
The representation space of the glued algebra consists of pairs of plane partitions whose asymptotic boundary conditions along the common direction are correlated as $[\lambda,\lambda]$ or $[\lambda,\lambda^{t}]$.
Demanding that the resulting algebra has sensible actions on these representations then fixes the entire algebra.
Finally, we have also extended the algebra with additional gluing operators that transform as $[\overline{\square},\overline{\square}]$ w.r.t.\ the two affine Yangians of $\mathfrak{gl}_1$.

\subsection{Comparison to the box-antibox construction}

First, we would like to compare the algebras constructed in this paper with the one in \cite{Li:2019nna}.
The extended algebra 
\begin{equation}\label{GCbbabab}
\textrm{box-box extended}:\qquad \mathcal{Y}(q)\,\oplus\, \tilde{\mathcal{Y}}(\tilde{q})\, \oplus\, [\square, \square] \,\oplus\, [\overline{\square}, \overline{\square}]
\end{equation}
(constructed in section~\ref{sec:extended}) is to be compared with the glued algebra in \cite{Li:2019nna}:
\begin{equation}\label{GCbab}
\textrm{box-antibox extended}:\qquad \mathcal{Y}(q)\,\oplus\, \hat{\mathcal{Y}}(\hat{q})\, \oplus\, [\square, \overline{\square}] \,\oplus\, [\overline{\square}, {\square}]
\end{equation}
The algebra  
\begin{equation}\label{GCbb}
\textrm{box-box}:\qquad \mathcal{Y}(q)\,\oplus\, \tilde{\mathcal{Y}}(\tilde{q})\, \oplus\, [\square, \square] \end{equation}
(constructed in section~\ref{sec:3tpp} and \ref{sec:algebra}) is to be compared with the glued algebra
\begin{equation}\label{GCbabsub}
\textrm{box-antibox}:\qquad \mathcal{Y}(q)\,\oplus\, \hat{\mathcal{Y}}(\hat{q})\, \oplus\, [\square, \overline{\square}] 
\end{equation}
which is a subalgebra of (\ref{GCbab}) in \cite{Li:2019nna}.
We have labeled the second affine Yangian algebra in the current paper as $\tilde{\mathcal{Y}}$ with parameter $\tilde{q}$, to distinguish it from the $\hat{\mathcal{Y}}(\hat{q})$ in the previous construction.
Since the algebra in (\ref{GCbb}) and (\ref{GCbabsub}) are subalgebras of (\ref{GCbbabab}) and (\ref{GCbab}), respectively, we can focus on the relation between (\ref{GCbbabab}) and (\ref{GCbab}).
\medskip

Comparing (\ref{GCbbabab}) and (\ref{GCbab}), we see that the main difference between the two is that for all the bimodules, the representation w.r.t.\ the $\tilde{\mathcal{Y}}$ is the conjugate to the representation w.r.t.\ the $\hat{\mathcal{Y}}$.
(We have assumed that the left affine Yangian $\mathcal{Y}(q)$ in (\ref{GCbbabab}) and (\ref{GCbab}) are identical.)
This suggests that $\tilde{\mathcal{Y}}(\tilde{q})$ and $\hat{\mathcal{Y}}(\hat{q})$ are conjugate to each other. 

In terms of the mode $W^{(s)}_n$ of the $\mathcal{W}_{\infty}$ algebra, where $s=\mathbb{N}_{\geq 2}$ and $n\in \mathbb{Z}$, the conjugation map is 
\begin{equation}\label{conjugateW}
W^{(s \textrm{ even})}_n \mapsto W^{(s \textrm{ even})}_n \qquad \textrm{and} \qquad W^{(s \textrm{ odd})}_n \mapsto -W^{(s \textrm{ odd})}_n \,.
\end{equation}
Using the translation between the operators in the affine Yangian of $\mathfrak{gl}_1$ and the modes of $\mathcal{W}_{1+\infty}$ in \cite{Gaberdiel:2017dbk}, one can show that the conjugation map (\ref{conjugateW}) corresponds to
\begin{equation}\label{conjugateY}
h_i \mapsto -h_i \,,\qquad
e_{j}\mapsto (-1)^{j-1}\, e_{j} \,, \qquad \psi_{j}\mapsto (-1)^j\,\psi_{j}\,, \qquad f_{j}\mapsto (-1)^{j-1}\, f_{j}\,,
\end{equation}
which is the automorphism of the affine Yangian of $\mathfrak{gl}_1$ (\ref{automodes}) with $\alpha=-1$.
To see this, we look at the relevant equations in \cite{Gaberdiel:2017dbk}: (3.1) and (3.6) for spin-$1$; (3.4), (3.5), and (3.8) for spin-$2$; (3.10)-(3.14) and (3.24) for spin-$3$; and finally (A.1)-(A.7) for spin-$4$.\footnote{
First, these $W^{(s)}_n$ modes in \cite{Gaberdiel:2017dbk} are related to the modes $V^{(s)}_m$ of $\mathcal{W}_{1+\infty}$ with $s=1,2,3,4$ by a shift of some non-local bilinear terms $\sigma_3\, \tilde{W}^{(s)}_n$: $V^{(s)}_n=W^{(s)}_n+\sigma_3\, \tilde{W}^{(s)}_n$, see (3.19) and (A.19) for spin $3$ and $4$.
Second, to compare to the modes in the $\mathcal{W}_{\infty}$ algebra, one needs to decouple the $\mathfrak{u}(1)$ current $J$, using (3.34), (3.37) and  (3.39) in  \cite{Gaberdiel:2017dbk} for spin $2$, $3$, and $4$. 
Namely it is the $\tilde{V}^{(s)}_m$ modes in \cite{Gaberdiel:2017dbk} that correspond to the $\mathcal{W}_{\infty}$ modes in (\ref{conjugateW}).
Both the non-local bilinear terms and the terms subtracted in the $\mathfrak{u}(1)$ decoupling process have the property that under the map (\ref{conjugateY}), the modes with even spins remain invariant while those with odd spins flip signs.
Finally, it is enough to check these up to spin-$4$ since the $\mathcal{W}_{\infty}$ algebra is uniquely determined by its two parameters, which can be parameterized by the central charge $c$ and the OPE coefficient $c^{4}_{33}$ \cite{Gaberdiel:2012ku, Linshaw:2017tvv}.
Therefore it is enough to examine these equations.}
They indeed have the property that under the transformation (\ref{conjugateY}), the modes with even spins remain invariant while those with odd spins flip signs. 

We can check that the conjugation map (\ref{conjugateY}) is indeed the transformation relating $\tilde{\mathcal{Y}}(\tilde{q})$ and $\hat{\mathcal{Y}}(\hat{q})$.
If we fix the parameters $q$ of the first affine Yangian algebra $\mathcal{Y}$, the parameters $\tilde{q}$ of $\tilde{\mathcal{Y}}$ in (\ref{GCbbabab}) and $\hat{q}$ in $\hat{\mathcal{Y}}$ in (\ref{GCbab}) are correlated.
Recall that in \cite{Li:2019nna}, the relation between the parameters of $\mathcal{Y}(q)$ and $\hat{\mathcal{Y}}(\hat{q})$ are determined by constraints from plane partitions connected by $[\square,\overline{\square}]$ and $[\overline{\square},\square]$ gluing operators, and are given in (\ref{hhbp}).
For the algebra in (\ref{GCbb}), the relation between the parameters of $\mathcal{Y}(q)$ and $\tilde{\mathcal{Y}}(\tilde{q})$ are determined by constraints from plane partitions connected by $[\square,\square]$ gluing operators and are given in (\ref{hthp}). 
For the algebra in (\ref{GCbbabab}), the constraints from additional plane partition pairs connected by $[\overline{\square},\overline{\square}]$ gluing operators impose the same constraints as those for (\ref{GCbb}), therefore the relations are still given by (\ref{hthp}).
Comparing (\ref{hhbp}) and (\ref{hthp}), we see that if we fix the parameters of the first affine Yangian algebra $\mathcal{Y}$, the relation between the parameters of $\tilde{\mathcal{Y}}$ in (\ref{GCbbabab}) and $\hat{\mathcal{Y}}$ in (\ref{GCbab}) is
\begin{equation}\label{hthh}
\tilde{h}_i=-\hat{h}_i \,,
\end{equation}
corroborating (\ref{conjugateY}).

In summary, identifying the left affine Yangian $\mathcal{Y}(q)$ in (\ref{GCbbabab}) and (\ref{GCbab}), one can relate the right affine Yangian $\tilde{\mathcal{Y}}(\tilde{q})$ in (\ref{GCbbabab}) and the right affine Yangian $\hat{\mathcal{Y}}(\hat{q})$ in (\ref{GCbab}) by
\begin{equation}\label{isoM}
\tilde{h}_i=-\hat{h}_i \,,\qquad \tilde{e}_j=(-1)^{j-1}\,\hat{e}_j\,,\qquad \tilde{\psi}_j=(-1)^{j}\,\hat{\psi}_j\,,\qquad \tilde{f}_j=(-1)^{j-1}\,\hat{f}_j\,,
\end{equation}
which can also be written as
\begin{equation}\label{isoF}
\tilde{h}_i=-\hat{h}_i \,,\qquad \tilde{e}(-u)=\hat{e}(u)\,,\qquad \tilde{\psi}(-u)=\hat{\psi}(u)\,,\qquad \tilde{f}(-u)=\hat{f}(u)\,,
\end{equation}
using the (tilde and hatted versions of the) mode expansions (\ref{generating}). 
\bigskip

However, although the right affine Yangian $\tilde{\mathcal{Y}}(\tilde{q})$ in (\ref{GCbbabab}) is isomorphic to the right affine Yangian $\hat{\mathcal{Y}}(\hat{q})$ in (\ref{GCbab}) by the map (\ref{isoF}), this does not mean that the full glued algebra 
(\ref{GCbbabab}) is isomorphic to the one in (\ref{GCbab}).
For (\ref{GCbbabab}) and (\ref{GCbab}) to be isomorphic, all the relations in (\ref{GCbbabab}) have to map to the corresponding ones in (\ref{GCbab}) under the putative isomorphism $\mathcal{I}$.
However, we will now show that it is impossible to achieve this, namely an isomorphism $\mathcal{I}$ between  (\ref{GCbbabab}) and (\ref{GCbab}) does not exist.

First let's focus on the left affine Yangian $\mathcal{Y}(q)$ in both (\ref{GCbbabab}) and (\ref{GCbab}).
The $\mathcal{Y}(q)$ is invariant under $\mathcal{I}$.
In (\ref{GCbbabab}), $\mathcal{Y}(q)$ has non-trivial algebraic relations with the operators $(r,Q,s)$ corresponding to $(\square, \square)$ and the operators $(\bar{r},\bar{Q},\bar{s})$ corresponding to $(\overline{\square}, \overline{\square})$; and in (\ref{GCbab}), $\mathcal{Y}(q)$ has non-trivial algebraic relations with the operators $(x,P,y)$ corresponding to $(\square, \overline{\square})$ and the operators $(\bar{x},\bar{P},\bar{y})$ corresponding to $(\overline{\square}, {\square})$.
Comparing these relations, we see that the triplet $(r,Q,s)$ in (\ref{GCbbabab}) and the triplet $(x,P,y)$ in (\ref{GCbab}) have exactly the same relations with the triplet $(e,\psi, f)$ in $\mathcal{Y}(q)$, and same for $(\bar{r},\bar{Q},\bar{s})$ and $(\bar{x},\bar{P},\bar{y})$.
Therefore, for these relations to correctly map to each other under $\mathcal{I}$, the identification of the operators in (\ref{GCbbabab}) and those in (\ref{GCbab}) need to be
\begin{equation}\label{isoLeft}
\begin{aligned}
r(u)=x(u)\,, \qquad Q(u)=P(u)\,, \qquad s(u)=y(u)\,;\\
\bar{r}(u)=\bar{x}(u)\,, \qquad \bar{Q}(u)=\bar{P}(u)\,, \qquad \bar{s}(u)=\bar{y}(u)\,.\\
\end{aligned}
\end{equation}
This is consistent with the fact that in the comparison between the two glued algebra (\ref{GCbbabab}) and (\ref{GCbab}), not only the two left affine Yangians $\mathcal{Y}(q)$ are identical, also the two pairs of gluing operators transform in the same way w.r.t.\  $\mathcal{Y}(q)$. 

Now let's focus on the right affine Yangians in (\ref{GCbbabab}) and (\ref{GCbab}).
It is enough to check the first line of (\ref{isoLeft}) to see whether it is consistent with the map demanded by considering the right affine Yangians in (\ref{GCbbabab}) and (\ref{GCbab}).

The relations of $(r,Q,y)$ with $(\tilde{e},\tilde{\psi},\tilde{f})$ in $\tilde{\mathcal{Y}}(\tilde{q})$ are given in (\ref{psirs}), (\ref{Qsingle}), (\ref{singlegluingOPE1}), and (\ref{singlegluingOPE2}), which we reproduce here:
\begin{equation}\label{rs-epsif-T}
\begin{aligned}
&\tilde{e}(z) \, r(w)  \sim \tilde{\varphi}_2(\Delta) \, r(w) \, \tilde{e}(z) \ , 
\qquad \,\,\,\tilde{e}(z) \, s(w)  \sim  \frac{\Delta}{\Delta} \, s(w) \, \tilde{e}(z) \ , \\
&\tilde{\psi}(z) \, r(w)  \sim  \tilde\varphi_2(\Delta) \, r(w)\, \tilde{\psi}(z)\,, 
\qquad  \tilde{\psi}(z) \, s(w)  \sim  \tilde\varphi^{-1}_2(\Delta) \, s(w)\, \tilde{\psi}(z)\,, \\
&\tilde{f}(z) \, r(w)  \sim \frac{\Delta}{\Delta} \, r(w) \, \tilde{f}(z) \,, 
\qquad\qquad \tilde{f}(z) \, s(w)  \sim \tilde{\varphi}^{-1}_2(\Delta) \, s(w) \, \tilde{f}(z) \,;
\end{aligned}
\end{equation}
and 
\begin{equation}\label{Q-epsif-T}
\begin{aligned}
\tilde{e}(z) \, Q(w)  &\sim \tilde{\varphi}_2(\Delta) \, Q(w) \, \tilde{e}(z) \,,\\
\tilde{\psi}(z) \, Q(w)  &\sim  Q(w) \, \tilde{\psi}(z) \,,\\
\tilde{f}(z) \, Q(w)  &\sim \tilde{\varphi}^{-1}_2(\Delta) \, Q(w) \, \tilde{f}(z) \,.
\end{aligned}
\end{equation}
On the other hand, the relations of $(x,P,y)$ with $(\hat{e},\hat{\psi},\hat{f})$ in $\hat{\mathcal{Y}}(\hat{q})$ are given in equations (8.1)-(8.4) and (8.7) of \cite{Li:2019nna}:
\begin{equation}\label{xy-epsif-H}
\begin{aligned}
&\hat{e}(z) \, x(w)  \sim  \frac{\Delta+\hat{\sigma}_3 \hat{\psi}_0-\hat{h}_2}{\Delta+\hat{\sigma}_3 \hat{\psi}_0-\hat{h}_2}  
 \, x(w) \,\hat{e}(z) \,,\qquad \quad \,\,  \hat{e}(z) \, y(w)  \sim  \hat{\varphi}_2(-\Delta-\hat{\sigma}_3 \hat{\psi}_0)  \, y(w) \,\hat{e}(z)  \,, \\
 &\hat{\psi}(z) \, x(w)  \sim  \hat\varphi^{-1}_2(-\Delta-\hat \sigma_3\hat{\psi}_0) \, x(w) \,\hat{\psi}(z) \,,  \qquad \hat{\psi}(z) \, y(w)  \sim  \hat\varphi_2(-\Delta-\hat \sigma_3\hat{\psi}_0) \, y(w) \,\hat{\psi}(z) \,,\\
&\hat{f}(z) \, x(w)  \sim  \hat{\varphi}^{-1}_2(-\Delta-\hat{\sigma}_3 \hat{\psi}_0)  \, x(w) \,\hat{f}(z)\,, \qquad  \hat{f}(z) \, y(w)  \sim  \frac{\Delta+\hat{\sigma}_3 \hat{\psi}_0-\hat{h}_2}{\Delta+\hat{\sigma}_3 \hat{\psi}_0-\hat{h}_2}   \, y(w) \,\hat{f}(z) \,;
\end{aligned}
\end{equation}
and 
\begin{equation}\label{P-epsif-H}
\begin{aligned}
\hat{e}(z) \, P(w)  &\sim \hat\varphi_2(-\Delta-\hat \sigma_3\hat{\psi}_0)  \, P(w) \, \hat{e}(z) \,,\\
\hat{\psi}(z) \, P(w)  &\sim  P(w) \, \hat{\psi}(z) \,,\\
\hat{f}(z) \, P(w)  &\sim \hat\varphi^{-1}_2(-\Delta-\hat \sigma_3\hat{\psi}_0)  \, P(w) \, \hat{f}(z) \,.
\end{aligned}
\end{equation}

Since the map from $(\tilde{e},\tilde{\psi},\tilde{f})$ to $(\hat{e},\hat{\psi},\hat{f})$ is given by (\ref{isoF}), for the relations (\ref{rs-epsif-T}) and (\ref{Q-epsif-T}) to map to (\ref{xy-epsif-H}) and (\ref{P-epsif-H}), we need
\begin{equation}\label{rQs-xPy-Right}
r(-u-\tilde{\sigma}_3\tilde{\psi}_0+\tilde{h}_2)=y(u) \,, \quad Q(-u-\tilde{\sigma}_3\tilde{\psi}_0+\tilde{h}_2)=\epsilon_p\,P(u)\,,\quad s(-u-\tilde{\sigma}_3\tilde{\psi}_0+\tilde{h}_2)=x(u)\,.
\end{equation} 
Finally, one can check that under the map (\ref{rQs-xPy-Right}), the relations among $(r,Q,s)$ (see (\ref{Pxy})) are correctly mapped to the relations among $(x,P,y)$ (see (8.8) of \cite{Li:2019nna}).
In particular, the factor $\epsilon_p$ in (\ref{rQs-xPy-Right}) is necessary for the $r-s$ relations to correctly map to the $x-y$ relation.

Now we have reached the self-contradiction that results from the assumption that there is an isomorphism between  (\ref{GCbbabab}) and (\ref{GCbab}).
The first line of (\ref{isoLeft}) is the map from $(r,Q,s)$ to $(x,P,y)$ demanded by the consistency viewed from the left corners of the glued algebras (\ref{GCbbabab}) and (\ref{GCbab}), whereas (\ref{rQs-xPy-Right}) is the same map but derived by demanding the consistency viewed from the right corners of  (\ref{GCbbabab}) and (\ref{GCbab}).
Clearly the two are different and irreconcilable.
As a final check, we reach the same conclusion when we repeat this calculation in terms of modes.
Therefore, we conclude that the algebra (\ref{GCbbabab}) constructed in this paper is not isomorphic to the corresponding one (\ref{GCbab}) constructed in \cite{Li:2019nna}.
By the same argument, as their subalgebras, the algebra (\ref{GCbb}) of this paper cannot be isomorphic to the algebra (\ref{GCbabsub}) in \cite{Li:2019nna}.
\medskip

This thus provides another motivation for studying the construction (\ref{GCbb}) or its extended version (\ref{GCbbabab}). 
The construction (\ref{GCbab}) only covers those algebras whose subalgebras have parameters satisfying (\ref{hhbp}) and (\ref{rhodefreview}).
However, there exist algebras in which the subalgebras do not belong to these classes.
For example, the affine Yangian of $\mathfrak{gl}_2$ can be constructed by gluing two affine Yangians of $\mathfrak{gl}_1$, but the parameters between the two $\mathcal{Y}$'s satisfy (\ref{hthp}) but not (\ref{hhbp}), which requires the construction (\ref{GCbb}) instead of (\ref{GCbab}) \cite{Pietro}.
\medskip

Finally, the fact that the algebra (\ref{GCbab}) of the box-antibox construction in  \cite{Li:2019nna} is not isomorphic to the algebra (\ref{GCbbabab}) of box-box construction in this paper  is reflected in the difference between the actions of the algebra (\ref{GCbab}) in \cite{Gaberdiel:2018nbs, Li:2019nna} on the pair of plane partitions and the actions of the algebra (\ref{GCbb}) or (\ref{GCbbabab}) constructed in the current paper.
The main difference is in the action of the gluing operators, in particular, in their bud conditions.
For example, for the algebra (\ref{GCbab}), the bud needs to have a minimal length, and all the boxes in it need to be grouped together into one bud; whereas for the algebra (\ref{GCbb}) or (\ref{GCbbabab}), the bud needs to have a precise length, but the boxes in it can be distributed between the left and right plane partitions. 
This results in the difference between the pole structures of the  gluing operators' actions for the algebra (\ref{GCbb}) or (\ref{GCbbabab}) and the one for (\ref{GCbab}).
Another interesting difference is that in (\ref{GCbab}), the action of a gluing operator might shift a box in the left plane partition to the right side, and vice versa. 
This can never happen in (\ref{GCbb}) or (\ref{GCbbabab}).
Given these microscopic differences in the mechanisms in which the algebras are fixed by their actions on the twin plane partitions, it is quite remarkable that the final results on the algebraic relations of these two different constructions (the one in \cite{Li:2019nna} and the current one) look so similar in structure.
\bigskip

\subsection{Future directions}

Given their close relation to the toric Calabi-Yau threefolds, it is natural to expect the glued algebras (\ref{GCbb}), (\ref{GCbbabab}), and (\ref{GCbab}) to serve as symmetry algebras of some physical systems from the toric threefolds, such as topological strings or the systems in \cite{Prochazka:2017qum}.
The microscopic difference between the actions (on twin plane partitions) of the algebra (\ref{GCbab}) constructed in \cite{Gaberdiel:2018nbs, Li:2019nna} and that of (\ref{GCbb}) and (\ref{GCbbabab}) constructed in the current paper then raises the question as to which algebra applies in a particular situation. 
The algebra (\ref{GCbb}) constructed in the current paper has the advantage of having representations that are geometrically simpler, i.e.\ just the gluing of two (complements of) crystals  --- namely, no ``high wall" is necessary. 
Hence it might be easier to relate to topological strings. 
It would be good to find systems in which one can match in detail the microscopic actions by the glued algebras on twin plane partitions with the more familiar physical or mathematical processes.
This certainly deserves future study.
 
In addition, it is possible to extend the procedure in this paper and construct bigger affine Yangian algebras using the affine Yangian of $\mathfrak{gl}_1$ as building blocks, which would correspond to toric diagrams with more than two vertices.
Take $n$ copies of the affine Yangian of $\mathfrak{gl}_1$, with parameters $q_i$ with $i=1,\cdots, n$. 
By definition, they are mutually commuting.
Fix a subset of pairs of $\mathcal{Y}_i$, and for each pair $<i j>$ introduce a set of gluing operators $\mathbf{g}_{i j}$ that interacts both with $ \mathcal{Y}_i$ and with $ \mathcal{Y}_j$: 
\begin{equation}\label{gluegeneral}
\bigoplus_i \mathcal{Y}_i \oplus \bigoplus_{<ij>}\mathbf{g}_{ij}\cdots\,,
\end{equation}
and the $\cdots$ denotes possible additional operators that interact with a cluster of $\mathcal{Y}_i$, e.g. $\mathbf{g}_{ijk}$.
By the isomorphism (\ref{iso1}), the construction (\ref{gluegeneral}) would produce a vertex operator algebra that contains $n$ (mutually commuting) $\mathcal{W}_{1+\infty}$ subalgebras, with parameters $(c,\lambda)_i$ (that corresponds to the $q_i$), and additional gluing operators that have non-trivial OPEs with pairs of $\mathcal{W}_{1+\infty}$ algebras.
For example, to construct the affine Yangian that corresponds to the large $\mathcal{N}=4$ supersymmetric $\mathcal{W}_{\infty}$ algebra, one would need to glue $4$ affine Yangians of $\mathfrak{gl}_1$ in a linear diagram.

Finally, for toric diagrams with loops, we need to generalize our techniques. 
Recall that for a toric diagram with two vertices (shown in Figure \ref{fig:room}), the condition (\ref{T4}) that the external lines cannot intersect --- namely, they cannot join and thus create the third vertex --- correspond precisely to the plane partition constraint that the bud has non-negative length (\ref{B4}), i.e.\ the boxes in the initial configuration have to all sit inside the room.
To construct the glued algebra for a toric diagram with loops, we need to relax this condition, i.e.\ $|p|$ can now be greater than two for some of the vertices. 
What condition can then replace (\ref{B4})? 
Can the representation space still be characterized as elegantly as the current twin plane partition? 
We leave these questions for future work. 

\subsection*{Acknowledgements}
We thank Pietro Longhi for initial collaboration on this project and for many helpful discussions.
We also thank Matthias Gaberdiel for helpful discussions, and for the hospitality of ETH Zurich, MPI-AEI, and Perimeter Institute during various stages of this project. 
Finally we are grateful for support from NSFC 11875064, the Max-Planck Partnergruppen fund, and the Simons Foundation through the Simons Foundation Emmy Noether Fellows Program at Perimeter Institute.

\appendix

\section{Deriving OPEs between 
single-box generators and $[\overline{\square},\overline{\square}]$ gluing operators
}
\label{sec:antiboxOPEs}

The OPEs between the single-box generators $\{e, f, \tilde{e}, \tilde{f}\}$ and the  $[\overline{\square},\overline{\square}]$ gluing operators $\{\bar{r}, \bar{Q}, \bar{s}\}$ can be computed by the same procedure used in deriving those for the $[{\square},{\square}]$ gluing operators $\{{r}, {Q}, {s}\}$.

\subsection{OPEs between single-box generators and $\{\bar{r},\bar{s}\}$: incomplete}
\label{sec:OPEsgbInc}

There are $8$ OPE relations between $\{e, f, \tilde{e}, \tilde{f}\}$ and $\{\bar{r},\bar{s}\}$, of which only two are independent:
\begin{equation}\label{2OPEbar}
\begin{aligned}
e(z) \, \bar{r}(w) &\sim \bar{G}(\Delta) \, \bar{r}(w) \, e(z) \qquad\textrm{and} \qquad f(z) \, \bar{r}(w) \sim \bar{H}(\Delta) \, \bar{r}(w) \, f(z)\,.
\end{aligned}
\end{equation}
The inverse of the two OPEs (\ref{2OPEbar}) are obtained by
\begin{equation}
(e,f, \bar{r})\rightarrow (f,e,\bar{s})\quad \Longrightarrow \quad(\bar{G},\bar{H})\rightarrow(\bar{G}^{-1},\bar{H}^{-1})\,.
\end{equation} 
The tilde versions of the $4$ OPEs above are obtained by 
\begin{equation}
(e,f)\rightarrow (\tilde{e},\tilde{f})\quad \Longrightarrow \quad(\bar{G},\bar{H})\rightarrow(\tilde{\bar{G}},\tilde{\bar{H}})\,.
\end{equation} 
Thus, we only need to fix the two OPEs in (\ref{2OPEbar}).

As in the derivation of $e\cdot r$ and $f\cdot x$ OPEs, we apply the ansatz (\ref{2OPEbar}) on various initial twin plane partition configurations.
For each initial state, demanding that the final states from the two sides of the ansatz coincide then produces constraints on the rational functions $\bar{G}(\Delta)$ and $\bar{H}(\Delta)$.

\subsubsection{$e\cdot \bar{\x}$ OPE and $\tilde{e}\cdot \bar{\x}$ OPE}
First, apply the ansatz
\begin{equation}\label{ebxOPE}
e(z)\,\bar{r}(w)\sim \bar{G}(\Delta)\, \bar{\x}(w)\,e(z)
\end{equation}
on the initial state with one single $\blacksquare$: $|\Lambda\rangle=|\blacksquare\rangle$.
For the l.h.s., there are two scenarios. 
First, after $\bar{r}$ removes the $\blacksquare$ and leaves some boxes behind, $e$ can add a $\square$ next to those boxes in the left plane partition along the $x_1$ or $x_3$ direction: 
\begin{equation}\label{erbarlhs1}
w=\sigma_3 \psi_0-h_2 +\bar{n}_{\textrm{left}} \, h_2 \qquad \textrm{and}\qquad z=h_1 \,\, \textrm{or}\,\, h_3\,,
\end{equation}
where we have used the pole of the $\bar{r}$ action in eq.\ (\ref{rbarpoleleft}) and that $g(\blacksquare)=0$ for the first $\blacksquare$.
The range of $\bar{n}_{\textrm{left}}$ has to start from $1$ instead of $0$:
\begin{equation}\label{nbarleftrange}
\bar{n}_{\textrm{left}}=1,\cdots,2+2\rho\,,
\end{equation}
in order for the $e$ action to happen. 
The second choice is that $e$ can add  a $\square$ next to those boxes in the left plane partition along the $x_2$ direction: 
\begin{equation}\label{erbarlhs2}
w=\sigma_3 \psi_0-h_2 +\bar{n}_{\textrm{left}}\,  h_2  \qquad \textrm{and}\qquad z=\bar{n}_{\textrm{left}} \, h_2
\end{equation}
with 
\begin{equation}\label{nbarleftrange0}
\bar{n}_{\textrm{left}}=0,\cdots,2+2\rho\,.
\end{equation}

For the r.h.s., there is only one scenario: $e$ adds a $\square$ next to the initial $\blacksquare$ along the $x_1$ or $x_3$ direction, and then $\bar{r}$ removes the $\blacksquare$ and leaves some boxes behind. 
The final states are identical to those in the first scenario of the l.h.s.\ in (\ref{erbarlhs1}) with (\ref{nbarleftrange}).

Therefore, $\bar{G}(\Delta)$'s denominator must contain a factor $\Delta+\sigma_3\psi_0-h_2$, in order to cancel those states in (\ref{erbarlhs2}), which only appear on the l.h.s.\ but not on the r.h.s..
And since $\bar{G}$ is homogenous, the minimal result for $\bar{G}$ is
\begin{equation}\label{Gbpartial}
\bar{G}(\Delta)=\frac{\Delta+b}{\Delta+\sigma_3\psi_0-h_2}\,,
\end{equation}
where $b$ is a constant to be fixed later.

\subsubsection{$f\cdot \bar{\x}$ OPE and $\tilde{f}\cdot \bar{\x}$ OPE}

First, apply the ansatz 
\begin{equation}\label{frbOPE}
f(z) \, \bar{r}(w) \sim \bar{H}(\Delta) \, \bar{r}(w) \, f(z)
\end{equation}
on the initial state with one single $\blacksquare$: $|\Lambda\rangle=|\blacksquare\rangle$.
For the l.h.s., applying $\bar{r}(w)$ and then $f(z)$ generates
$B(\blacksquare)=2+2\rho$ states: $\bar{n}_{\textrm{left}}-1$ number of $\square$'s lined up along the $x_2$ direction at  $(x_1,x_3)=(0,0)$ and $\bar{n}_{\textrm{right}}=2+2\rho-n_{\textrm{left}}$ number of $\tilde{\square}$'s lined up along the $\tilde{x}_2$ direction at  $(\tilde{x}_1,\tilde{x}_3)=(0,0)$, created by the action of $\bar{r}(w)$ followed by $f(z)$, with poles
\begin{equation}\label{wzpole}
w=\sigma_3\psi_0+(\bar{n}_{\textrm{left}}-1)\, h_2 \qquad \textrm{and}\qquad z=(\bar{n}_{\textrm{left}}-1)\, h_2\,,
\end{equation}
where we have used the pole of the $\bar{r}$ action in eq.\ (\ref{rbarpoleleft}) and that $g(\blacksquare)=0$ for the first $\blacksquare$.
The range of $\bar{n}_{\textrm{left}}$ is again given by (\ref{nbarleftrange}). 
Note that $\bar{n}_{\textrm{left}}$ has to start from $1$ instead of $0$, in order for the $\bar{f}$ to be able to remove a $\square$.
On the other hand, the r.h.s.\ vanishes since $f(z)|\blacksquare\rangle=0$.
Taking the difference between the two poles in (\ref{wzpole}), we see that $\bar{H}(\Delta)$ has to contain a factor of $\Delta+\sigma_3\psi_0$ in the denominator, in order to remove these $2+2\rho$ states that only appear on the l.h.s. but not on the r.h.s..

Next, consider the initial state $|\blacksquare+\square_i\rangle$, where the additional $\square$ is at the position $(x_1,x_2,x_3)=(1,0,0)$ for $i=1$ or $(0,0,1)$ for $i=3$. 
Now for the l.h.s.,  there are two scenarios. 
In the first case, $f(z)$ removes the initial $\square$.
\begin{equation}\label{wzpoleNl1}
w=\sigma_3\psi_0+(\bar{n}_{\textrm{left}}-1)\, h_2 \qquad \textrm{and}\qquad z=h_i\,,
\end{equation}
with the range of $\bar{n}_{\textrm{left}}$ still given by (\ref{nbarleftrange}).
Note that here $\bar{n}_{\textrm{left}}$ has to start from $1$ instead of $0$, in order for the initial $\square$ not to ``float" in the plane partition. 
The second possibility is that $f(z)$ doesn't remove the initial $\square$ but removes the box furthest to the right of 
the $\bar{n}_{\textrm{left}}$ $\square$'s:
\begin{equation}\label{wzpoleNl2}
w=\sigma_3\psi_0+(\bar{n}_{\textrm{left}}-1)\, h_2 \qquad \textrm{and}\qquad z=(\bar{n}_{\textrm{left}}-1)\, h_2\,,
\end{equation}
where the range of $\bar{n}_{\textrm{left}}$ is again given by (\ref{nbarleftrange}).
On the other hand, for the r.h.s., the non-vanishing states have   
\begin{equation}\label{zwpoleNr}
z=h_i \qquad \textrm{and} \qquad w=\sigma_3\psi_0+(\bar{n}_{\textrm{left}}-1)\, h_2 \,,
\end{equation}
with $\bar{n}_{\textrm{left}}$ given by (\ref{nbarleftrange0}).

Now we compare the two sides.
The $2+2\rho$ states in the second scenario (\ref{wzpoleNl2}) only appear on the l.h.s.\ but not on the r.h.s..
However, they are already taken care of by the factor $\Delta+\sigma_3\psi_0$ in the denominator of $\bar{H}(\Delta)$ that was fixed just now.
Now compare the first scenario (\ref{wzpoleNl1}) for the l.h.s.\ with the states (\ref{zwpoleNr}) for the r.h.s.
There is only one state that appears only one on side: namely, (\ref{zwpoleNr}) with $n_{\textrm{left}}=0$ for the r.h.s.
To cancel this state, we need a factor $\Delta+\sigma_3\psi_0-h_2-h_i$ in the numerator of $\bar{H}(\Delta)$.
Since $i=1$ or $i=3$ are on equal footings, we need  $(\Delta+\sigma_3\psi_0+h_1)(\Delta+\sigma_3\psi_0+h_1)  $.

To summarize, this analysis shows that 
\begin{equation}
	\bar{H}(\Delta) \sim \frac{ (\Delta + \sigma_3\psi_0 +   h_1) (\Delta + \sigma_3\psi_0 +   h_3)}{(\Delta + \sigma_3\psi_0)}\ . 
\end{equation}
Using the fact that $\bar{H}(\Delta)$ is homogeneous and $\bar{G}(\Delta)\bar{H}(\Delta)=\varphi_2^{-1}(-\Delta-\sigma_3\psi_0)$ where $\bar{G}(\Delta)$ is partially fixed in (\ref{Gbpartial}), we have 
\begin{equation}\label{Hbarincomplete}
	\bar{H}(\Delta) = \frac{ (\Delta + \sigma_3\psi_0 +   h_1) (\Delta + \sigma_3\psi_0 +   h_3)}{(\Delta + \sigma_3\psi_0)(\Delta + b)}\ . 
\end{equation}

As in the cases of $G$ and $H$ in (\ref{eq:e-x-OPE-coeff-incomplete}) and (\ref{Hincomplete}), there is one constant $b$ that cannot be fixed.
It will be determined in the next subsection, together with the single boxes' contributions to the $ \bar{\textbf{Q}}$ charge function.

\subsection{Single boxes' contribution to $\bar{\textbf{Q}}$ charge function }
\label{sec:Qbchargebox}

The single boxes' contribution to the $\bar{\textbf{Q}}$ charge function are related to the following two OPEs between single-box annihilators and gluing operators
\begin{equation}\label{exOPE4}
\begin{aligned}
\bar{Q}_{\,{\square}}(u):&\qquad &f(z) \, \bar{\x}(w) \sim \bar{H}(\Delta) \, \bar{r}(w) \, f(z)\,, \\
\textrm{and}\quad \bar{Q}_{\,{\tilde{\square}}}(u):&\qquad &\tilde{f}(z) \, \bar{\x}(w) \sim \tilde{\bar{H}}(\Delta) \, \bar{r}(w) \, \tilde{f}(z)\,, \\
\end{aligned}
\end{equation}
respectively.
Again we only need to study the first one, since the second one is the tilde version of the first.

For the first line in (\ref{exOPE4}), applying the $f\cdot\bar{\x}$ OPE on a generic twin plane partition $|{\Lambda}\rangle$ gives\footnote{
For this computation, it is most convenient to choose the figuration in which all the boxes in the bud are in the right plane partition, so as to have directly $\bm{\Psi}_{\Lambda-{\blacksquare}}(u)$. 
However, the final result doesn't depend on the choice of the initial configuration.}
\begin{equation}
\begin{aligned}
	&
	{
	\Big[ +  \frac{1}{\sigma_3} {\rm Res}_{u =  h( \square)} { \bm{\Psi}}_{\Lambda-{\blacksquare}}(u) \Big]^{\frac{1}{2}}
	}
	{
		\Big[ \textrm{Res}_{u=\bar{p}_{-}({{\blacksquare}})} \, \bar{\textbf{\P}}_\Lambda(u) \Big]^{\frac{1}{2}} 
	}
	\\
	&\qquad =
	\bar{H}( h(\square) -\bar{p}_-({\blacksquare})) \ 
	{
		\Big[ \textrm{Res}_{u=\bar{p}_{-}({\blacksquare})} \, \bar{\textbf{Q}}_{\Lambda - \square}(u) \Big]^{\frac{1}{2}} 
	}
	{
		\Big[ +  \frac{1}{ \sigma_3} {\rm Res}_{u =  h( \square)}  {\bm{\Psi}}_{\Lambda}(u) \Big]^{\frac{1}{2}}
	} \,,
\end{aligned}
\end{equation}
which gives
\begin{equation}\label{Pfromfxbar}
\begin{aligned}
	\bar{Q}^{1/2}_{\square}(\bar{p}_-({\blacksquare}) ) 
		& = \bar{H}( h( \square )  - \bar{p}_-({\blacksquare}) ) \ \Big[  \psi_{{\blacksquare}}(h( {\square} ) ) \Big]^{1/2} \,.
\end{aligned}
\end{equation}
Namely, $\bar{Q}_{\square}(u)$ and $\bar{H}(\Delta)$ contain the same information. 
Similarly for the tilde version.

Recall that the $\bar{H}(\Delta)$ was fixed by the pole structures of the $f$ and $\bar{\x}$ action up to one constant $b$ in (\ref{Hbarincomplete}).
Taking the result of $\psi_{{\blacksquare}}$ in (\ref{chargebbox}), we see the most natural solution for (\ref{Pfromfxbar}) is
\begin{equation}
b=\sigma_3\psi_0-h_2
\end{equation}
in  (\ref{Hbarincomplete}), which would give
\begin{equation}\label{Pbox1b}
\bar{Q}_{{\square}}(u)=\varphi^{-1}_2(-u+h({\square})+\sigma_3\psi_0-h_2)
\qquad \textrm{and}\qquad
\bar{H}(\Delta)=\varphi^{-1}_2(-\Delta-\sigma_3\psi_0)\,.
\end{equation}
Similarly for the tilde version
\begin{equation}
\bar{Q}_{\tilde{\square}}(u)=\tilde{\varphi}^{-1}_2(-u+\tilde{h}({\tilde{\square}})+\tilde{\sigma}_3\tilde{\psi}_0-\tilde{h}_2)\qquad \textrm{and}\qquad
\tilde{\bar{H}}(\Delta)=\tilde{\varphi}^{-1}_2(-\Delta-\tilde{\sigma}_3\tilde{\psi}_0)\,.
\end{equation}
This also fixes the $\bar{G}$ and $\tilde{\bar{G}}$:
\begin{equation}\label{GbGtb}
\begin{aligned}
\bar{G}(\Delta)= \frac{(\Delta+\sigma_3{\psi}_0-h_2)}{(\Delta+\sigma_3 {\psi}_0-h_2)} \qquad \textrm{and} \qquad  \tilde{\bar{G}}(\Delta)= \frac{(\Delta+\tilde \sigma_3 \tilde{\psi}_0 -  \tilde{h}_2) }{(\Delta +\tilde \sigma_3 \tilde{\psi}_0 -  \tilde{h}_2) }  \,.
\end{aligned}
\end{equation}

\subsection{OPEs between single-box generators and $\{\bar{r}, \bar{s}\}$: final}
\label{sec:OPEsgbC}
Plugging (\ref{Pbox1b})-(\ref{GbGtb}) into the ansatz (\ref{2OPEbar}) and their tilde versions, we get the OPEs between the single-box generators $\{e,f,\tilde{e},\tilde{f}\}$ and the $[\overline{\square},\overline{\square}]$ gluing operators $\bar{r}$ and $\bar{s}$:
\begin{equation}
\begin{aligned}
e(z) \, \bar{\x}(w) & \sim  \frac{(\Delta+\sigma_3{\psi}_0-h_2)}{(\Delta+\sigma_3 {\psi}_0-h_2)}  \, \bar{\x}(w) \, e(z) \ ,  
& \quad 
f(z) \, \bar{\x}(w) & \sim \varphi_2^{-1}(-\Delta - \sigma_3{\psi}_0 ) \, \bar{\x}(w) \, f(z) \ , \\
\tilde{e}(z) \, \bar{\x}(w) & \sim \frac{(\Delta+\tilde \sigma_3 \tilde{\psi}_0 -  \tilde{h}_2) }{(\Delta +\tilde \sigma_3 \tilde{\psi}_0 -  \tilde{h}_2) }  \, \bar{\x}(w) \, \tilde{e}(z) \ , 
& \quad
\tilde{f}(z) \, \bar{\x}(w) & \sim\tilde \varphi_2^{-1}(-\Delta - \tilde \sigma_3 \tilde{\psi}_0)   \, \bar{\x}(w) \, \tilde{f}(z) \ ;
\end{aligned}
\end{equation}
and
\be
\begin{aligned}
e(z) \, \bar{\y}(w) & \sim  \varphi_2(-\Delta - \sigma_3{\psi}_0 )   \, \bar{\y}(w) \, e(z) \ ,  
& \quad 
f(z) \, \bar{\y}(w) & \sim \frac{(\Delta+\sigma_3{\psi}_0-h_2)}{(\Delta+\sigma_3 {\psi}_0-h_2)} \, \bar{\y}(w) \, f(z) \ , \\
\tilde{e}(z) \, \bar{\y}(w) & \sim \tilde \varphi_2(-\Delta - \tilde \sigma_3 \tilde{\psi}_0)  \, \bar{\y}(w) \, \tilde{e}(z) \ , 
& \quad
\tilde{f}(z) \, \bar{\y}(w) & \sim  \frac{(\Delta+\tilde \sigma_3 \tilde{\psi}_0 -  \tilde{h}_2) }{(\Delta +\tilde \sigma_3 \tilde{\psi}_0 -  \tilde{h}_2) }   \, \bar{\y}(w) \, \tilde{f}(z) \ .
\end{aligned}
\ee

\subsection{OPEs between single-box operators 
and $\bar{Q}$}
\label{sec:QbeOPE}

The results on the single boxes' contributions to the $\bar{\textbf{Q}}$ charge function immediately give us the OPEs between $\bar{Q}$ and the four single-box generators $\{\e,f,\tilde{e},\tilde{f}\}$:
\begin{equation}
\begin{aligned}
&\bar{Q}(z) \,e(w) \sim \varphi^{-1}_2(\Delta-\sigma_3\psi_0) \, e(w)\,\bar{Q}(z)\,, \qquad  \bar{Q}(z) \,f(w) \sim \varphi_2(\Delta-\sigma_3\psi_0) \, f(w)\,\bar{Q}(z)\,,\\
&\bar{Q}(z) \,\tilde{e}(w) \sim \tilde{\varphi}^{-1}_2(\Delta-\tilde\sigma_3\tilde\psi_0) \, \tilde{e}(w)\,\bar{Q}(z) \,,\qquad \bar{Q}(z) \,\tilde{f}(w) \sim \tilde\varphi_2(\Delta-\tilde\sigma_3\tilde\psi_0) \, \tilde{f}(w)\,\bar{Q}(z)\,.
\end{aligned}
\end{equation}

The derivation is parallel to the one for the $\psi \cdot e$ OPE in section~\ref{sec:psiOPEs}. 
Take the $Q\cdot e$ OPE for example. 
Applying  
\begin{equation}\label{Qbe}
\bar{Q}(z)\, e(w)\sim \bar{K}(z-w) \,e(w)\,  \bar{Q}(z)
\end{equation} 
 on an arbitrary twin plane partition $|\Lambda \rangle$ gives
\begin{equation}
\begin{aligned}
&\qquad\qquad \qquad \sum_{\square \in \textrm{Add}(\Lambda)} \frac{E(\Lambda\rightarrow \Lambda+\square)}{w-h(\square)} \, \bar{\textbf{Q}}_{\Lambda+\square}(z) \, |\Lambda+\square\rangle \\
&\qquad 
= \sum_{\square \in \textrm{Add}(\Lambda)} \bar{K}(z-w) \frac{E(\Lambda\rightarrow \Lambda+\square)}{w-h(\square)} \, \bar{\textbf{Q}}_{\Lambda}(z)\, |\Lambda+\square\rangle\,.
\end{aligned}
\end{equation}
Since this is a vector equation, it needs to hold for each final state $|\Lambda+\square\rangle
$, which gives
\begin{equation}
\bar{K}(z-h(\square))=\frac{\bar{\textbf{Q}}_{\Lambda+\square}(z)}{\bar{\textbf{Q}}_{\Lambda}(z)}=\bar{Q}_{\square}(z)=\varphi^{-1}_2(z-h(\square)-\sigma_3\psi_0)\,,
\end{equation}
where in the last step we have used the result for $\bar{Q}_{\square}(z)$, see (\ref{Pbox1b}). 
This fixes the  OPE (\ref{Qbe}) to be
\begin{equation}
\bar{Q}(z) \,e(w) \sim \varphi^{-1}_2(\Delta-\sigma_3\psi_0) \,e(w)\,\bar{Q}(z)\,.
\end{equation}
The other three cases are completely parallel.

\bibliographystyle{utphys}
\bibliography{yangian}

\end{document}